%% file: main.tex
\documentclass[journal=chreay,manuscript=review,etalmode=truncate,maxauthors=10]{achemso}

\usepackage[version=3]{mhchem} 

\usepackage{lipsum} 
\usepackage{color} 
\floatstyle{plaintop}
\restylefloat{scheme}
\floatstyle{plain}
\usepackage{graphicx}
\usepackage{amsmath}
\usepackage{amssymb}
\usepackage{amsthm}
\usepackage{multirow}
\usepackage{xcolor} 
\usepackage{booktabs}
\usepackage{algorithm}
\usepackage{algorithmic}
\usepackage{dblfloatfix}
\usepackage{tikz,lipsum,lmodern}
\usepackage[most]{tcolorbox}
\usepackage{rotating}

\usepackage{array}
\usepackage{rotating}
\usepackage{pifont}
\usepackage{bbding}
\usepackage{multirow}
\usepackage{wasysym}
\usepackage{booktabs}
\usepackage{longtable}
\usepackage{adjustbox}
\usepackage{array}

\newcolumntype{R}[2]{%
    >{\adjustbox{angle=#1,lap=\width-(#2)}\bgroup}%
    l%
    <{\egroup}%
}

\newcommand\numberthis{\addtocounter{equation}{1}\tag{\theequation}}

\newenvironment{packed_enum}{
\begin{enumerate}
  \setlength{\itemsep}{1pt}
  \setlength{\parskip}{0pt}
  \setlength{\parsep}{0pt}
}{\end{enumerate}}

\usepackage{tikz}
\newcommand*\circled[1]{\tikz[baseline=(char.base)]{
\node[shape=circle,draw,inner sep=2pt] (char) {#1};}}

\author{John A. Keith}
\email{jakeith@pitt.edu}
\affiliation[Pitt]{Department of Chemical and Petroleum Engineering
Swanson School of Engineering, University of Pittsburgh, Pittsburgh, USA}

\author{Valentin Vassilev-Galindo}
\affiliation[Lux]{Department of Physics and Materials Science, University of Luxembourg, L-1511 Luxembourg City, Luxembourg}

\author{Bingqing Cheng}
\affiliation[Camb]{Accelerate Programme for Scientific Discovery, 
Department of Computer Science and Technology, 15 J.~J.~Thomson Avenue, Cambridge CB3 0FD, United Kingdom, and Cavendish Laboratory, University of Cambridge, J.~J.~Thomson Avenue, Cambridge CB3 0HE, United Kingdom}

\author{Stefan Chmiela}
\affiliation[TUBerlin]{Department of Software Engineering and Theoretical Computer Science, Technische Universit\"at Berlin, 10587, Berlin, Germany}

\author{Michael Gastegger}
\affiliation[TUBerlin]{Department of Software Engineering and Theoretical Computer Science, Technische Universit\"at Berlin, 10587, Berlin, Germany}

\author{Klaus-Robert M\"{u}ller}
\email{klaus-robert.mueller@tu-berlin.de}
\affiliation[TUBerlin]{Machine Learning Group, Technische Universit\"at Berlin, 10587, Berlin, Germany; Department of Artificial Intelligence, Korea University, Anam-dong, Seongbuk-gu, Seoul, 02841, Korea; Max-Planck-Institut f\"{u}r Informatik, Saarbr\"{u}cken, Germany; Google Research, Brain team, Berlin, Germany}

\author{Alexandre Tkatchenko}
\email{alexandre.tkatchenko@uni.lu}
\affiliation[Lux]{Department of Physics and Materials Science, University of Luxembourg, L-1511 Luxembourg City, Luxembourg}

\title{Combining Machine Learning and Computational Chemistry for Predictive Insights Into Chemical Systems} 

\begin{document}

\input{Sections/0-topMatter}


\tableofcontents

\input{Sections/1-intro}
\input{Sections/2-compChem}
\input{Sections/3-ML-sc}
\input{Sections/4-whyML4Chem}

\input{Sections/5-applications}

\input{Sections/6-conclusions}

\input{Sections/7-acknowledgement}
\input{Sections/8-bios}

 \typeout{}

\footnotesize
\bibliography{all_references}

\end{document}

%% file: Sections/0-topMatter.tex
\begin{tocentry}
\centering
\includegraphics[width=5cm]{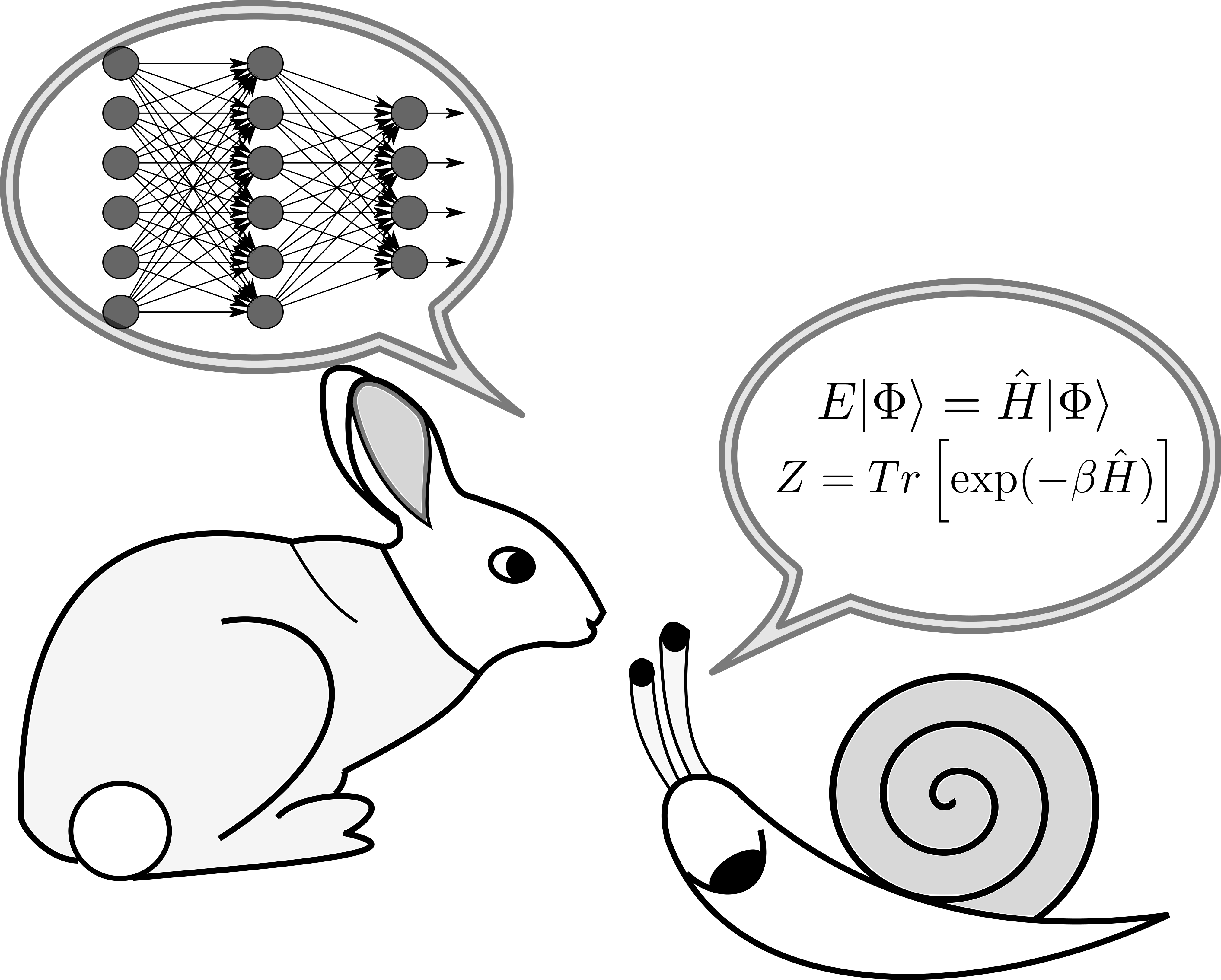}
\centering
\end{tocentry}

\begin{abstract}
Machine learning models are poised to make a transformative impact on chemical sciences by dramatically accelerating computational algorithms and amplifying insights available from computational chemistry methods. However, achieving this requires a confluence and coaction of expertise in computer science and physical sciences. This review is written for new and experienced researchers working at the intersection of both fields. We first provide concise tutorials of computational chemistry and machine learning methods, showing how insights involving both can be achieved. We then follow with a critical review of noteworthy applications that demonstrate how computational chemistry and machine learning can be used together to provide insightful (and useful) predictions in molecular and materials modeling, retrosyntheses, catalysis, and drug design.  
\end{abstract}

%% file: Sections/1-intro.tex
\newpage
\section{Introduction}

\subsection{Background}
A lasting challenge in applied physical and chemical sciences has been to answer the question: how can one \emph{identify and make} chemical compounds or materials that have optimal properties for a given purpose? A substantial part of research in physics, chemistry, and materials science concerns the discovery and characterization of novel compounds that can benefit society, but most advances still are generally attributed to trial-and-error experimentation, and this requires significant time and cost. Current global challenges create greater urgency for faster, better, and less expensive research and development efforts. Computational chemistry (CompChem) methods have significantly improved over time, and they promise paradigm shifts in how compounds are fundamentally understood and designed for specific applications. 

Machine learning (ML) methods have in the last decades witnessed an unprecedented technological evolution enabling a plethora of applications, some of which have become daily companions in our lives.~\cite{lecun2015deep,schmidhuber2015deep,goodfellow2016deep}. Applications of ML include technological fields such as web search, translation, natural language processing, self-driving vehicles, control architectures, and in the sciences, e.g. medical diagnostics~\cite{capper2018dna,klauschen2018scoring,jurmeister2019machine,ardila2019end}, particle physics~\cite{baldi2014searching}, nano sciences,\cite{Leineneabb6987} bioinformatics,~\cite{lengauer2007bioinformatics,senior2020improved} brain-computer interfaces,~\cite{blankertz2008optimizing} social media analysis,~\cite{perozzi2014deepwalk} robotics,~\cite{thrun2005probabilistic,won2020adaptive} and team, social, or board games.~\cite{Moneyball,IBMWatson,silver2016mastering} 
These methods have also become popular for accelerating the discovery and design of new materials, chemicals, and chemical processes.~\cite{Tkatchenko2020NatCommun}
At the same time, we have witnessed hype, criticism, and misunderstanding about how ML tools are to be used in chemical research. 
From this, we see a need for researchers working at the intersection of CompChem+ML to more critically recognize the true strengths and weaknesses of each component in any given study. 
Specifically, we wanted to review why and how CompChem+ML can provide useful insights into the study of molecules and materials.

While developing this review, we polled the scientific community with an anonymous online survey that asked for questions and concerns regarding the use of ML models with chemistry applications. Respondents raised excellent points including:
\begin{packed_enum}
    \item ML methods are becoming less understood while they are also more regularly used as black box tools.
    \item Many publications show inadequate technical expertise in ML (e.g. inappropriate splitting of training, testing, and validation sets).
    \item It can be difficult to compare different ML methods and know which is the best for a particular application, or whether ML should even be used at all.
    \item Data quality and context are often missing from ML modeling, and data sets need to be made freely available and clearly explained. 
\end{packed_enum} 
Additionally, when asked about the most exciting active and emerging areas of ML in the next five years, respondents mentioned a wide range of topics from catalysis discovery, drug and peptide design, ``above the arrow'' reaction predictions, and generative models that promise to fundamentally transform chemical discovery. 
When asked about challenges that ML will not surmount in the next five years, respondents mentioned modeling complex photochemical and electrochemical environments, discovering exact exchange-correlation functionals, and completely autonomous reaction discovery. 
This review will give our perspective on many of these topics.  

As context for this review, Figure \ref{fig:heatmap} shows a heatmap depicting the frequency of ML keywords found in scientific articles that also have keywords associated with different American Chemical Society (ACS) technical divisions. 
Preparing this figure required several steps. First, lists of ML keywords were chosen. Second, lists of keywords were created by perusing ACS division symposia titles from over the past five years. Third, Python scripts used Scopus Application Programming Interfaces (APIs) to identify the number of scientific publications that matched sets of ML and division symposia keywords. 
Figure \ref{fig:heatmap} elucidates several interesting points. 
First, the most popular ML approaches across all divisions are clearly neural networks, followed by genetic algorithms and  support vector machines/kernel methods. 
Second, divisions such as physical (PHYS), analytical (ANYL), and environmental (ENVR) are already using diverse sets of ML approaches, while divisions such as inorganic (INOR), nuclear (NUCL), and carbohydrate (CARB) are primarily employing more distinct subsets of approaches, while other divisions such as educational (CHED), history (HIST), law (CHAL), and business-oriented divisions (BMGT and SCHB), i.e. divisions that produce much fewer scholarly journal articles, are not linking to publications that mention ML. 
Third, ML has more prevalence across practically all divisions over time.  
For further insight, Table \ref{tab:heatmap} lists the top four keywords obtained from recent ACS symposium titles as well as their respective contribution percentage reflected in Figure \ref{fig:heatmap}. There, one sees that a handful of keywords can dominate matches in some of the bins, e.g.: `electro\*', `sensor', `protein', and `plastic'.
With any ML application, there will be a risk of imperfect data and/or user bias, but this is a useful launch point to appreciate how and where ML is being used in chemical sciences.  

\begin{figure*}[thbp!]
\includegraphics[width=5.5in]{figures/heatmaps.png}
  \caption{Heatmaps illustrating the extent that ML terms appear in scientific papers aligned by American Chemical Society (ACS) technical divisions from 2000-2010 a) and from 2010-present b). c) A line graph showing the number of occurrences of any ML term being found in papers attributed to the ACS PHYS division, from 2000-present. Figures were made by Charles D. Griego. Python scripts used to generate these figures and corresponding Table \ref{tab:heatmap} are freely available with a creative commons attribution license. Readers are welcome to use, adapt, and share these scripts with appropriate attribution: https://github.com/keithgroup/scopus\_searching\_ML\_in\_chem\_literature)
  }
  \label{fig:heatmap}
\end{figure*}

\begin{sidewaystable}[htbp!]
\caption{List of top ranked keywords (per ACS division) with corresponding percentage of matches for any ML term.}
\label{tab:heatmap}
\scriptsize
\begin{tabular}{lllll}
Division & Rank 1                               & Rank 2                                 & Rank 3                                & Rank 4                                \\ \hline
PHYS     & electro* (56.3\%)               & spectroscopy (9.7\%)              & ion* (6.0\%)                     & nano* (5.5\%)                    \\ 
ANYL     & sensor* (55.4\%)                & spectroscopy (13.1\%)             & characterization* (11.6\%)       & spectrometry (4.3\%)             \\ 
ENVR     & *sensor* (60.9\%)               & soil* (14.5\%)                    & water quality (4.2\%)            & environmental monitor* (3.0\%)   \\ 
AGFD     & protein* (31.9\%)               & agricultur* (18.2\%)              & food (10.8\%)                    & fruit* (5.7\%)                   \\ 
ENFL     & fuel* (19.2\%)                  & petroleum (11.6\%)                & energy efficiency (11.1\%)       & batter* (10.7\%)                 \\ 
AGRO     & soil (43.3\%)                   & crop* (25.4\%)                    & groundwater (11.5\%)             & developing countr* (4.4\%)       \\ 
ORGN     & protein* (64.6\%)               & amino acid* (19.7\%)              & peptide* (8.4\%)                 & aromatic* (3.2\%)                \\ 
POLY     & plastic* (51.1\%)               & polymer* (37.7\%)                 & polymeriz* (5.0\%)               & polymeric (2.8\%)                \\ 
PMSE     & *polymer* (50.4\%)              & *peptide* (30.4\%)                & thin film* (8.9\%)               & tissue engineering (4.3\%)       \\ 
BIOT     & biochemi* (37.9\%)              & biophysic* (18.3\%)               & systems biology (10.3\%)         & biotechnology (9.9\%)            \\ 
GEOC     & groundwater (33.4\%)            & mining (31.6\%)                   & *geochem* (12.4\%)               & anthropogenic (10.9\%)           \\ 
MEDI     & protein interaction* (25.7\%)   & drug discovery (19.1\%)           & drug design (19.1\%)             & antibiotic* (11.3\%)             \\ 
COMP     & drug discovery (18.7\%)         & drug design (18.6\%)              & molecular model* (14.3\%)        & protein database* (13.1\%)       \\ 
COLL     & nanoparticle* (21.2\%)          & adsorption (19.6\%)               & thin film* (14.9\%)              & tribolog* (9.6\%)                \\ 
BIOL     & drug discovery (41.9\%)         & protein folding (19.3\%)          & biosynthesis (12.2\%)            & cytochrome* (12.2\%)             \\ 
TOXI     & toxi* (99.2\%)                  & chemical exposure* (0.6\%)        & antibody drug conjugate* (0.1\%) &                              \\ 
CATL     & cataly* (64.0\%)                & metal oxides (20.2\%)             & photocataly* (5.3\%)             & surface chemistry (2.8\%)        \\ 
CINF     & drug discovery (51.7\%)         & computational chemistry (17.5\%)  & bio* modeling (7.8\%)            & chem* database* (7.6\%)          \\ 
INOR     & electrochem* (67.1\%)           & nanomaterial* (14.9\%)            & organometallic* (5.8\%)          & metal organic framework* (4.0\%) \\ 
NUCL     & nuclear fuel* (28.6\%)          & isotope* (27.3\%)                 & radioisotope* (15.0\%)           & nuclear medicine* (9.0\%)        \\ 
CARB     & carbohydrate* (43.0\%)          & glycoprotein* (42.4\%)            & glycan* (6.6\%)                  & oligosaccharide* (5.6\%)         \\ 
RUBB     & rubber* (100.0\%)               &                               &                              &                              \\ 
CELL     & cellulose (41.7\%)              & polysaccharide* (26.3\%)          & lignin (16.3\%)                  & lignocellulos* (9.7\%)           \\ 
I\&EC & water purification (38.1\%) & industrial chem* (23.7\%)  & rare earth element* (11.9\%)             & industrial and engineering chemistry (8.3\%) \\ 
FLUO     & fluorine* (99.8\%)              & radiopharmaceutical chem* (0.2\%) &                              &                              \\ 
CHED     & chem* class* (76.4\%)           & chem* communication* (8.5\%)      & chem* educat* (5.5\%)            & lab* safety (5.5\%)              \\ 
CHAS  & chem* safety (51.5\%)       & lab* safety (16.7\%)       & environmental health and safety (16.7\%) & chem* regulations (15.2\%)                   \\ 
BMGT  & chem* compan* (65.4\%)      & chem* enterprise* (28.8\%) & chem* business* (3.8\%)                  & chem* research and development (1.9\%)       \\ 
SCHB  & commercial chem* (50.0\%)   & chem* sector* (28.6\%)     & academic entrepreneur* (14.3\%)          & science advoca* (7.1\%)                      \\ 
HIST     & chem* histor* (53.8\%)          & evolution of chem* (30.8\%)       & history of chem* (15.4\%)        &                              \\ 
PROF     & chem* education (100.0\%)       &                               &                              &                              \\ 
CHAL     & pharmaceutical patent* (60.0\%) & chem* in commerce (30.0\%)        & chem* patent* (10.0\%)           &                              \\ \hline
\end{tabular}
\end{sidewaystable}

\subsection{Motivation for this review}
\begin{figure*}[h!]
\includegraphics{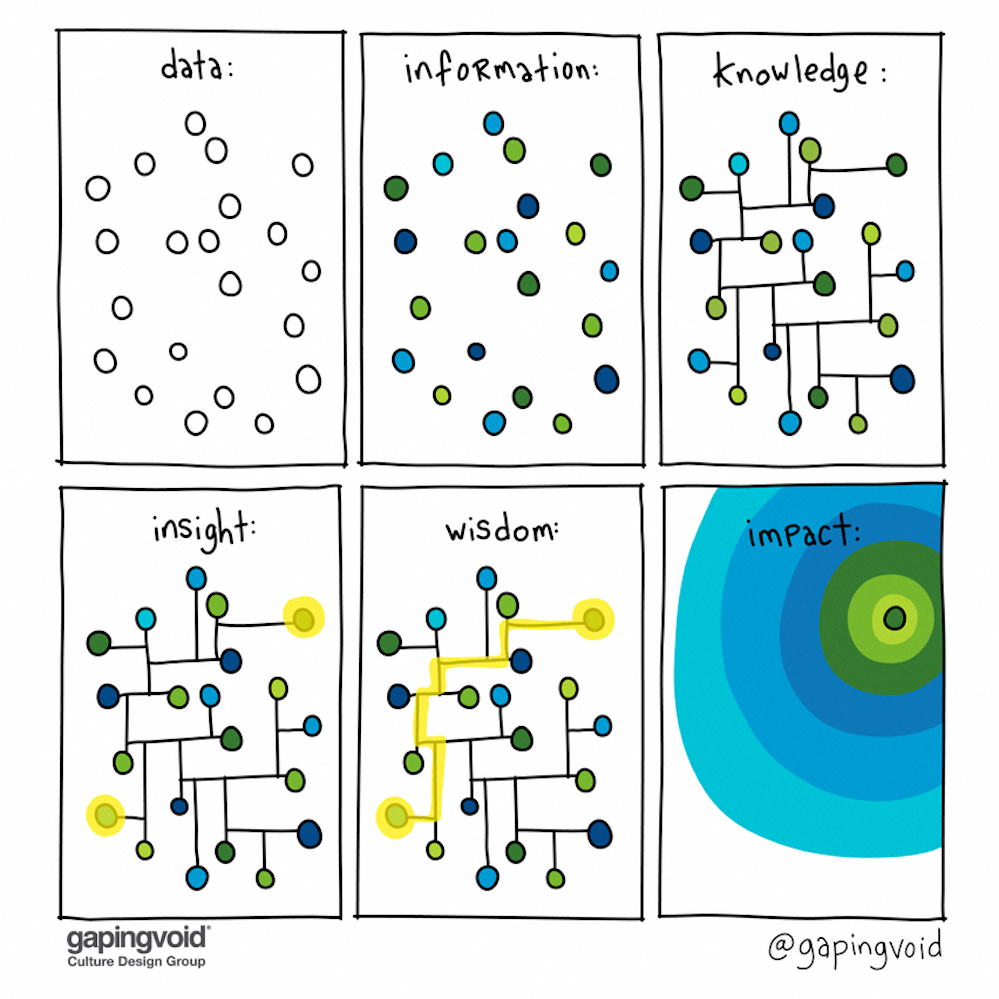}
  \caption{A data-information-knowledge-insight-wisdom-impact hierarchy courtesy of the Gapingvoid Culture Design Group, https://www.gapingvoidart.com.}
  \label{GapingVoid}
\end{figure*}

The survey results and literature analysis above showed an opportunity for a tutorial reference to help readers address future research challenges that will require joint applications of CompChem, ML, and chemical and physical intuition (CPI).
We will classify concepts in this review using a popular rendition of a ``data to wisdom'' hierarchy, Figure \ref{GapingVoid}. Scholars have noted shortcomings with similar constructs,\cite{rowley2007dikw} but we mean for this figure to reflect scientific progress, starting from data and ending in impact. 
CompChem, ML, and CPI each bring something to the table since all have different strengths and weaknesses.  
CPI can lead to knowledge, insight, and wisdom from data and information, but the applicability of CPI may be limited when one is faced with large data sets. Alternatively, CompChem is extraordinarily well-suited for generating high quality data that contain useful information (\emph{vide infra}). Subsequently, non-linear ML models can provide numerical representations of partial or complete sets of data. The application of CPI to these ML models can be used to recognize relationships with respect to chemical and physical concepts to produce knowledge and then ideally insightful predictions. 
At the time of writing, none of the authors of this review would say that a combination of CompChem, ML, and CPI has reached its full potential of consistently providing wisdom (nor impact), since computational predictions must still be tested by experiment to ensure their validity.

ML brings many reasons to be optimistic about increased knowledge and insights. 
ML models are extremely well-suited for recognizing and accurately quantifying non-linear relationships (\emph{vide infra}), a task that is especially difficult for even the most expert-level CPI alone. 
However, useful ML requires robust datasets that can be provided by CompChem as long as the CPI component is selecting and correctly interpreting appropriate methods for the task at hand. 
We furthermore note that the knowledge generation process shown in Fig.~\ref{GapingVoid} is by no means a linear one --- on the contrary, it contains many loops and dead ends. 
As we show later, within the troika of CompChem+ML+CPI, ML acts as a catalyst  since it can be used in place of explorative data-driven hypotheses generation. 
Automatically generated hypotheses are then validated and calibrated with CompChem and CPI to yield further improved ML modeling (enriched by more physical prior knowledge), which then loops back with improved hypotheses. 
This feedback loop is the key to the modern knowledge discovery leading to insight, wisdom and hopefully a positive impact to society. 

%% file: Sections/2-compChem.tex
\newpage
\section{CompChem and Notable Intersections with ML}
\label{CCTutorial}

\subsection{Computational modeling, data, and information across many scales}

We consider quantum mechanics as described by the non-relativistic time-independent Schr{\"o}\-dinger equation as our ``standard model'' because it accurately represents the physics of charged particles (electrons and nuclei) that make up almost all molecules and materials. 
Indeed, this opinion has been held by some for almost a century:
\begin{quote}
    The fundamental laws necessary for the mathematical treatment of a large part of physics and the whole of chemistry are thus completely known, and the difficulty lies only in the fact that application of these laws leads to equations that are too complex to be solved.  -- P. A. M. Dirac, 1929
\end{quote}

Any theoretical method for predicting molecular and/or material phenomena must first be rooted in quantum mechanics theory and then suitably coarse-grained and approximated so that it can be applied in a practical setting. CompChem, or more precisely, computational quantum chemistry defines computationally-driven numerical analyses based on quantum mechanics. 
In this section we will explain how and why different CompChem methods capture different aspects of underlying physics.  
Specifically, this section provides a concise overview of the broad range of CompChem methods that are available for generating datasets that would be useful for ML-assisted studies of molecules and/or materials. 

\subsubsection{Models and levels of abstraction}

A traditional example of a \emph{``good model''} is the ideal gas equation: $PV = nRT$, which can be considered `simple', `useful', and `insightful'.\cite{Box1976}
The ideal gas equation relates macroscopic pressure ($P$), volume ($V$), number of molecules ($n$), and temperature ($T$) of gases under idealized conditions, without requiring explicit knowledge of the processes occurring on an atomic scale. 
Its simple functional form needs just one parameter, the ideal gas constant $R$, and this makes it possible to formulate useful insights, such as how at constant pressure a gas expands with rising temperature.
On the other hand, this elegant equation only holds for conditions where the gas behaves as an ideal gas.
The derivation of more accurate models of gases requires more mathematically complicated equations of state that rely on more free parameters\cite{mcquarrie1997physical} that in turn obfuscate physical insights, require more computational effort to solve, and thus make the model less ``good''.
This example also offers a convenient connection to ML models that will be discussed later in Section \ref{section:ML}. 
As mathematical models for complex phenomena become more complicated and less intuitive to derive, ML models that can infer non-linear relationships from data become more applicable when increasing amounts of empirical data become available. 

\begin{figure*}[h!]
\includegraphics[width=3.33in]{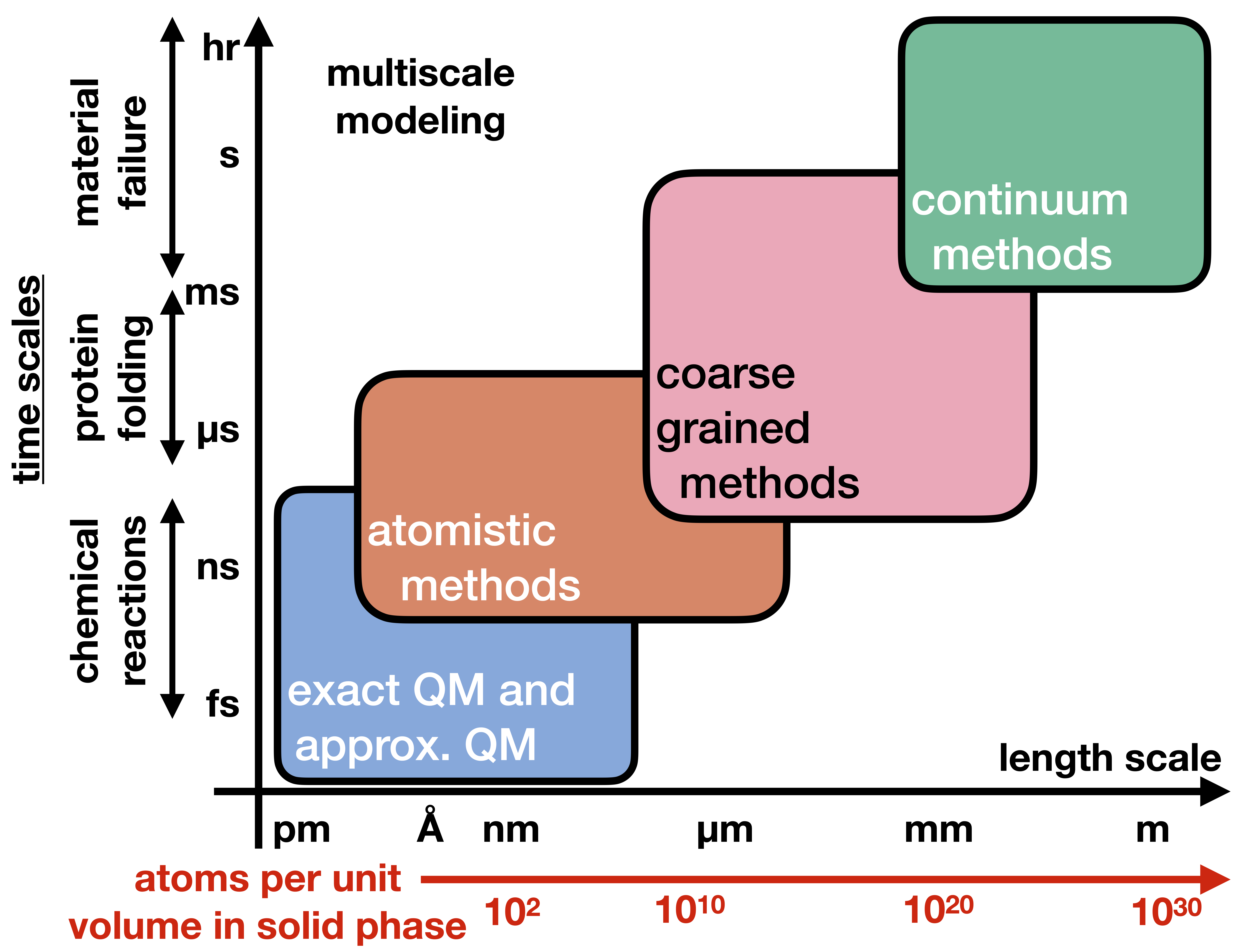}
  \caption{Hierarchy of computational methods and corresponding time and length scales}
  \label{fig:multiscale}
\end{figure*}

Alternatively, the conventional CompChem treatment entails first determining the system's relevant geometry and its total ground state energy, and from that physical properties of interest (e.g. pressure, volume, band gap, polarizability, etc) can be obtained using quantum and/or statistical mechanics. In this Section, we discuss the relevant CompChem methods for these.
While the mathematical physics for these methods might occasionally be too complicated for a user to fully understand, many algorithms exist so that they can still be easily run in a `black-box' way with modern computational chemistry software and accompanying tutorials.\cite{cramer2013essentials,frenkel2001understanding,foresman1996exploring,eastman2017openmm}
CompChem thus serves as an invaluable tool to generate data and information for knowledge and insights across many length and time scales. 
Fig. \ref{fig:multiscale} is an adaption of a multiscale hierarchy\cite{frenkel2001understanding} of different classes of CompChem methods, showing their applicability for modeling different length and time scales, and a depiction of how large scale models may be developed based on smaller scale theories.  

\subsubsection{CompChem representations}
Integral to every CompChem study is the user's \emph{representation} for the system, i.e. how the user chooses to describe the system. 
CompChem representations can range from simple and lucid (e.g. a precise chemical system such as a water molecule isolated in a vacuum) to complex and ambiguous (e.g. a putative but unproved depiction of a solid-liquid interface under electrochemical conditions). 
Approximate wavefunctions (expressed on a basis set of mathematical functions) or approximate Hamiltonians (referred to as levels of theory) as described below in this section can also be considered representations.
One might then say that many representations for different components of a system will constitute an overall representation, and this is true.
The point we make is that the validity of any computational result depends on the overall representation, and sometimes an incorrect representation may provide a correct result due to ``fortuitous error cancellation''.
In CompChem studies, a valid representation is one that captures the nature of the physical phenomena of a system. 
For a molecular example, if one is determining the bond energy of a large biodiesel molecule using CompChem methods,\cite{oyeyemi2014accurate} it may or may not be justified to approximate a nearby long-chain alkyl group (-C$_n$H$_{(2n+1)}$) simply as a methyl (-CH$_3$) or even a hydrogen atom. 
Indeed, choosing such a representation can sometimes be a useful example of CPI since alkyl bonds usually exhibit relatively short-ranged interactions (a feature that will be discussed in the context of ML in more detail in Section~\ref{sec:locality}). 
An atomic scale geometry with fewer atoms would reduce the computational cost of the study or allow a more accurate but more computationally expensive calculation to be run.  
On the other hand, it might also be a poor choice if the chemical group, e.g. a substituted alkyl group participated in physical organic interactions such as subtle steric, induction, or resonance effects.\cite{anslyn2006modern}
For a solid-state example, a user might exercise good CPI by assuming that a relatively small unit cell under periodic boundary conditions would capture salient features of a bulk material or a material surface (as is often the case for many metals). 
On the other hand, subtle symmetry-breaking effects in materials (e.g. distortions arising from tilting octahedra groups in perovskites\cite{glazer1972classification}, or surface reconstruction phenomena that occur on single crystals)\cite{giessibl1995atomic} might only be observed when considering larger and more computationally expensive unit cells.
Relevant to both examples, it may also be that the CompChem method itself brings errors that obfuscate phenomena that the user intends to model.
In general, CompChem errors may be due to errors in the initial set up the CompChem application, or if they were due to errors in how the CompChem method is treating the physics of the system, and both factors reflect the representation used in the study.  
In Section \ref{section:ML}, we will discuss how the choice of ML representation also plays similarly critical roles in determining whether and to what extent an ML model is useful.

\subsubsection{Method accuracy}
The quantitative \emph{accuracy} of a CompChem model stems from its suitability in describing the system. 
As explained above, an observed accuracy will depend on the representation being used. 
High quality CompChem calculations have traditionally been benchmarked against datasets that consist of well-controlled and relatively precise thermochemistry experiments on small, isolated molecules.\cite{Curtiss2000, Haunschild2012}
The error bars for standard calorimetry experiments are approximately 4 kJ/mol (or 1 kcal/mol or 0.04 eV), and computational methods that can provide greater accuracy than this are stated as achieving `chemical accuracy'.
Note that this term should be used when describing the accuracy of the method compared to the most accurate data possible; for example, if one CompChem method was found to reproduce another CompChem method within 1 kJ/mol, but both methods reproduce experimental data with errors of 20 kJ/mol, then neither method should be called chemically accurate.  
There are many well-established reasons why CompChem models can bring errors.
For example, errors may be due to size consistency\cite{taylor1994lecture} or size extensivity\cite{bartlett1981mp} problems that are intrinsic within the CompChem method, larger systems sometimes embody significant medium and long range interactions (e.g. van der Waals forces)\cite{Stohr2019} or self-interaction errors\cite{Lundberg2005} that might not be noticeable in small test cases.   
The recommended path forward is to consider which fundamental interactions are in play in the system of interest, and then to use a CompChem model that is adequate at describing those interactions.  
Besides this, users should make use of existing tutorial references that provide practical knowledge about which parameters in a CompChem calculation should be carefully noted, for example Ref. ~\citenum{Morgante2020}.
Historically the most popular CompChem methods for molecular and materials modeling (the B3LYP\cite{beck93jcp} and PBE\cite{perd+96prl} exchange correlation functionals, see section \ref{sec:DFT}) are often said to have an expected accuracy of about 10-15 kJ/mol (or 2-4 kcal/mol or 0.1-0.2 eV) when modeling differences between the total energies of two similar systems, and errors are expected to be somewhat larger when considering transition state energies.
Though this is used as a simple rule, it is obviously an oversimplification and actual accuracy is only assessed by thoughtful benchmarking of the case being considered.\cite{Goerigk2017,Zhao2005benchmark,Sim2018,Riley2007,Maldonado2021}

\subsubsection{Precision and reproducibility}
In CompChem, one normally assumes that that any two users using the same representation for the system with the same code on the same computing architecture will obtain the exact same result within the numerical precision of the computers being used. 
This is not always the case, especially for molecular dynamics (MD) simulations that often rely on stochastic methods.\cite{Abraham2019} 
Computational precision also becomes more concerning when there are different versions of codes in circulation, errors that might arise from different compilers and libraries, and a lack of consensus in the community about which computational methods and which default settings should be used for specific application systems, e.g. grid density selections,\cite{Wheeler2010} or standard keywords for molecular dynamics simulations.\cite{Abraham2019,PLUMEDConsortium2019}
There have been efforts to confirm that different codes can reproduce energies for the same system representation,\cite{PLUMEDConsortium2019,Lejaeghere2016} but some commercial codes hold proprietary licenses that restrict publications that critically benchmark calculation accuracy and timings across different codes. 
A path forward to benefit the advancement of insight is the development of (open) source codes\cite{sonnenburg2007need} that perform as well if not better than commercial codes. 
While increased access to computational algorithms is beneficial, it also raises the need for enforcing high standards of quality and reproducibility.\cite{Durrani2020,Perkel2020}
We are also glad to see active developments to more lucidly show how any set of computational data is generated, precisely with which codes, keywords, and auxiliary scripts and routines.\cite{Govoni2019,Kitchin2015,Alvarez-Moreno2015,Huber2020}
We are now in an era where truly massive amounts of data and information can be generated for CompChem+ML efforts. 
To go forward, one needs to know what constitutes good and useful data, and the next section provides an overview of how to do this using CompChem.

\subsection{Hierarchies of methods}

\begin{figure*}[h!]
\includegraphics[width=5.0in]{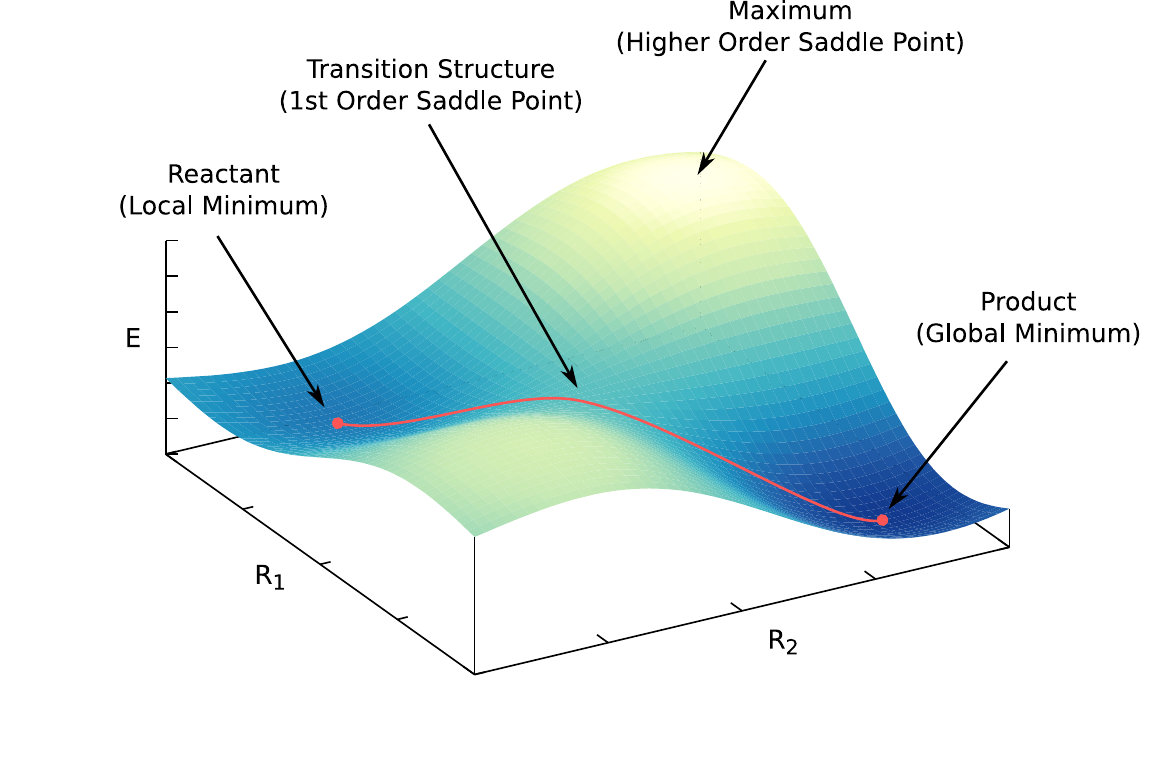}
  \caption{Potential energy surface (PES) of a fictional system with the two coordinates $R_1$ and $R_2$. The minima of the PES correspond to stable states of a system, such as equilibrium configurations and reactants or products. Minima can be connected by paths (red line), along which rearrangements and reactions can occur. The maximum along such a path is called a transition state. Transition states are first order saddle points, a maximum in one coordinate and minima in all others.
  They correspond to the minimum energy required to transition between two PES minima and play a crucial role in the description of chemical transformations.}
  \label{fig:pes}
\end{figure*}

Earlier we mentioned that a usual task in CompChem is to calculate the ground state energy of an atomic scale system. 
Indeed, CompChem methods can determine the energy for a hypothetical configuration of atoms, and this constitutes the potential energy surface (PES) of the system (Fig.~\ref{fig:pes}).
The PES is a hypersurface spanning $3N$ dimensions, where $N$ is the number of atoms in the system.
Since the PES is used to analyze chemical bonding between atoms within the system, the PES can also be simplified by ignoring translational and rotational degrees of freedom for the entire system. 
This reduces the dimensionality of the PES from $3N$ to $3N-5$ for linear systems (e.g. diatomic molecules or perfectly linear molecules such as acetylene) or $3N-6$ for all other non-linear systems.  
Furthermore, since visualization is difficult beyond three dimensions, PES drawings will show a 1-D or 2-D projection of this hypersurface where the $z$-axis is conventionally used to represent the scale for system energy.

Any arbitrary PES will contain several interesting features.
Minima on the PES correspond to mechanically stable configurations of a molecule or material, for example reactant and product states of a chemical reaction or different conformational isomers of a molecule. 
Because they are minima, the second derivative of the energy given by the PES with respect to any dimension will be positive.
Minima can also be connected by pathways, which indicate chemical transformations (Fig.~\ref{fig:pes}, red line).
Along such pathways, the second derivative can be positive, zero, or negative, but all other second derivatives must be positive. 
Transition states are first order saddlepoints, and thus represent a maximum in one coordinate and a minimum along all others.
They correspond to the lowest energy barriers connecting two minima on the PES and are hence important for characterizing transitions between PES minima (e.g. chemical reactions).
Second order saddle points\cite{Heidrich1986} and bifurcating pathways\cite{Ess2008} can also exist, but these are not discussed further here.

\begin{figure*}[h!]
\includegraphics[width=7.0in]{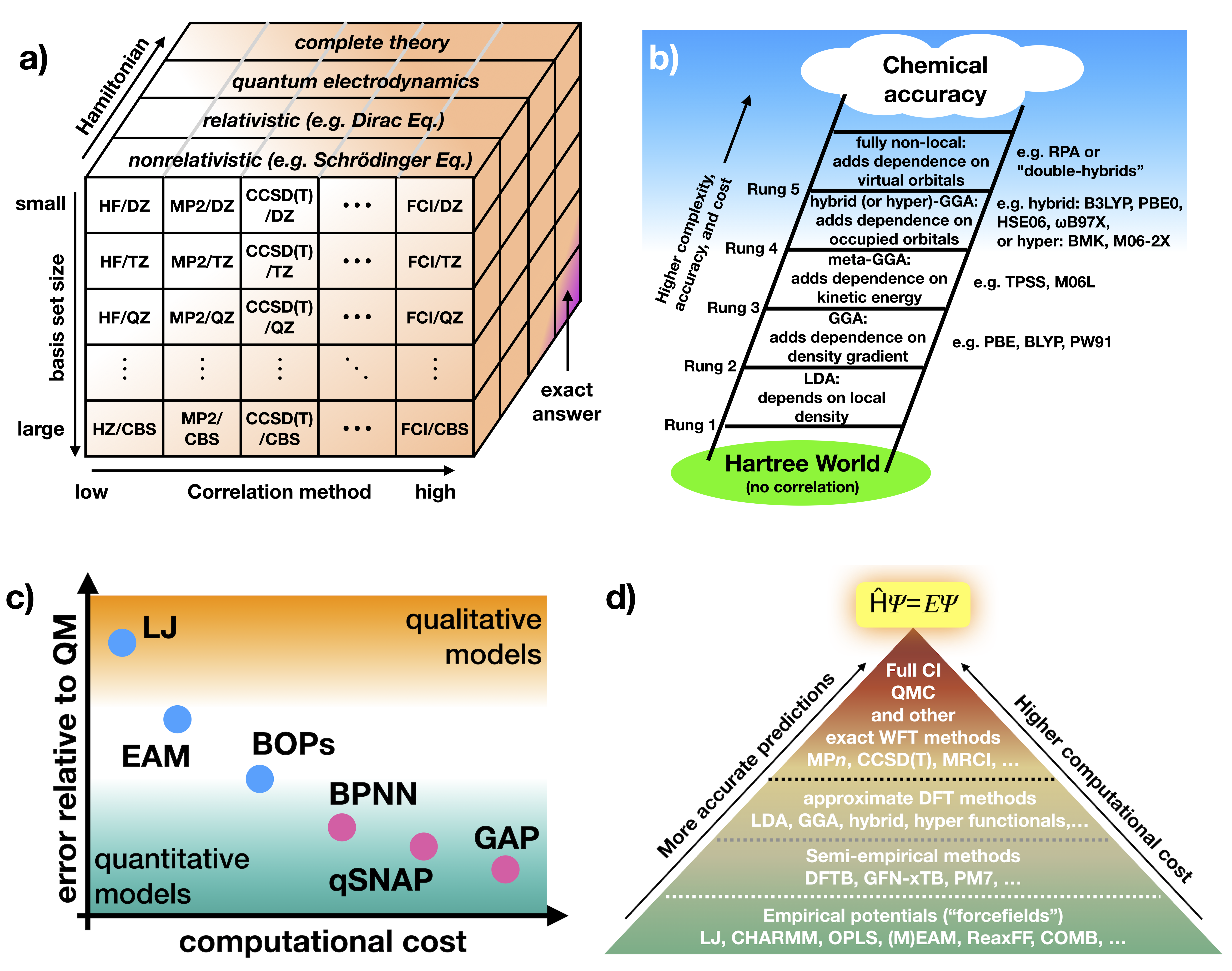}
  \caption{a) A `magic cube'\cite{Tarczay2001} depiction of hierarchies of correlated wavefunction approaches. b) A `Jacob's Ladder'\cite{perdew2001jacob} depiction of hierarchies of Kohn-Sham DFT approaches. c) An illustration of expected hierarchies of atomistic potentials adapted from Ref.~\bibpunct{}{}{,}{n}{,}{,}\cite{osti_1639258}. d) An illustration of overall hierarchies in predictive atomic scale modeling methods.}
  \label{fig:CompChemHierarchies}
\end{figure*}

A wide range of higher-level properties of the system can be predicted and/or derived using the PES, including predicted thermodynamic binding constants, kinetic rate constants for reactions, or properties based on dynamics of the system. 
The task then to choose an appropriate CompChem method that can carry out energy and gradient calculations on the system's PES. 
Fig. \ref{fig:CompChemHierarchies} shows several different hierarchies for CompChem methods capable of doing this.
Note that all of these methods mentioned in this figure fall in the categories of the bottom two regions in the multiscale hierarchy Fig. \ref{fig:multiscale}. 
All of these methods in principle could be used to develop coarse-grained or continuum models as well.
Also note that methods in Fig. \ref{fig:CompChemHierarchies} will bring very different computational costs and opportunities for methods involving ML.

\subsubsection{Wavefunction theory methods}
In standard computational quantum chemistry, a system's energy can be computed in terms of the Schr\"{o}dinger equation.\bibpunct{}{}{,}{s}{,}{,}\cite{schroedinger1926quantisierung1,schroedinger1926quantisierung2,schroedinger1926quantisierung3} 
The wavefunction that will be used to represent the positions of electrons and nuclei in the system ($\Psi(\mathbf{r}, \mathbf{R})$) is hard to intuit since it can be complex valued. 
However, its square describes the real probability density of the nuclear ($\mathbf{R}$) and electronic positions ($\mathbf{r}$).
In a real system, the position and interactions of a single particle in the system with respect to all other particles will be \emph{correlated}, and this makes exactly solving the Schr\"{o}dinger equation impossible for almost all systems of practical interest.
To make the problem more tractable, one may exploit the Born-Oppenheimer approximation;\cite{born-opp27adp} since nuclei are expected to move much slower than the electrons they can be approximated as stationary at any point along the PES.
This allows the energy to be calculated using the time-independent Schr\"{o}dinger equation and solving the eigenvalue problem: 
\begin{equation}
     \label{TISE}
     \hat{H} \Psi = (\hat{T} + \hat{V}) \Psi = E \Psi,
\end{equation}
Here, the Hamiltonian operator ($\hat{H}$) is the sum of the kinetic ($\hat{T}$) and potential ($\hat{V}$) operators, $\Psi$ is the wavefunction (i.e. an eigenfunction) that represents particles in the system, and $E$ is the energy (i.e. an eigenvalue).
In this way, nuclei can be treated as fixed point charges, and then Eq. \ref{TISE} can be transformed into the so-called electronic Schr{\"o}dinger equation, where the Hamiltonian $\hat{H}_\mathrm{el}$ and wavefunction $\Psi_\mathrm{el}(\mathbf{r};\mathbf{R})$ now only depend on the nuclear coordinates $\mathbf{R}$ in a parametric fashion:
\begin{align*}
    \label{TISE_mb}
    \hat{H}_\mathrm{el} \Psi_\mathrm{el}(\mathbf{r;R}) &= \left[ \hat{T}_\mathrm{e}(\mathbf{r}) + \hat{V}_\mathrm{eN}(\mathbf{r;R}) + \hat{V}_\mathrm{NN}(\mathbf{R}) + \hat{V}_\mathrm{ee}(\mathbf{r}) \right] \Psi_\mathrm{el}(\mathbf{r;R})  \\
    & = E_\mathrm{el} \Psi_\mathrm{el}(\mathbf{r;R}) \numberthis
\end{align*}
The above expression has $\hat{H}_\mathrm{el}$ composed of single electron ($\mathrm{e}$), electron-nuclear ($\mathrm{eN}$), nuclear-nuclear ($\mathrm{NN}$), and electron-electron ($\mathrm{ee}$) terms. 
Here, we will now implicitly assume the Born-Oppenheimer approximation throughout and leave off the subscript indicating the electronic problem.
However, we note that the Born--Oppenheimer approximation is not always sufficient and the computationally intensive nonadiabatic quantum dynamics may be required.\cite{curchod2018ab}
In certain cases, semi-classical treatments are appropriate,
for example, nonadiabatic effects between electrons and nuclei can be considered using nuclear-electronic orbital methods.\cite{Pavosevic2020}

A second common approximation is to expand the total electronic wavefunction in terms one-electron wave functions (i.e. spin orbitals): $\phi(\mathbf{r}_i)$.
Electrons are fermions and therefore exhibit antisymmetry, which in turn results in the Pauli exclusion principle. 
Antisymmetry means that the interchange of any two within the system should bring an overall sign change to the wavefunction (i.e. from $+$ to $-$, or \emph{vice versa}). 
This property is conveniently captured mathematically by combining one electron spin orbitals into the form of a Slater determinant:
\begin{equation}
\Psi(\mathbf{r}_1,\ldots,\mathbf{r}_n) = \frac{1}{\sqrt{n!}}
\begin{vmatrix}
\phi_1(\mathbf{r}_1) & \ldots & \phi_n(\mathbf{r}_1) \\
\vdots      & \ddots & \vdots      \\
\phi_1(\mathbf{r}_n) & \ldots & \phi_n(\mathbf{r}_n) \\
\end{vmatrix}
\label{eq:slaterdet}
\end{equation}

Note that a determinant's sign changes whenever two columns or rows are interchanged, and in a Slater determinant this corresponds to interchanging electrons and thus the physically appropriate sign change for the overall wavefunction. 
Additionally, $\frac{1}{\sqrt{n!}}$ is a normalizing factor to ensure the wavefunction is unitary. 

The spin orbitals can be treated as a mathematical expansion using a basis set of $\mu$ functions $\chi_\mu$, each having coefficients $c_{\mu i}$, which are generally Gaussian basis functions,\cite{Peterson2002,Hehre1969,Schafer1992} Slater-type hydrogenic orbitals,\cite{VanLenthe2003} or plane waves under periodic boundary conditions:~\cite{Slater1964,MacDonald1980,Louie1979}
\begin{equation}
    \phi_i = \sum_\mu c_{\mu i} \chi_\mu
    \label{BasisSet}
\end{equation}
The different types of mathematical functions bring different strengths and weaknesses, but these will not be discussed further here. 
A universal point is that larger basis sets will have more basis functions and thus give a more flexible and physical representation of electrons within the system. 
On one hand this can be crucial for capturing subtle electronic structure effects due to electron correlation. 
On the other hand, larger basis sets also necessitate significantly higher computational effort. 
A standard technique to avoid high computational effort in electronic structure calculations is to replace non-reacting core electrons with analytic functions using effective core potentials (ECPs, i.e. pseudopotentials).\cite{Goedecker1996,Melius1974,Wadt1985,Hay1985,Cao2002,Metz2000,Dolg2016,Shaw1967,Kahn1976,Christiansen1979,Hamann1979,Vanderbilt1985,Garrity2014,Kresse1994,Joubert1999,Troullier1990}
This requires reformulating the basis sets that describe the valence space of the atoms, for example see Refs.~\bibpunct{}{}{,}{n}{,}{,}\cite{Peterson2003,Roy2008}. 
Larger nuclei that bring higher atomic numbers and larger numbers of electrons will also exhibit relativistic effects,\bibpunct{}{}{,}{s}{,}{,}\cite{Pyykko2012} 
and relativistic Hamiltonians are based on the Dirac equation\cite{dirac1928quantum,Tecmer2016} or quantum electrodynamics.\cite{feynman1998quantum} 
These methods can range from reasonably cost-effective methods\cite{Nakajima2012,VanLenthe2000} to those bringing extremely high computational cost.\cite{Visscher2002} 
Practical applications have traditionally used standard non-relativistic Hamiltonian methods along with ECPs (or pseudopotentials) that have been explicitly developed to account for compressed core orbitals that result from relativistic effects.

Using the Born-Oppenheimer approximation (Eq.~\ref{TISE_mb}) together with a Slater determinant wavefunction (Eq.~\ref{eq:slaterdet}) expressed in a finite basis set (Eq.~\ref{BasisSet}) brings about the simplest wavefunction based method, the Hartree-Fock (HF) approach (for historical context see Refs.~\bibpunct{}{}{,}{n}{,}{,}\cite{Hartree1935,Slater1951,Fock1930}).
The HF method is a mean field approach, where each electron is treated as if it moves within the average field generated by all other electrons. 
It is generally considered inaccurate when describing many chemical systems, but it continues to serve as a critical pillar for CompChem electronic structure calculations since it either establishes the foundation for all other accurate methods or provides energy contributions (i.e. exact exchange) that is not provided in some CompChem methods. 
CompChem methods that achieve accuracy higher than HF theory are said to contain \emph{electron correlation}, a critical component for understanding molecules and materials (as described in more detail in section \ref{section:correlatedWF}). 
Expressing $\Psi$ as a Slater determinant and rearranging Eq. \ref{TISE_mb} while temporarily neglecting nuclear-nuclear interactions allows one to define the HF energy in terms of integrals of the electronic spin orbitals:  
\begin{align*}
E_{\text{HF}} =& -\sum_i \int d \mathbf{r}_1 \phi^*_i (\mathbf{r}_1)%
\underbrace{\frac{1}{2} \nabla^2}_{\hat{T}_\mathrm{e}}%
\phi_i (\mathbf{r}_1)%
- \sum_i \int d \mathbf{r}_1 \phi^*_i (\mathbf{r}_1)%
\underbrace{\sum^N_\alpha \frac{Z_\alpha}{|\mathbf{r}_1-\mathbf{R}_\alpha|}}_{\hat{V}_\mathrm{eN}}%
\phi_i (\mathbf{r}_1) \\
& + \sum_{i,j>i} \int \int d \mathbf{r}_1 d \mathbf{r}_2 \phi^*_i (\mathbf{r}_1) \phi_i(\mathbf{r}_1)%
\underbrace{\frac{1}{|\mathbf{r}_1-\mathbf{r}_2|}}_{\hat{V}_\mathrm{ee} (\mathrm{Coulomb})}%
\phi^*_j(\mathbf{r}_2) \phi_j(\mathbf{r}_2) \\
& - \sum_{i,j>i} \int \int d \mathbf{r}_1 d \mathbf{r}_2 \phi^*_i (\mathbf{r}_1) \phi_j(\mathbf{r}_1)%
\underbrace{\frac{1}{|\mathbf{r}_1-\mathbf{r}_2|}}_{\hat{V}_\mathrm{ee} (\mathrm{exchange})}%
\phi^*_j(\mathbf{r}_2) \phi_i(\mathbf{r}_2) \numberthis
\label{HFenergy}
\end{align*} 
where the first two terms are referred to as one-electron integrals and represent the kinetic energy due to the electrons and the potential energy contributions from electron-nuclei interactions.
The remaining terms are two-electron integrals that describe the potential energy arising from electron-electron interactions and are called Coulomb and exchange integrals.
Using Lagrange multipliers, one can express the Hartree--Fock equation in a compact matrix form, the so-called Roothan--Hall equations,\bibpunct{}{}{,}{s}{,}{,}\cite{jensen2007introduction,Roothaan1951,Hall1951} which allow for an efficient solution:
\begin{equation}
    \label{HF_eqns}
    \mathbf{F} \mathbf{C} = \boldsymbol{\epsilon} \mathbf{S} \mathbf{C}
\end{equation}
Each matrix has a size of $\mu \times \mu$, where $\mu$ the number basis functions used to express the orbitals of the system.
$\mathbf{C}$ is a coefficient matrix collecting the basis coefficients $c_{\mu i}$ (see Eq.~\ref{BasisSet}), while $\mathbf{S}$ is the overlap matrix measuring the degree of overlap between individual basis functions and $\boldsymbol{\epsilon}$ is a diagonal matrix of the spin orbital energies.
Finally, $\mathbf{F}$ is the Fock matrix, with elements of a similar form as in Eq.~\ref{HFenergy}, but expressed in terms of basis functions $\chi_\mu$.
One important detail not readily apparent in Eq. \ref{HF_eqns} is that the Fock matrix depends on the orbital coefficients that must be provided before Eq. \ref{HF_eqns} can be solved.
As such, Eq.~\ref{HF_eqns} cannot be solved in closed form, but instead requires a so-called self-consistent field approach.
Starting from an arbitrary set of trial (i.e. initial guess) functions, one iteratively solves for optimal molecular orbital coefficients which are then used to construct a new Fock matrix, until a minimum energy is reached in accordance with the variational principle of quantum mechanics. 
Evaluating and transforming the two-electron integrals in Eq.~\ref{HFenergy} are a significant bottleneck for these calculations and thus the computational effort of the HF methods formally scales as $\mathcal{O}(\mu^4)$ with the number of basis functions. 
This means, that a calculation on a system twice as large will require at least $2^4 = 16$ times as much computing time.
The electronic exchange interaction resulting from the antisymmetry of the wavefunction imposes a strong constraint on the mathematical form of ML models for electronic wavefunctions. Construction of efficient and reliable antisymmetric ML models for the many-body wavefunction is an important area of current research.~\cite{hermann2020deep,pfau2020ab}

\subsubsection{Correlated wavefunction methods}
\label{section:correlatedWF}
The system's \emph{correlation energy} is defined as sum of electron-electron interactions that originate beyond the mean-field approximation for electron-electron interactions that is provided by HF theory. 
While correlation energy makes up a rather small contribution to the overall energy of a system (usually about 1\% of the total energy), because internal energies in molecular and material systems are so enormous, this contribution becomes rather significant. 
As an example, most molecular crystals would be unstable as solids if calculated using the HF level of theory. 
The missing component is attractive forces that are obtained from levels of theory that account for correlation energy. 
Correlation energies are obtained by calculating additional electron-electron interaction energies that arise from different arrangements of electron configurations (i.e. effectively, different possible excited states) that are not treated with the mean field approach of HF theory. 

The most complete correlation treatment is the full configuration interaction (FCI) method, which is the exact numerical solution of the electronic Schr{\"o}dinger equation (in the complete basis limit) that considers interactions arising from all possible excited configurations of electrons.
The FCI wavefunction takes the form of a linear combination of all possible excited Slater determinants which can be generated from a single HF reference wavefunction by electron excitations:
\begin{equation}
	\Psi_\mathrm{FCI} = a_0 \Psi_{HF} + \sum_{\alpha, \beta} a^\alpha_\beta \Psi^\alpha_\beta + \sum_{\alpha, \beta, \gamma, \delta} a^{\alpha\gamma}_{\beta\delta} \Psi^{\alpha\gamma}_{\beta\delta} + \ldots 
	\label{eq:fci}
\end{equation}
where $\Psi^\alpha_\beta$ represents the Slater determinant obtained by exciting an electron from orbital $\alpha$ into an unoccupied orbital $\beta$ and the $a$s are expansion coefficients determining the weight of the different contributing configurations.
Expectedly, FCI calculations scale extremely poorly with the number of electrons in the system ($\mathcal{O}(n!)$), as the number of possible configurations grows rapidly, making them feasible only for small molecules. 
For an example of the state of the art, FCI calculations have been used to benchmark highly accurate methods on calculations on a benzene molecule.\cite{Eriksen2020} 

Most correlated wavefunction methods use a subset of the possible configurations in Eq.~\ref{eq:fci} to be computationally tractable.
The configuration interaction (CI)\cite{helgaker2013molecular} method for example only includes determinants up to a certain permutation level (e.g. `s'ingle and `d'ouble excitations in CISD). 
Alternatively,  MP$n$\cite{bartlett1981mp} (e.g. MP2) recovers the correlation energy by applying different orders of perturbation theory. 
Coupled cluster theory, another widely used post-HF method, includes additional electron configurations via cluster operators.\cite{bartlet2007cc}
One coupled cluster method that involves single, double, and perturbative triples excitations, CCSD(T), is referred to as the ``gold-standard'' approach for CompChem electronic structure methods since it brings high accuracy for molecular energies. 
However, there are many newer advances that improve upon CCSD(T).\cite{Rezac2013,Eriksen2020}  
Note that just because a method has a reputation for being accurate does not mean that it will be for all systems. 
For example, consider again the benzene molecule, which is best illustrated having dotted resonance bond depicting a planar molecule with equal C--C bond lengths. 
Such a geometry will not be found to be stable with many different CompChem methods, in part because of subtle chemical bonding interactions and/or errors that arise from specific choices of basis sets used with different levels of theory.\cite{Moran2006,Samala2017} 

A key point to reiterate is that correlated wavefunction methods are founded on the HF theory, and so they are even more computationally demanding than HF calculations, e.g. $\mathcal{O}(n^5)$ for MP2, $\mathcal{O}(n^6)$ for CCSD and CISD and $\mathcal{O}(n^7)$ for CCSD(T).
However, this computational expense is alleviated by continually improving computing resources (e.g. the usability of graphics processing units (GPUs)) \cite{Titov2013,Seritan2020,Anderson2007,Andrade2013} and the development of efficiency enhancing algorithms such as pseudospectral methods,\bibpunct{}{}{,}{s}{,}{,}\cite{Friesner1985,Martinez1992,Friesner2002} resolution of the identity (RI),\cite{sierka2003fast} domain-based local pair natural orbital methods (DLPNO),\cite{riplinger2013efficient} and explicitly correlated R12/F12 methods.\cite{kong2012explicitly}
There are also ongoing efforts to develop other CompChem methods based on quantum Monte Carlo\cite{austin2012quantum} and density matrix renormalization group theory (DMRG)\cite{marti2010density} to provide high accuracy with competitive scaling with other computational methods.
Efforts are beginning to become implemented that use ML to accelerate these types of calculations.\cite{schnorb,0274_McGibbon2017,townsend2020transferable,coe2018machine,jeong2020automation,hermann2020deep,pfau2020ab}

Schemes have also been developed to exploit systematic errors between different levels of theory with different basis sets so that approximations can be extrapolated toward an exact result.
Examples include the complete basis set (CBS),\cite{Montgomery2000} Gaussian G$n$,\cite{Curtiss1998} Weizmann (W-)$n$\cite{Karton2006} methods, and high accuracy extrapolated \emph{ab initio} thermochemistry (HEAT)\cite{Tajti2004} methods. 
For a recent review on these and other methods see Ref.~\citenum{Karton2016}.
These schemes are also becoming a target of recent work using ML methods.\cite{Zaspel2019}

HF determinants provide good baseline approximations of the ground state electronic structure of many molecules, but they may describe poorly more complicated bonding that arises during bond dissociation events, excited states, and conical intersections.\cite{Park2020,Hirao1992,Buenker1978,Levine2008}  
Some many-body wavefunctions are best described as a superposition of two or more configurations, e.g. when other configurations in Eq.~\ref{eq:fci} can have similar or higher expansion coefficients $a$ than the HF determinant.
For this reason, high quality single reference methods like CCSD(T) fail because the theory assumes that salient electronic effects are captured by the initial single HF configuration. (In fact, methods such as CCSD(T) have been implemented with diagnostic approaches available that let users know when there may be cause for concern).\cite{Jiang2012,Lee2003,Duan2020}
In these cases, it may no longer be trivial to find reliable black-box or automated procedures (e.g. in situations involving resonance states, chemical reactions, molecular excited states, transition metal complexes, and metallic materials, etc.).\cite{Park2020}
So-called multiconfiguration approaches,\cite{Park2020} such as the generalized valence bond (GVB) method\cite{Bobrowicz1977} or the complete active space self consistent field (CASSCF),\cite{Roos1980} the multireference CI (MRCI) methods,\cite{Szalay2012} complete active space perturbation theory (CASPT2),\cite{Pulay2011} or multireference coupled cluster (MRCC),\cite{Lyakh2012,Evangelista2018} can more physically model these systems since they employ several suitable reference configurations with different degrees of correlation treatments.
These methods are not black-box and should be expected to require an experienced practitioner with CPI to choose the reference states that can substantially influence the quality of results.\cite{Jensen2005erratum} 
This is an area though where ML can bring progress in automating the selections of physically justified active spaces.\cite{jeong2020automation} 

In closing, there are a large number of available correlated wavefunction methods but many are even more costly than HF theory by virtue of requiring an HF reference energy expression shown in Eq. \ref{HFenergy}. 
Fig. \ref{fig:CompChemHierarchies}a depicts a so-called `magic cube' (that is an extension beyond a traditional `Pople diagram'\cite{Pople1965diagram,Zaspel2019}) that concisely shows a full hierarchy of computational approaches across different Hamiltonians, basis sets, and correlation treatment methods.
This makes it easy to identify different wavefunction methods that should be more accurate and more likely to provide useful atomic scale insights (as well as those that would be more computationally intensive).
Another important aspect highlighted in the `magic cube' is that higher level wavefunction methods require larger basis sets to successfully model electron correlation effects.  
A CCSD(T) computation carried out with a small basis set for example might only offer the same accuracy as MP2 while being two orders of magnitude more expensive to evaluate.\cite{helgaker2013molecular}
As was mentioned earlier with the benzene system, spurious errors with different basis sets might still be found that indicate problems with specific combinations of levels of theory and basis sets. 
The deep complexity of correlated wavefunction methods makes this a promising area for continued efforts in CompChem+ML research.

\subsubsection{Density-functional theory}\label{sec:DFT}
Density-functional theory (DFT)\cite{parr1994density} is another method to calculate the quantum mechanical internal energy of a system using an energy expression that relies on functionals (i.e., a function of a function) of electronic density $\rho$:
\begin{equation}
    \label{DFT_eqn}
    E[\rho] = T[\rho] + V[\rho] 
\end{equation}
Compared to wavefunction theory, DFT should be far more efficient since the dimensionality of a density representation for electrons will always be in three rather than the 3$n$ dimensions for any $n$-electron system described by a many-body wavefunction method.
DFT brings an important drawback that the exact expression for the energy functional is currently unknown, all approximations bring some degree of uncontrollable error, and this has precipitated disagreeable opinions from purists in chemical physics, especially those who are developing correlated wavefunction methods. 
However, there is also substantial evidence that DFT approximations are reasonably reliable and accurate for many practical applications that bring information, knowledge, and sometimes insight. 
We now provide a bird's-eye view of DFT-based methods.

One thrust of DFT developments since its inception has focused on designing accurate expressions strictly in terms of a density representation, and these approaches are referred to as `kinetic energy (KE-)' or `orbital-free (OF-)' DFT.\cite{Witt2018}
Some energy contributions (e.g. nuclear-electron energy and classical electron-electron energy terms) can be expressed exactly, but other terms such as the kinetic energy as a function of the density are not known and must be approximated. 
OF-DFT is very computationally efficient (these methods should scale linearly with system size\cite{Hung2010,Mi2016}) but these formulations have not yet been developed to rival the accuracy or transferability of wavefunction methods, though they have been used for studying different classes of chemical and materials systems.\cite{Mi2018Nonlocal,Ayers2005,Huang2010Nonlocal}
OF-DFT methods are also used in exciting applications modeling chemistry and materials under extreme conditions.\cite{Burakovsky2013,Sjostrom2015,Kang2020two}  
One should expect that once highly accurate forms are developed and matured, accurate CompChem calculations on electronic structures on systems having more than a million of atoms might become commonplace.
Indeed, there are efforts to use ML to develop more physical OFDFT methods.\cite{0363_Snyder2013,0060_Brockherde2017}  

The most commonly used form of DFT (which is also one of the most widely used CompChem methods in use today) is called Kohn--Sham (KS-)DFT.\cite{kohn1965ks}
In KS-DFT, one assumes a fictitious system of non-interacting electrons with the same ground state density as the real system of interest.
This makes it possible to split the energy functional in Eq.~\ref{DFT_eqn} into a new form that involves an exact expression of the kinetic energy for non-interacting electrons:
\begin{equation}
E[\rho] = T_\mathrm{ni}[\rho] + V_\mathrm{eN}[\rho] + V_\mathrm{ee}[\rho] + \Delta T_\mathrm{ee}[\rho] + \Delta V_\mathrm{ee}[\rho] \label{eq:ks}
\end{equation}
Here, $T_\mathrm{ni}[\rho]$ is the kinetic energy of the non-interacting electrons, $V_\mathrm{eN}[\rho]$ is the exact nuclear-electron potential and $V_\mathrm{ee}[\rho]$ is the Coulombic (classical) energy of the non-interacting electrons.
The last two terms are corrections due to the interacting nature of electrons and non-classical electron-electron repulsion.
KS-DFT also expands the three dimensional electron density into a spin orbital-basis $\phi$ similar to HF theory to define the one-electron kinetic energy in a straightforward manner.
This allows the $T_\mathrm{ni}$, $V_\mathrm{eN}$ and $V_\mathrm{ee}$ expressions to be evaluated exactly and one arrives at the KS energy:
\begin{align*}
\label{KS_DFT_eqn}
E_{\text{KS}}[\rho] =& -\underbrace{\sum_i \int d \mathbf{r}_1 \phi^*_i (\mathbf{r}_1)%
\frac{1}{2} \nabla^2
\phi_i (\mathbf{r}_1)}_{{T}_\mathrm{ni}}%
-\underbrace{\sum_i \int d \mathbf{r}_1 \phi^*_i (\mathbf{r}_1)%
\sum^N_\alpha \frac{Z_\alpha}{|\mathbf{r}_1-\mathbf{R}_\alpha|}%
\phi_i (\mathbf{r}_1)}_{V_\mathrm{eN}} \\
& + \underbrace{\sum_{i} \int d \mathbf{r}_1 \phi^*_i (\mathbf{r}_1) 
\left[ \frac{1}{2} \int d \mathbf{r}^{'} \frac{\rho(\mathbf{r}^{'})}{|\mathbf{r}_1-\mathbf{r}^{'} |} \right]%
\phi_i(\mathbf{r}_1)}_{\hat{V}_\mathrm{ee} (\mathrm{Coulomb})} + E_{\text{xc}}[\rho] \numberthis
\end{align*}
The last two correction terms in Eq.~\ref{eq:ks} arise from electron interactions, and these are combined into the so-called 'exchange-correlation' term ($E_{\text{xc}}$), which uniquely defines which scheme of KS-DFT is being used.
In theory, an exact $E_{\text{xc}}$ term would capture all differences between the exact FCI energy and the system of non-interacting electrons for a ground state.

The KS-DFT equations can be cast in a similar form as the Roothan--Hall equations (Eq.~\ref{HF_eqns}), which allows for a computationally efficient solution.
Moreover, the elements of the Kohn-Sham matrix (which replaces the Fock matrix $\mathbf{F}$) are easier to evaluate due to the fact that several of computationally intensive integrals are now accounted for via $E_{\text{xc}}$.
Hence, the formal scaling for KS-DFT is $\mathcal{O}(n^3)$ with respect to the number of electrons.
Even though this is much poorer scaling than ideally linear scaling OF-DFT, the exact treatment of non-interacting electrons makes KS-DFT more accurate. 
Furthermore, there are several modern exchange-correlation functionals that routinely achieve much higher accuracy than HF theory with less computational cost, and thus KS-DFT is a competitive alternative with many correlated wavefunction methods in many modern applications.

A remaining problem is constructing a practical expression for the exchange-correlation functional, as its exact functional form remains unknown. This has spawned a wealth of approximations that have been founded with different degrees of first principles and/or empirical schemes.
Classes of KS-DFT functionals are defined by whether the exchange-correlation functional is based on just the homogeneous electron gas (i.e. the `local density approximation', LDA), that and its derivative (i.e. the `generalized gradient approximation', GGA), as well as other additional terms that should result in physically improved descriptions and/or error cancellations. 
The resulting hierarchy of KS-DFT functionals is often referred to as a `Jacob's Ladder' of DFT (Fig. \ref{fig:CompChemHierarchies}b). 
Generally, the higher up the ladder one goes, the more accurate but more computationally demanding the calculation.\cite{Maurer2019}
While the intrinsic inexactness in DFT makes it difficult to assess which functionals are physically better than others.\cite{Jacobsen2012,BurkeWebPage} 
The Jacob's Ladder hierarchy is useful for clearly designating how and why newer methods should perform in specific applications (for perspective see Refs.~\citenum{Tran2016,Kozuch2013,Janesko2017}), though there remains substantial benefit to those that bring CPI to the development efforts.

Indeed, by being based on a ground-state representation for homogeneous electron gas, DFT calculations can sometimes bring more easily  physical insight into some systems that are very challenging for wavefunction theory to examine (e.g. metals\bibpunct{}{}{,}{s}{,}{,}\cite{Gerber2007,Pisani1988} where HF theory provides divergent exchange energy behaviors). 
On the other hand, DFT is also generally not well-suited for studying physical phenomena involving localized orbitals or band structures such as those found in semiconducting materials with small band gaps, molecular or material excited charge transfer states, or interaction forces that can arise due to excited states, e.g. dispersion (or London) forces. The former features can normally be treated using Hubbard-corrected DFT+U models that require a system-specific $U-J$ parameter\cite{Shishkin2019,Petukhov2003} or more generalizable but much more computationally expensive hybrid DFT approaches. 
Dispersion forces (i.e. van der Waals interactions) are non-existent in semilocal DFT approximations, and it is now commonplace to introduce them into DFT calculations using a variety of different methods.\cite{Stohr2019}  

There is also growing interest in using embedded QC calculation schemes that can partition systems into discrete regions that could be treated with highly accurate correlated wavefunction theory and computationally efficient KS-DFT schemes separately.\cite{Sun2016Quantum,Cortona1991,Huang2008Advances,Manby2012,Libisch2014} 
DFT has also been extended to the modeling of excited states in the form of time-dependent (TD-)DFT.\cite{casida2012tddft}
Similar to ground state DFT, TDDFT is a computationally inexpensive alternative to excited state wavefunction-based methods.
The approach yields reasonable results where excitations induce only small changes in the ground state density, e.g. low lying excited states.\cite{casida2012tddft,casida1998excited}
However, due to its single reference nature, TDDFT tends to break down in situations where more than one electronic configuration contribute significantly to the excited state. 
Just as with correlated wavefunction methods, there are already signs of CompChem+ML efforts to improve the applicability of DFT-based methods.\cite{0362_Snyder2012,schmidt2019machine,meyer2020machine,bogojeski2020quantum,nagai2020completing}

\subsubsection{Semi-empirical methods}

Correlated wavefunctions and, to a lesser degree, KS-DFT are still very computationally demanding and only of limited use for large scale simulations.
Further approximations based on wavefunctions and DFT methods have been developed to simplify and accelerate energy calculations.
These so-called semi-empirical methods still explicitly consider the electronic structure of a molecule, but in a more approximate way than methods described above.

Semi-empirical approaches based on wavefunction theory include methods like extended H\"{u}ckel theory and neglect of diatomic differential overlap (NDDO).\cite{thiel2014semiempirical}
Both approaches are simplifications of the Hartree--Fock equations (Eq.~\ref{HFenergy}) by introducing approximations to the different integrals.
In the NDDO approach,\cite{pople1970approximate} only the two-electron integrals in Eq.~\ref{HFenergy} are considered, where the two orbitals on the right and left hand side of the $\frac{1}{|\mathbf{r}_i - \mathbf{r}_j |}$ operator are located on the same atom.
The remaining two-center (and one-center) integrals are then approximated by introducing a set of empirical functions, one for each unique type of integral.
Moreover, the overlap matrix in Eq. \ref{HF_eqns} is assumed to be diagonal, which greatly simplifies the energy evaluation.
This reduces the required computational effort tremendously and allows the scaling of these approaches to be reduced to $\mathcal{O}(N^2)$.
NDDO serves as a basis for more sophisticated semi-empirical schemes, such as AM1,\cite{dewar1985development} PM7\cite{stewart2013optimization} and MNDO,\cite{dewar1977ground} where the energy is usually determined self-consistently using a minimally sized basis set.
Inadequacies in theory can be compensated by different empirical parametrization schemes that can allow these calculations to rival the accuracy of higher level theory for some systems. For example Dral et al.\cite{Dral2016} provided a recent `big-data' analysis of the performance of several semi-empirical methods with large datasets.

Semi-empirical schemes are also carried over to approximate KS-DFT with so-called density functional tight binding (DFTB).\cite{koskinen2009density}
DFTB simplifies the Kohn--Sham equations (Eq.~\ref{KS_DFT_eqn}) by decomposing the total electron density $\rho$ into a density of free and neutral atoms $\rho_0$ and a small perturbation term $\delta \rho_0$ ($\rho = \rho_0 + \delta\rho_0$).
Expanding Eq.~\ref{KS_DFT_eqn} in the perturbation $\delta \rho_0$ makes it possible to partition the total energy into three terms amendable to different approximation schemes:
\begin{equation}
E[\delta \rho_0] = E_\mathrm{rep} + E_\mathrm{Coul}[\delta \rho_0] + E_\mathrm{BS}[\delta \rho_0].
\label{eq:DFTB}
\end{equation}
$E_\mathrm{rep}$ is a repulsive potential containing interactions between the nuclei and contributions from the exchange correlation functional (these are typically approximated via pairwise potentials).
The charge fluctuation term $E_\mathrm{Coul}$ is modeled as a Coulomb potential of Gaussian charge distributions computed from the approximate density.
Finally, $E_\mathrm{BS}$ refers to the `band structure' term which considers the electronic structure and contains contributions from $T_\mathrm{ni}$, $V_\mathrm{eN}$, and the exchange correlation functional (see Eq.~\ref{KS_DFT_eqn}).
To compute $E_\mathrm{BS}$, the density is expressed in a minimal basis of atomic orbitals, similar as in NDDO.
The necessary Hamiltonian and overlap integrals are then evaluated via an approximate scheme based on Slater--Koster transformations.
In addition to the energy, atomic partial charges are also computed in this step, which are then used in $E_\mathrm{Coul}$.
As a consequence, DFTB equations can also be solved self-consistently.
DFTB methods are parameterized by finding suitable forms for the repulsive potential and adjusting the parameters used in the Slater--Koster integrals.
Non-self consistent and self-consistent tight-binding DFT methods\cite{Elstner1998, Bannwarth2019} have been developed for simulating large scale systems.
Semi-empirical methods have also been a target of different ML schemes, yielding improved parametrization schemes and more accurate functional approximations.\cite{dral2015machine,hegde2017machine,stoehr2020accurate,li2018density}

\subsubsection{Nuclear quantum effects}\label{sec:nqes}

The quantum nature of lighter elements such as H--Li and even heavier elements that form strong chemical bonds (C--C bond in graphene for example~\cite{poltavsky2018quantum}) gives rise to significant nuclear quantum effects (NQEs). 
Such effects are responsible for large differences from the Dulong-Petit limit of the heat capacity of solids, isotope effects, and the deviations of the particle momentum distribution from the Maxwell-Boltzmann equation.~\cite{ceri+16cr}
To capture NQEs, path-integral molecular dynamics (PIMD)~\cite{marx1996ab,chandler1981exploiting} or centroid molecular dynamics (CMD)~\cite{cao1994formulation,hele2015communication} can be used,
but these methods are associated with much higher computational costs (usually about 30 times higher) compared with classical MD simulations using point nuclei.
Moreover, because systems may be influenced by competing NQEs,
the extent of NQEs is sensitive to the potential energy surface assumed.
(Semi-)local DFT approaches may not even qualitatively predict isotope fractionation ratios, and usually hybrid DFT is needed to reach quantitative accuracy.~\cite{wang2014quantum}
However, employing hybrid DFT calculations or beyond in PIMD/CMD simulations can accrue extremely high computational costs.
For this reason, ML force fields have been proposed as efficient means to carry out PIMD simulations, enabling essentially exact quantum-mechanical treatment of both electronic and nuclear degrees of freedom, at least for small molecules with dozens of atoms.~\cite{sauceda2020dynamical,Chmiela2018}

\subsubsection{Interatomic Potentials}\label{sec:FF}

\begin{table}
\scriptsize
\caption{Types of interatomic potentials and their areas of application.}
\label{tab:potential}
\begin{tabular}{p{0.15\linewidth} p{0.1\linewidth} p{0.25\linewidth} p{0.4\linewidth}}
\toprule
Potential      & Reactive & Typical applications                                    & Examples \\
\midrule
\raggedright Pairwise-distance-based & sometimes & \raggedright materials, liquids & Lennard-Jones~\cite{Wang2020LJ,Li2015LJ},
Morse~\cite{girifalco1959application}, Coulomb interactions, Buckingham~\cite{buckingham1938classical}\\ \\

\raggedright Distance and angle-based & usually no & \raggedright materials, liquids &  many water potentials (e.g. SPC, TIP4P, mW)\cite{Jorgensen1983Comparison}, Stillinger-Weber~\cite{stillinger1985computer}\\ \\

\raggedright Class I (non-polarizable) force fields & no       & Proteins, lipids, polymers, nucleic acids, carbohydrates, organic molecules, liquids & AMBER,\cite{weiner1981amber, Salomon-Ferrer2013} GAFF,\cite{Wang2004GAFF} CHARMM,\cite{brooks2009charmm} GROMOS,\cite{oost+04jcc,Schmid2011GROMOS,daura1998parametrization} OPLS,\cite{jorg+96jacs,jorgensen1984optimized} DREIDING,\cite{Mayo1990} MMFF94,\cite{Halgren1996} UFF,\cite{Rappe1992} COMPASS,\cite{Sun1998Compass} INTERFACE\cite{Heinz2013}, interatomic potentials for ionic systems\cite{Gale1996} \\ \\
\raggedright Class II (polarizable) & no       & Proteins, lipids, polymers, nucleic acids, carbohydrates, organic molecules, liquids & AMOEBA,\cite{pond+10jpcb} classical Drude oscillator models,\cite{Lemkul2016} fluctuating charge (FQ) models,\cite{Banks1999} MB-Pol,\cite{Babin2013} distributed point polarizable models (DPP2),\cite{Kumar2010} and many more.\cite{Xu2018Perspective} \\ \\
\raggedright Embedded atom method (EAM)-like & yes      & Reactions within solid materials & EAM,\cite{daw+93msr} MEAM,\cite{baskes1992modified} Finnis-Sinclair,\cite{Finnis1984} Sutton-Chen\cite{Sutton1990}  \\ \\
Bond-order potentials (BOPs)            & yes      & Reactions within solids, liquids, gases & Brenner,\cite{brenner1990empirical} Tersoff,\cite{tersoff1988new,ters89prb} REBO,\cite{brenner1990empirical,Brenner2002} COMB,\cite{Liang2013COMB,yu2007charge} ReaxFF,\cite{Senftle2016,van2001reaxff} APT\cite{Rappe2007} \\ \\
\raggedright Other quantum mechanics-derived forcefields & yes & Reactions within liquids and gases & EVB\cite{Warshel1980} and related models\cite{Wu2008an,Hartke2015}  \\ 
\bottomrule
\end{tabular}
\end{table}

Interatomic potentials introduce an additional level of abstraction compared to methods described above.
Instead of using exact quantum mechanical expressions to create the PES for the system, analytic functions are used to model a pre-supposed PES that contains explicit interactions between atoms, while electrons are treated in an implicit manner (sometimes using partial charge schemes).\cite{Singh1984,Storer1995,Mehler1991,Chen2007QTPIE,Poier2019,Rappe1991}
Interatomic potentials thus are (often times dramatically) more computationally efficient than correlated wavefunction, DFT, and semi-empirical approaches.
This efficiency makes it possible to study even larger systems of atoms (e.g. biomolecules, surfaces, and materials) than is possible with other computational methods.
Note that different empirical potentials bring substantially different computational efficiencies; for example LJ potentials are more efficient than classical forcefields like AMBER and CHARMM, while those are more efficient than most bond-order potentials such as ReaxFF.\cite{Senftle2016,van2001reaxff}
The degree of efficiency arises from the balance of using accurate and/or physically justified functional forms, approximations, and model parameterizations.
There are many different formulations (see Fig.~\ref{fig:CompChemHierarchies}c), and we will discuss the most general classes.
An overview of the different types of potentials and their features is provided in Tab.~\ref{tab:potential}.
For extensive discussions on these methods including semi-empirical approaches, we refer to the extensive review by Akimov and Prezhdo (Ref.~\citenum{akimov2015large}).
An excellent review for interatomic potentials is provided by Harrison \emph{et. al.} (Ref.~\citenum{harrison2018review}), and an excellent overview of modern methods can be found in a special issue of J. Chem. Phys.\cite{Piquemal2017}

The distinctions between different types of forcefields can be blurry sometimes, and we will differentiate categories in ascending complexity.
One of the simplest interatomic potentials is the Lennard--Jones potential:\cite{Lennard_Jones_1931}
\begin{equation}
E_\mathrm{LJ} = \sum_{i, j>i} 4 \varepsilon_{ij} \left[ \left(\frac{\sigma_{ij}}{r_{ij}}\right)^{12} - \left(\frac{\sigma_{ij}}{r_{ij}}\right)^{6}\right].
\label{eq:lj}
\end{equation}
It models the total energy as the sum of all pairwise interaction between atoms $i$ and $j$ using an attractive and repulsive term depending on the interatomic distance $r_{ij}$.
$\varepsilon_{ij}$ modulates the strength of the interaction function, while $\sigma_{ij}$ defines where it reaches its minimum.
The Lennard--Jones potential is a prototypical ``good model'' of interatomic potentials, as it has a sufficiently simple physical form with only two parameters while still yielding useful results.

For covalent systems such as bulk carbon or silicon, just pairwise distances are not sufficient to capture the local coordination of the atoms, and many empirical potentials~\cite{Jorgensen1983Comparison,stillinger1985computer,tersoff1988empirical} for these systems were expressed as a function of the pairwise distances and three-body terms within a certain cutoff distance.
The pairwise term can take the form of LJ-type, electrostatic, or harmonic potentials,
and the three-body term is usually a function of the angles formed by sets of three atoms.

So-called Class I classical force fields introduce a more complicated energy expression:
\begin{align*}
E_\mathrm{tot} = &\sum_{\mathrm{bonds}} k_{ij} \left(r_{ij} - \bar{r}_{ij} \right)^2 + \\
&\sum_{\mathrm{angles}} k_{ijk} \left(\theta_{ijk} - \bar{\theta}_{ijk} \right)^2 + \\
&\sum_{\mathrm{dihedral}} \sum_\gamma k_{ijkl}^{(\gamma)} \left[ 1 - \cos(\gamma\phi_{ijkl} - \bar{\phi}_{ijkl}^{(\gamma)} )\right] + \\
&\sum_{i, j>i} \frac{q_i q_j}{r_{ij}} + E_\mathrm{LJ} \numberthis
\label{eq:blff}
\end{align*}
The first three terms are the energy contributions of the distances ($r_{ij}$), angles ($\theta_{ijk}$) and dihedral angles ($\phi_{ijkl}$) between bonded atoms.
Because of this, they are also referred to as bonded contributions.
Bond and angle energies are modeled via harmonic potentials, with the $k_{ij}$ and $k_{ijk}$ parameters modulating the potential strength and $\bar{r}_{ij}$ and $\bar{\theta}_{ijk}$ being the equilibrium distances and angles.
The dihedral term is modeled with a Fourier series to capture the periodicity of dihedral angles, with $k_{ijkl}$ and $\phi_{ijkl}$ as free parameters.
The last two terms account for non-bonded interactions.
The long range electrostatics are modeled as the Coulomb energy between charges $q_i$ and $q_j$ and the van der Waals energy is treated via a Lennard--Jones potential (Eq.~\ref{eq:lj}).
In Class I/II force fields, empirical parameters are tabulated for a variety of elements in wide ranges of chemical environments (for example Ref.~\citenum{Vanommeslaeghe2010}). 
Parameters for any one system should not necessarily be assumed to transfer well to other systems, and reparametrizations may be needed depending on the application.
Different sets of parametrization schemes give rise to different types of classical FFs, with CHARMM,\cite{brooks2009charmm} Amber,\cite{weiner1981amber,Salomon-Ferrer2013} GROMOS\cite{oost+04jcc,Schmid2011GROMOS,daura1998parametrization} and OPLS\cite{jorg+96jacs,jorgensen1984optimized} being a few of many examples.

An extension beyond these FFs are Class II (i.e. ``polarizable'') force fields, where the static charges are replaced by environment dependent functions (e.g. AMOEBA\cite{zhang2018amoeba}).
A significant advantage to Class I and II types of forcefields is that they are computationally efficient, which makes them will suited for molecular dynamics simulations of complex and extended (bio)molecules, such as proteins, lipids or polymers. 
Implementations of forcefield calculations on GPUs makes these simulations extremely productive.\cite{Gotz2012,Salomon-Ferrer2013GPU,Stone2010,Glaser2015,Lagardere2018} 
A disadvantage of Class I and II types of interatomic potentials is that they rely on predefined bonding patterns to compute the total energy, and this limits their transferability.
In general, bonds between atoms are defined at the beginning of the simulation run and cannot change. Furthermore, bonding terms make use of harmonic potentials that are not suitable for modeling bond dissociation.

Reactive potentials, which eschew harmonic potential dependencies and thus can describe the formation and breaking of chemical bonds, include the embedded atom method (EAM, Fig.~\ref{fig:CompChemHierarchies}c), which is used widely in materials science.\cite{daw+93msr}
EAM is a type of many-body potential primarily used for metals, where each atom is embedded in the environment of all others.
The total energy is given by
\begin{equation}
E_\mathrm{tot} = \sum_i^N \left[ F_i(\tilde{\rho}_i) + \frac{1}{2} \sum_{j \neq i} V_{ij}(r_{ij}) \right].
\label{eq:eam}
\end{equation}
$F_i$ is an embedding function and $\tilde{\rho}_i$ an approximation to the local electron density based on the environment of atom $i$.
$F_i(\tilde{\rho}_i)$ can be seen as a contribution due to non-localized electrons in a metal.
$V_{ij}$ is a term describing to the core-core repulsion between atoms.
An EAM potential is determined by the functional forms used for $F_i$ and $V_{ij}$ as well as how the density is expressed.
Its dependence on the local environment without the need for predefined bonds make EAM well suited for modeling material properties of metals.
An extension of EAM is modified EAM (MEAM)\cite{baskes1992modified} which includes directional dependence in the description of the local density $\tilde{\rho}_i$, but this brings greater computational cost.
EAMs also form the conceptual basis of the embedded atom neural network (EANN) MLPs.\cite{zhang2019embedded}

Another common type of reactive potentials are bond order potentials (BOPs).
In general, BOPs model the total energy of a system as interactions between the neighboring atoms: 
\begin{equation}
E_\mathrm{tot} = \sum_{i, j>i} \left[ V_\mathrm{rep}(r_{ij}) - b_{ij(k)} V_\mathrm{att}(r_{ij}) \right] f_\mathrm{cut}(r_{ij}).
\label{eq:bop}
\end{equation}
$V_\mathrm{rep}$ and $V_\mathrm{att}$ are repulsive and attractive potentials depending on the interatomic distance $r_{ij}$.
A cutoff function $f_\mathrm{cut}$ restricts all interactions to the local atomic environment.
$b_{ij(k)}$ is the bond order term, from which the potential takes its name.
This term measures the bond order between atoms $i$ and $j$ (i.e., `1' for a single bond, `2' for a double bond, `0.6' for partially dissociated bond). 
Bond orders can also depend on neighboring atoms $k$ in some implementations.
BOPs are typically used for covalently bound systems, such as bulk solids and liquids containing hydrogen, carbon or silicon (e.g. carbon nanotubes and graphene).
Depending on the exact form of the expressions in Eq.~\ref{eq:bop}, different types of BOPs are obtained, such as Tersoff\cite{ters89prb,tersoff1988new} and REBO\cite{brenner1990empirical,Brenner2002} potentials.
BOPs can also be extended to incorporate dynamically assigned charges, yielding potentials like COMB\cite{Liang2013COMB,0001_Agrawal2016} or ReaxFF.\cite{Senftle2016,van2001reaxff}
As with EAMs, BOPs have also been used as a starting point for constructing more elaborate machine learning potentials (MLPs)\cite{pun2019physically,guo2020intelligent,narayanan2017machine} that will also be discussed in more detail in Section 4.  

While efficient and versatile, all interatomic potentials described above are inherently constrained by their functional forms.
A different approach is pursued by machine learned potentials (MLPs), such as Behler-Parinello Neural Networks,\cite{0032_Behler2007a} q-SNAP,\cite{wood2018extending} and GAP potentials\cite{0025_Bartok2010} (Fig. \ref{fig:CompChemHierarchies}c).
In MLPs, suitable functional expressions for interactions and energy are determined in a fully data-driven manner and ultimately only limited by the amount and quality of available reference data.
One can then use substantially more data to generate a much more accurate MLP than would be possible when using, for instance, a ReaxFF potential trained on similar data sets.\cite{Boes2016} 

For the sake of completeness, we note that all approaches described here are fully atomistic -- each atom is modeled as an individual entity.
It is also possible to combine groups of atoms into pseudo-particles giving rise to so-called coarse grained methods.
On an even higher level of abstraction, whole environments can be modeled as a single continuum.
As such approaches are not subject of the present review, we refer the interested reader e.g. to Refs.~\citenum{ingolfsson2014power} and \citenum{mennucci2012polarizable}.

\subsection{Response properties}
\label{sec:responseProps}

Once an energy calculation is completed by one of the CompChem methods above, many other interesting molecular properties can be calculated.
Most of these properties can be obtained as the response of the energy to a perturbation, e.g. changes in nuclear coordinates $\mathbf{R}$, external electric ($\boldsymbol{\epsilon}$) or magnetic ($\mathbf{B}$) fields or the nuclear magnetic moments $\{\mathbf{I}_i\}$.
Given an expression for the energy, which depends on the above quantities, so-called response properties can be computed via the corresponding partial derivatives of the energy.
A general response property $\boldsymbol{\Pi}$ then takes the form
\begin{equation}
\boldsymbol{\Pi}(n_\mathbf{R}, n_{\boldsymbol{\epsilon}}, n_\mathbf{B}, n_{\mathbf{I}_i}) = \frac{\partial^{n_\mathbf{R} + n_{\boldsymbol{\epsilon}} + n_\mathbf{B} + n_{\mathbf{I}_i}} E(\mathbf{R}, \boldsymbol{\epsilon}, \mathbf{B}, \mathbf{I}_i)}{\partial \mathbf{R}^{n_\mathbf{R}} \partial \boldsymbol{\epsilon}^{n_{\boldsymbol{\epsilon}}} \partial \mathbf{B}^{n_{\mathbf{B}}} \partial \mathbf{I}_i^{n_{\mathbf{I}_i}}},\label{eq:response}
\end{equation}
where the $n$s indicate the $n$-th order partial derivative with respect to the quantity in the subscript.\cite{jensen2007introduction}

A common response property are nuclear forces $\mathbf{F} = -\boldsymbol{\Pi}(1,0,0,0)$ that are the negative first derivative of the energy with respect to the nuclear positions. 
Such calculations allow a plethora of different geometry optimization schemes for chemical structures on the PES.
Hessian calculations corresponding to the second derivative of energy with respect to nuclear positions are necessary to confirm the location of first order saddle points on the PES and identify normal modes and their frequencies for vibrational partition functions that are useful for modeling temperature dependencies based on statistical thermodynamics.
Hessian calculations are computationally costly, since they normally involve calculations based on finite differences methods involving many nuclear force calculations. 
Many methods have been developed to allow CompChem algorithms to sample minimum energy regions of the PES\cite{Jager2018,Dieterich2010,Wales1997,Zhang2015ABCcluster,Goedecker2004} or precisely locate points of interest.\cite{Schlegel2011,Sheppard2008} 
Historically, many of these techniques have relied on approximate or full Hessian calculations,\cite{BernhardSchlegel1984} but other approaches such as the nudged-elastic band\cite{henk+00jcp,Sheppard2012} and string\cite{Zimmerman2013,Samanta2013,Burger2006} methods are popular alternatives that do not require a Hessian calculation. 
There have also been efforts using different forms of ML to accelerate procedures or overcome long-standing challenges in efficient sampling of and optimization on the PES.\cite{Peterson2016,DelRio2019,GarridoTorres2019,Meyer2019,Koistinen2019,noe2019boltzmann}

The general expression above can provide a wealth of other quantities, some of which are relevant for molecular spectroscopy and/or provide a direct connection to experiment (see Tab.~\ref{tab:response_properties}).
\begin{table}[hbtp!]
\scriptsize
	\caption{Response properties of the potential energy.}
	\label{tab:response_properties}
	\begin{tabular}{lllll}
		\toprule
		$n_\mathbf{R}$ & $n_{\boldsymbol{\epsilon}}$ & $n_\mathbf{B}$ & $n_\mathbf{I}$ & \textbf{Property} \\
		\midrule
		0 & 0 & 0 & 0 & Energy\\
		1 & 0 & 0 & 0 & Forces\\
		2 & 0 & 0 & 0 & Hessian (harmonic frequencies)\\
		0 & 1 & 0 & 0 & Dipole moment (IR)\\
		1 & 1 & 0 & 0 & Infrared absorption intensities (IR)\\
		0 & 2 & 0 & 0 & Polarizability (Raman)\\
		0 & 0 & 1 & 1 & Nuclear magnetic shielding (NMR)\\
		0 & 0 & 0 & 2 & Nuclear spin-spin coupling (NMR)\\
		0 & 1 & 1 & 0 & Optical rotation (circular dichroism)\\
	\end{tabular}
	
	\centering
	$\vdots$
\end{table}	
Infrared spectra can be simulated based on dipole moments $\boldsymbol{\mu} = -\boldsymbol{\Pi}(0,1,0,0)$, while
molecular polariziabilities $\boldsymbol{\alpha} = -\boldsymbol{\Pi}(0,2,0,0)$ offer access to polarized and depolarized Raman spectra.
A central response property of the magnetic field are nuclear magnetic shielding tensors $\boldsymbol{\sigma}_i = \boldsymbol{\Pi}(0,0,1,1)$.
These allow the computation of chemical shifts recorded in nuclear magnetic resonance (NMR) spectroscopy via their trace $\sigma_i = \frac{1}{3}\mathrm{Tr}[\boldsymbol{\sigma}_i]$.
The beauty of this formalism lies in the fact, that a single energy calculation method provides access to a wide range of quantum chemical properties in a highly systematic manner.
A large number of modern MLPs use the response of the potential energy with respect to nuclear positions to obtain energy conserving forces.
However, far fewer applications model perturbations with respect to electric and magnetic fields.
Ref.~\citenum{christensen2019operators} extends the descriptor used in the FCHL Kernel by an explicit field dependent term making it possible to predict dipole moments across chemical compound space. Ref.~\citenum{gastegger2020machine} introduces a general NN framework to model interactions of a system with vector fields, which was then used to predict dipole moments, polarizabilities and nuclear magnetic shielding tensors as response properties.

\subsection{Solvation models}
An important aspect of CompChem is molecular descriptions from with a solution environment.
Simulating a dynamical environment composed of many surrounding molecules is usually not feasible with electronic-structure methods. 
To circumvent this problem, solvation modeling schemes have been devised (see Refs.~\citenum{varghese2019origins,Basdogan2020,Tomasi2005,Cramer2008universal,Klamt2011,hirata2003molecular} for discussions on this topic).

The most popular approach are so-called polarizable continuum solvent models (PCM).\cite{mennucci2012polarizable} They model the electrostatic interaction of a solute molecule with its environment by representing the charge distribution of the solvent molecules as a continuous electric field, the reaction field. This dielectric continuum can be interpreted as a thermally averaged representation of the environment and is typically assigned a constant permittivity depending on the particular solvent to be modeled ($\varepsilon=80.4$ for water). The solute is placed inside a cavity embedded in this continuum.
The charge distribution of the molecule then polarizes the continuous medium, which in turn acts back on the molecule. To compute the electrostatic interactions arising from this mutual polarization with electronic structure theory, a self consistent scheme is employed. After constructing a suitable molecular cavity, a Poisson problem of the following form is solved: 
\begin{equation}
-\nabla\left[ \epsilon(\mathbf{r}) \nabla V(\mathbf{r}) \right] = 4 \pi \rho_\mathrm{m}(\mathbf{r})\label{eq:pb}
\end{equation}
Here, $\rho_\mathrm{m}(\mathbf{r})$ is the charge distribution of the solute and $\epsilon(\mathbf{r})$ is the position dependent permittivity, which usually is set to one within the cavity and the $\varepsilon$ of the solvent on the outside.
$V(\mathbf{r})$ is the electrostatic potential composed of the two terms
\begin{equation}
V(\mathbf{r}) = V_\mathrm{m}(\mathbf{r}) + V_\mathrm{s}(\mathbf{r}),
\end{equation}
where $V_\mathrm{m}(\mathbf{r})$ is the solute potential and $V_\mathrm{s}(\mathbf{r})$ is the apparent potential due to the surface charge distribution $\sigma(\mathbf{s})$,
\begin{equation}
V_\mathrm{s}(\mathbf{r}) = \int_{\Gamma} \frac{\sigma(\mathbf{s})}{|\mathbf{r}-\mathbf{s}|} d \mathbf{s}.\label{eq:pot_solv}
\end{equation}
$\Gamma$ indicates the surface of the cavity.
Eq.~\ref{eq:pb} is solved numerically to obtain the surface charge distribution $\sigma(\mathbf{s})$.
Once $\sigma(\mathbf{s})$ has been determined in this fashion, the potential is computed according to Eq.~\ref{eq:pot_solv} and used to construct an effective Hamiltonian of the form
\begin{equation}
\hat{H}_\mathrm{eff} = \hat{H} + V_\mathrm{s}(\mathbf{r}),
\end{equation}
where $\hat{H}$ is the vacuum Hamiltonian.
These equations are then solved self consistently in a Roothan--Hall or Kohn--Sham approach, yielding the electrostatic solvent-solute interaction energy. This scheme is also called the self-consistent reaction field approach (SCRF).

Continuum models differ in how the cavities are constructed and how Eq.~\ref{eq:pb} is solved to obtain the surface charge distribution.
Variants include the original PCM model, also refered to as dielectric PCM (D-PCM),\cite{miertuvs1981electrostatic} the integral equation formulation of PCM (IEFPCM),\cite{cances1997new} SMD,\cite{marenich2009universal} conductor PCM (C-PCM)\cite{barone1998quantum} or the conductor-like screening model (COSMO).\cite{klamt1993cosmo}
The latter two approaches replace the dielectric medium by a perfect conductor to allow for a particularly efficient computation of $\sigma(\mathbf{s})$.
PCMs can be further extended with statistical thermodynamics treatments to account for solutes having different size and concentration effects, and this leads to models such as COSMO-RS.\cite{klamt1995conductor} 

A drawback of most PCM-like approaches is that they neglect local solvent structures. 
Thus, they cannot reliably account for situations where explicit solvent interactions are important, e.g. when for stabilizing specific sites for a transition state through hydrogen bonding.\cite{varghese2019origins}
Furthermore, while implicit models might be parametrized to fit bulk-like properties of mixed or ionic solvents (e.g. Ref.~\citenum{Bernales2012}), the complex local solvent environment presented by these systems are treatable by other means.
For mixed solvent systems a range of hybrid schemes such as COSMO-RS,\cite{Klamt2011} reference interaction site models (RISMs)\cite{Truchon2014,Nishihara2017} or QM/MM\cite{Kamerlin2009,Boereboom2018,Gregersen2004} approaches have been developed. As an in-depth discussion of these alternative schemes exceeds the scope of this review, we instead refer to other references.\cite{Maldonado2020,skyner2015review}

ML models are becoming used to describe solvent effects.
Ref.~\citenum{gastegger2020machine} introduces a continuum ML model based on a reaction field that can predict energies and  response properties for continuum solvents, it can extrapolate to solvents not seen during training, and it can be extended to operate in a QM/MM fashion to account for explicit solvents effects in a Claisen rearrangement reaction.
Ref.~\citenum{Basdogan2020Machine} implemented automatable calculation schemes and unsupervised ML to allow predictions of single ion solvation energies for monovalent and divalent cations and anions based on physically rigorous quasi-chemical theory.\cite{Pratt1998,Rempe2000} 
Ref.~\citenum{Chew2020} used convolutional neural networks and molecular dynamics simulations to carry out high-throughput screening of mixed solvent systems. 
Ref.~\citenum{Zhang2019solvation} implemented efficient ways to carry out ML-based QM/MM molecular dynamics simulations.

\subsection{Insightful predictions for molecular and material properties}
By solving for electronic structures, by whatever means is appropriate, one obtains molecular energies and energy spectrum (typically corresponding to \emph{quasiparticles} given by Kohn-Sham or Hartree-Fock orbitals). 
From these, one can then compute molecular or material properties that arise from quantum mechanical and statistical operators, e.g. thermodynamic energies, response properties, highest and lowest occupied molecular orbital energies, band gaps, among other properties. 
Many properties are defined by the characters of the orbitals, and having knowledge of these should always be helpful and aid in deriving useful insight into designing molecules and materials for a particular function.
Furthermore, one is often interested in how these molecules behave over time (i.e. the dynamics given some statistical ensemble that depends on temperature, pressure, etc) over all possible degrees of freedom. 
By understanding how energies and forces change over time, one can predict thermal and pressure dependencies as well as spectroscopic properties for advanced knowledge that builds toward insightful predictions. 

Molecular and materials chemistry is vastly complex and variable, and one often faces a question of whether to span wider chemical spaces versus take deeper explorations of a specific phenomenon. 
A key problem is that even after the effort of either approach, it is also not as clear how information for one system might be related to another to provide more knowledge. 
For instance, one may decide to calculate all possible properties of ethanol with a  CompChem method, but understanding how any calculated property would be correlated to an analogous property of isopropanol is still usually difficult to do. 
There is great interest in understanding chemical and materials space through applications of quantitative structure activity/property relationships,\cite{Katritzky2010,0439_Muratov2020} cheminformatics,\cite{0123_Fourches2010} conceptual DFT,\cite{Geerlings2020} and alchemical perturbation DFT.\cite{VonRudorff2020}
All these applications benefit from greater access to CompChem data, and all have promise as being interfaced with ML for transformative applications to catalyze wisdom and impact.

%% file: Sections/3-ML-sc.tex
\newpage
\section{Machine Learning Tutorial and Intersections with Chemistry}
\label{section:ML}

Machine learning (ML) has had a dramatic impact on many aspects of our daily lives and has arguably become one of the most far-reaching technologies of our era. 
It is hard to overstate its importance in solving long standing computer science challenges such as image classification~\cite{hinton2012improving,szegedy2015going,russakovsky2015imagenet, krizhevsky2017imagenet} or natural language processing~\cite{blei2003latent,bengio2003neural,wu2016google,mikolov2013efficient,vaswani2017attention} -- tasks that require knowledge that is hard to capture in a traditional computer program~\cite{friedman2001elements,rasmussen2003gaussian,bishop2006pattern}. 
Previous classical artificial intelligence (AI) approaches relied on very large sets of rules and heuristics, but these were unable to cover the full scope of these complex problems. 
Over the past decade, advances in ML algorithms and computer technology made it possible to learn underlying regularities and relevant patterns from massive datasets that enable automatic constructions of powerful models that can sometimes even outperform humans at those tasks.

This development inspired researchers to approach challenges in science with the same tools, driven by the hope that ML would revolutionize their respective fields in a similar way. 
Here, we give an overview of these developments in chemistry and physics to serve as an orientation for newcomers to ML.
We will first explain what tasks ML is good at and when it might not be the best solution to a problem. We will start by introducing the field of ML in general terms and dissect its strengths and weaknesses. 

\subsection{What is ML?}

In the most general sense, ML algorithms estimate functional relationships without being given any explicit instructions of how to analyze or draw conclusions from the data. Learning algorithms can recover mappings between a set of inputs and corresponding outputs or just from the inputs alone. Without output labels, the algorithm is left on its own to discover structure in the data.

\emph{Universal approximators}~\cite{hornik1989multilayer,tran2015variational} are commonly used for that purpose. These reconstruct any function that fulfills a few basic properties, such as continuity and smoothness, as long as enough data is available.
Smoothness is a crucial ingredient that makes a function learnable, because it implies that neighboring points are correlated to $\mathcal{Y}$ in similar ways. 
That property means that one can draw successful conclusions about unknown points as long as they are close to the training data (coming from the same underlying probability distribution).\cite{rasmussen2003gaussian} In contrast, completely random processes in the above sense allow no predictions.

An association that immediately springs to mind is traditional regression analysis, but ML goes a step further. 
Regression analyses aim to reconstruct the function that goes through a set of known data points with the lowest error, but ML techniques aim to identify functions to predict \emph{interpolations between data points} and thus minimize the prediction error for new data points that might later appear.\cite{vapnik1995} 
Those contrasting objectives are mirrored in the different optimization targets: In traditional regression the optimization task
\begin{equation}
    \label{eq:loss_min}
\hat{f}=\underset{f \in \mathcal{F}}{\arg \min }\left[ \sum_{i}^{M} \mathcal{L}\left(f\left(\mathbf{x}_{i}\right), y_{i}\right)\right],
\end{equation} 
only measures the fit to the data, but learning algorithms typically aim to find models $\hat{f}$ that satisfy
\begin{equation}
    \label{eq:empirical_risk_min}
\hat{f}=\underset{f \in \mathcal{F}}{\arg \min }\left[\sum_{i}^{M} \mathcal{L}\left(f\left(\mathbf{x}_{i}\right), y_{i}\right)+\|\Gamma \Theta\|^{2}\right].
\end{equation} 
Both optimization targets reward a close fit, often using the squared loss $\mathcal{L}\left(\hat{f}(\mathbf{x}), \mathbf{y}\right)=\left(\hat{f}(\mathbf{x})-\mathbf{y}\right)^{2}$.
However, the key difference is an additional \emph{regularization} term in Eq. \ref{eq:empirical_risk_min}, which influences the selection of candidate models  by introducing additional properties that promote generalization. To understand why this is necessary, it is helpful to consider that Eq.~\ref{eq:empirical_risk_min} is only a proxy for the optimization problem
\begin{equation}
    \label{eq:generatization}
\hat{f}=\underset{f \in \mathcal{F}}{\arg \min }\left[\int \mathcal{L}\left(f\left(\mathbf{x}_{i}\right), y_{i}\right) dp(\mathbf{x},\mathbf{y})\right],
\end{equation}
that we would actually like to solve. 
In an ideal world, we would minimize the loss function over the \emph{complete} distribution of inputs and labels $p(\mathbf{x},\mathbf{y})$. 
However, this is obviously impossible in practice, so we apply the principle of Occam’s razor that presumes that simpler (parsimonious) hypotheses are more likely to be correct. With this additional consideration we hope to be able to recover a reasonably general model, despite only having seen a finite training set.
A common way to favor simpler models is via an additional term in the cost function, which is what $\|\Gamma \Theta\|^{2}$ in Eq.~\ref{eq:empirical_risk_min} expresses. Here, $\Gamma$ is a matrix that defines ``simplicity'' with regard to the model parameters $\Theta$. Usually, $\Gamma = \lambda \mathbb{I}$ (with $\lambda > 0$) is chosen to simply favor a small L\textsuperscript{2}-norm on the parameters, such that the solution does not rely on individual input features too strongly. This particular approach is called Tikhonov regularization,~\cite{poggio1990networks, smola1998connection,Rasmussen2015} but other regularization techniques also exist.~\cite{caruana2001overfitting, srivastava2014dropout}

A model that is heavily regularized (i.e. using a large $\lambda$) will eventually become \emph{biased} in that it is too simplistic to fit the data well. 
In contrast, a lack of regularization might yield an overly complex model with high \emph{variance}. 
Such an ``overly fit'' model will follow the data exactly to the point that it also models the noise components and consequently fail to generalize (see Fig.~\ref{fig:cross_validation}).
Finding the appropriate amount of regularization $\lambda$ to manage under- and over-fitting is known as attaining a good \emph{bias-variance trade-off}.\cite{geman1992neural} 
We will introduce a process called \emph{cross-validation} to address this challenge further below (see Section~\ref{sec:model_selection}).

\subsubsection{What is ML good at?}

\paragraph{Implicit knowledge from data}
ML algorithms can infer functional relationships from data in a statistically rigorous way without detailed knowledge about the problem at hand. 
ML thus captures implicit knowledge from a dataset -- even aspects where CPI might not be available. 
Traditional modeling approaches, like classical forcefields discussed in Section~\ref{sec:FF}, rely on preconceived notions about the PES it is modeling and thus the way the physical system behaves. 
In contrast, ML algorithms start from a loss function and a much more general model class. 
Within the limits permitted by the noise inherent to the data, generalization can be improved to arbitrary accuracy given increasingly larger informative training data sets. 
This process allows us to explore a problem even before there is a reasonably full understanding. 
An ML predictor can serve as starting point for theory building and be regarded as a versatile tool in the modelling loop: building predictive models, improving them, enriching them by formal insight, improving further and ultimately extracting a formal understanding. 
More and more research efforts start to combine data-driven learning algorithms with rigorous scientific or engineering theory to yield novel insights and applications.\cite{0336_schutt2017quantum,won2020adaptive,Leineneabb6987}

\paragraph{Redundancy in first principles calculations}

For a quantum chemical property for compounds in a dataset, first principles calculations need to be repeated independently for each input, even if they are very similar. No formally rigorous method exists to exploit redundancies in the calculations in such a scenario.
The empiricism of learning algorithms however does provide a pathway to extract information based on compound structure similarity. A data-driven angle allows one to ask questions in new ways and can give rise to new perspectives on established problems. For example, unsupervised algorithms like clustering or projection methods group objects according to latent structural patterns and provide insights that would remain hidden when only looking at individual compounds.

\subsubsection{What is ML not good at?}

\paragraph{Lack of generality and precision}
Some difficult problems in chemistry and physics can be solved accurately with CompChem, but doing so would require significant resources. 
For example, enumerating all pair-wise interactions in a many-body system will inevitably scale quadratically, and there is no obvious path around this. 
One might ask if empirical approaches can address such fundamental problems more efficiently, but this is unfortunately not possible since ML is more suited for finding solutions in general function spaces rather than in deterministic algorithms where constraints guide the solution process.
However, if we were not as interested in finding a full solution but rather some aspect of it, the stochastic nature of ML can be beneficial. 
For instance, a traditional ML approach might not be the best tool for explicitly calculating the Schr{\"o}dinger equation, but it might be a far more useful tool for developing a forcefield that returns the energy of a system without the need for a cumbersome wavefunction and a self-consistent algorithm. 
As an example, Hermann and No\'{e}~\cite{hermann2020deep} used deep neural networks to show how ML methods may be suitable for overcoming challenges faced by traditional CompChem approaches.

\paragraph{Reliance on high-quality data}
ML algorithms require a large amount of high quality data, and it is hard to decide \emph{a priori} when a data set is sufficient.
Sometimes, a data set may be large, but it does not adequately sample all the relevant systems one intends to model, e.g. a molecular dynamics simulation might generate many thousands of molecular confirmations used to train an ML forcefield, but perhaps that sampling only occurred in a local region of the PES.
In this case, the ML forcefield would be effective at modeling regions of the PES it was trained to but useless in other regions until more data and broader sampling occurred.
This feature is general to all empirical models that are generally limited in their extrapolation abilities.  

\paragraph{Inability to derive high-level concepts}

Standard ML algorithms cannot conceptualize knowledge from a data set.  
Two of the main reasons are the non-linearity and excessive parametric complexity of most models that allow many equally viable solutions for the same problem.\cite{bietti2019inductive, montavon2017explaining} 
It can be hard to gain insight into the modeled relationship because it is not based on a small set of simple rules.
Techniques have emerged to make ML models interpretable (explainable AI - XAI\cite{DBLP:series/lncs/11700}). 
While helpful, drawing scientific insight clearly still requires human expertise.\cite{baehrens2010explain,bach2015pixel,0336_schutt2017quantum,montavon2018methods,holzinger2018machine,lapuschkin2019unmasking,samek2020toward,DBLP:series/lncs/11700} 
Furthermore, the path from an ML model back to a physical set of equations is being explored, but it is far from being fully established automatically.\cite{bongard2007automated,schmidt2009distilling,brunton2016discovering,boninsegna2018sparse,hoffmann2019reactive,watters2017visual,raissi2018deep} 

\paragraph{Prone to artifacts}
Despite following the rules of best practice, ML algorithms can give unexpected and undesired results. Instead of extracting meaningful relationships, they may occasionally exploit nuisance patterns within the underlying experimental design, like the model architecture, the loss function or artifacts in the dataset.
This results in a ``clever Hans'' predictor,\cite{lapuschkin2019unmasking} which technically manages the learning problem but uses a trivial solution that is only applicable within the narrow scope of the particular experimental setup at hand. 
The predictor will appear to be performing well, while actually harvesting the wrong information, and therefore not allowing any generalization or transferable insights.

For example, a recently proposed random forest predictor for the success of Buchwald-Hartwig coupling reactions~\cite{ahneman2018predicting} was later revealed to give almost the same performance when the original inputs were replaced by Gaussian noise.~\cite{Chuangeaat8603, Estradaeaat8763} 
This finding strongly suggested that the ML algorithm exploited some hidden underlying structure in the input data, irrespective of the chemical knowledge that was provided through the descriptor. 
Even though the model might appear quite useful, any conclusions that rely on the importance of the chemical features used in the model were thus rendered questionable at best.
This example demonstrates that out-of-sample validation alone is often not sufficient to establish that a proposed model has indeed learned something meaningful. Therefore, the hypothesis described by the model must be challenged in extensive testing in practically relevant scenarios like actual physical simulations. In other words the ML model needs to lead to a better understanding of the modeling itself and the underlying chemistry.

\subsection{Types of learning}
\label{ssec:learning_types}

ML models are classified by the type of learning problem they solve. Consider for instance a data scientist who develops an ML model that can predict acidity constants (p$K_{\text{a}}$'s) for any molecule. 
A researcher with knowledge of physical organic chemistry might be aware of the empirical Taft equation\cite{anslyn2006modern} that provides a linear free energy relationship between molecules on the basis of empirical parameters that account for a molecules fundamental field, inductive, resonance, and steric effects (e.g. values related to Hammett $\rho$ and $\sigma$ values).
There are several ways the data scientist might develop an ML model to do this. Examples mentioned here include supervised, unsupervised, and reinforcement learning. 

\subsubsection{Supervised learning}
\emph{Supervised learning} addresses learning problems where the ML model $\hat{f}: \mathcal{X} \stackrel{\mathrm{ML}}{\longrightarrow} \mathcal{Y}$ connects a set of known inputs $\mathcal{X}$ and outputs $\mathcal{Y}$, either to perform a regression or classification task. While the former maps onto a continuous space (e.g. energy, polarizability), the latter outputs a categorical value (e.g. acid or base; metal or insulator) for each data point.

Using the p$K_{\text{a}}$ predictor example, a supervised learning algorithm could be trained to correlate recognizable chemical patterns or structures to experimentally known p$K_{\text{a}}$s. 
The goal would be to deduce the relationship between these inputs and outputs, such that the model is able to generalize beyond the known training set. 
A standard universal approximator has to accomplish this learning task without any preconceived notion about the problem at hand and will therefore likely require many examples before it can make accurate predictions. 
Recently, a lot of research is being carried out that investigates ways to  incorporate high-level concepts into the learning algorithm in the form of prior knowledge.\cite{0080_chmiela2017machine, Chmiela2018} 
In this vein, one could take into account chemically relevant parameters such as Hammett constants so that the parameterized ML model incorporates the modified Hammett or Taft equation.
An example of a classification problem in materials science is the categorization of materials, where identifying characteristics of the electronic structure can be used to distinguish between insulators and metals.\cite{ghiringhelli2015big}

\subsubsection{Unsupervised learning}
\emph{Unsupervised learning} describes problems were only the inputs $\mathcal{X}$ are known, with no corresponding labels. In this setting, the goal is to recover some of the underlying structure of the data to gain a higher-level understanding. Unsupervised learning problems are not as rigorously defined as supervised problems in the sense that there can be multiple correct answers, depending on the model and objective function that is applied.

For example, one might be interested in separating conformers of a molecule from a molecular dynamics trajectory, given exclusively the positions of the atoms. A \emph{clustering} algorithm (like the k-means algorithm) could identify those conformers by grouping the data based on common patterns.\cite{cheng2020mapping, reinhardt2020predicting} Alternatively, a \emph{projection} technique could reveal a low-dimensional representations of the dataset.\cite{meila2018regression} Often data is represented in high dimension, despite being intrinsically low-dimensional. With the right projection technique, it is possible to retain the meaningful properties in a representation with less degrees of freedom. A conceptually simple embedding method is principal component analysis (PCA) in which the relationship that is sought to be preserved is the scalar product between the data points.\cite{friedman2001elements} 
There are many other linear and non-linear projection methods, such as multi-dimensional scaling,\cite{cox2008multidimensional} kernel PCA,\cite{scholkopf1997kernel,scholkopf1998nonlinear} t-distributed stochastic neighbor embedding (t-SNE),\cite{maaten2008visualizing} sketch-map,\cite{ceriotti2011simplifying} and the uniform manifold approximation and projection (UMAP).\cite{mcinnes2018umap} Finally, anomaly detection is a further variant of unsupervised learning, where 'outliers' to the available data can be discovered.\cite{ruff2020unifying} 
However, without knowing the labels (in this example, the potential energy associated with each geometry), there is no way to conclusively verify that the result is correct.
The literature is gradually seeing more instances of unsupervised learning, particular to reveal important chemical properties to efficiently explore chemical/materials spaces.

\subsubsection{Reinforcement learning}
\emph{Reinforcement learning} describes problems that combine aspects of supervised and unsupervised learning. Reinforcement learning problems often involve defining an \emph{agent} within an environment that learns by receiving feedback in the form of punishments and rewards. The progress of the agent is characterized by a combination of explorative activity and exploitation of already gathered knowledge.\cite{kaelbling1996reinforcement}
For chemistry applications, reinforcement learning techniques are being increasingly used for finding molecules with desired properties in large chemical spaces.\cite{Leineneabb6987} 

\subsection{Universal approximators}

Universal approximators have their origins in the 1960s, where the hope was to construct ``learning machines'' that have similar capabilities as the human brain. 
An early mathematical model of a single simplified neuron emerged that was called a \emph{perceptron} (Eq.~\ref{perceptron}). \cite{rosenblatt1958perceptron,rosenblatt1961principles} 
\begin{equation}
    \label{perceptron}
    f({\bf x}) = {\rm sign} (\sum_{i=1}^N w_i x_i - b)
\end{equation} 
Here, ${\bf x}$ denotes the $N$-dimensional input to the perceptron. It has $N+1$ parameters consisting of $w_i$ (so-called weights) and a single $b$ (a so-called threshold) that are adapted to the data. This adaption process is typically called ``learning'' (\emph{vide infra}) and it amounts to minimizing a predefined loss function.

In the 1960s, this simple neural network had very limited use, as it was only able to model a linear separating hyperplane. Even simple non-linear functions like the XOR were out of reach.\cite{minsky2017perceptrons} Thus, excitement waned but then reappeared two decades later with the emergence of novel models consisting of more neurons and their arrangement in multi-layer neural network structures~\cite{rumelhart1986learning} (see Eq. \ref{NN_eq}). Recent algorithmic and hardware advances now allow deep and increasingly complex architectures.\cite{lecun2015deep,schmidhuber2015deep}

\begin{equation}
    f_k({\bf x}) = {\rm g} \left(\sum_{j=1}^H w_{kj}{\rm g} \left[ \sum_{i=1}^N w_{ji} x_i -b \right] \right),
\label{NN_eq}
\end{equation}

In Eq.(\ref{NN_eq}), $g(\cdot)$ denotes an activation function that is a non-linear transformation that allows complex mappings between input and output. 
As with the perceptron, the parameters of multi-layer NNs can be learned efficiently using iterative algorithms that compute the gradient of the loss-function using the so-called back-propagation (BP) algorithm.\cite{rumelhart1986learning,lecun1985procedure,bishop1995neural}  
In the late 1980s, artificial neural networks were then proven to be universal approximators of smooth nonlinear functions, \cite{hornik1989multilayer,funahashi1989approximate,cybenko1989approximation} and so they gained broad interest even outside the ML community that then was still relatively small.

In 1995, a novel technique called Support Vector Machines (SVM) \cite{cortes1995support,vapnik1995} and kernel-based learning were then proposed, \cite{scholkopf1998nonlinear,muller2001introduction,scholkopf2002learning, scholkopf1999input} which came with some useful theoretical guarantees. SVMs implement a nonlinear predictor:
\begin{equation}
    \label{SVM_eq}
    f({\bf x}) =  \sum_{j=1}^N  y_j \alpha_{j} K({\bf x}_j,{\bf x}) -b,
\end{equation}
where $K$ is the so-called kernel. The kernel implicitly defines an inner product in some feature space and thus avoids an explicit mapping of the inputs. This ``kernel trick''\cite{muller1997predicting} makes it possible to introduce non-linearity into any learning algorithm that can be expressed in terms of inner products of the input.\cite{scholkopf1998nonlinear} It has since been applied to many other algorithms beyond SVMs,\cite{muller2001introduction} such as Gaussian Processes (GP),\cite{Rasmussen2015} PCA,\cite{scholkopf1997kernel,scholkopf1998nonlinear} independent component analysis (ICA).\cite{harmeling2003kernel}

The most effective kernels are tailored to the specific learning task at hand, but there are many generic choices, such as the polynomial kernel $K ({\bf x}_j,{\bf x}) = ({\bf x}_j\cdot{\bf x} -b)^d$, which describes inner products between degree $d$ polynomials. Another popular choice is the Gaussian kernel $K ({\bf x}_j,{\bf x}) = \exp({-{1 / 2 \sigma^2}({\bf x}_j-{\bf x})^2})$. It is one of the most versatile kernels because it only imposes smoothness assumptions on the solution depending on the width parameter $\sigma$. \cite{smola1998connection,scholkopf2002learning}

As seen in Eq.~(\ref{SVM_eq}), an SVM can also be understood as a shallow neural network with a fixed set of non-linearities. In other words, the kernel explicitly defines a similarity metric to compare data points, whereas neural networks have some freedom to shape this transformation during training because they nest parameterizable  non-linear transformations on multiple scales. This difference gives both techniques some unique strengths and drawbacks. Despite that, there exists a duality between both approaches that allows neural networks to be translated into kernel machines and analyzed more formally (see Refs.~\citenum{braun2008relevant,montavon2011kernel,montavon2013analyzing}).

In the context of computational chemistry, both NNs and kernel-based methods are the most used ML approaches. Simpler learners, such as nearest neighbors models or decision trees can still be surprisingly effective. Those have also been successfully used to solve a wide spectrum of problems including drug design, chemical synthesis planning, and crystal structure classification.\cite{Mitchell2014,0022_Ballester2010, 0013_Mizoguchi2020,0067_Carr2009,Legrain2018,LamPham2017} 

\subsection{The ML workflow}

In the following, we summarize the overall ML process, starting from a dataset all the way to trained and tested model. The ML workflow typically includes the following stages:

\begin{itemize}
\item Gathering and preparing the data
\item Choosing a representation
\item Training the model
\begin{itemize}
\item Train model condidates
\item Evaluate model accuracy
\item Tune hyperparameters
\end{itemize}
\item Testing the model out of sample
\end{itemize}

Note, that the progression to a good ML model is not necessarily linear and some steps (except the out of sample test) may require reiteration as we learn about the problem at hand.

\subsubsection{Datasets}

On a fundamental level, ML models could be simply regarded as sophisticated parametrizations of datasets. While the architectural details of the model matter, the reference dataset forms the backbone that ultimately determines its effectiveness. If the dataset is not representative of the problem at hand, the model will be incomplete and behave unpredictably in situations that have been improperly captured. The same applies to any other shortcomings of the dataset, such as biases or noise artifacts that will also be reflected in the model.
Some of these dataset issues are likely to remain unnoticed when following the standard model selection protocol since training and test datasets are usually sampled from the same distribution. If the sampling method is too narrow, errors seen during the cross-validation procedure may appear to be encouragingly small, but the ML model will fail catastrophically when applied to a real problem. If the training and test sets come from different distributions, then techniques to compensate this covariate shift can be used.\cite{sugiyama2007covariate,sugiyama2012machine}

Robust models can generally only be constructed from comprehensive datasets, but it is possible to incorporate certain patterns into  models to make them more data-efficient. Prior scientific knowledge or intuition about specific problems can be used to reduce the function space from which a ML algorithm has to select a solution. If some of the unphysical solutions are removed a priori, less data are necessary to identify a good model.
This is why NNs and kernel methods, despite both being broad universal function classes, bring different scaling behaviors. The choice of the kernel function provides a direct way to include prior knowledge such as invariances, symmetries or conservation laws, whereas NNs are typically used if the learning problem cannot be characterized as specifically.\cite{zien2000engineering,0080_chmiela2017machine,Chmiela2018} In general, without prior knowledge NNs often require larger datasets to produce the same accuracy as well-constrained kernel methods that embody problem knowledge. This consideration is particularly important if the data is expensive, e.g. if it comes from high quality experiments and/or expensive computations.

\subsubsection{Descriptors}

In order to apply ML, the dataset needs to be encoded into a numerical representation (i.e.~features/descriptors) that allow the learning algorithm to extract meaningful patterns and regularities.\cite{behler2011atom,0444_Bartok2013,rupp2012fast,faber2015crystal,0161_Hansen2015,christensen2020fchl,faber2018alchemical,huo2017unified,drautz2019atomic} This is particularly challenging for unstructured data like molecular graphs that have well-defined invariable or equivariable characteristics that are hard to capture in a vectorial representation. For example, atoms of the same type are indistinguishable from each other, but it is hard to represent them without imposing some kind of order (which inevitably assigns an identity to each atom). Furthermore, physical systems can be translated and rotated in space without affecting many attributes. Only a representation that is adapted to those transformations can solve the learning problem efficiently.

It turned out to be a major challenge to reconcile all invariances of molecular systems in a descriptor without sacrificing its uniqueness or computability. Some representations cannot avoid collisions, where multiple geometries map onto the same representation. Others are unique, but prohibitively expensive to generate. Many solutions to this problem have been proposed, based on general strategies such as invariant integration,\cite{Chmiela2018} parameter sharing,\cite{0140_Gilmer2017, 0336_schutt2017quantum, 0337_Schutt2017a, 0033_Kocer2019} density representations,\cite{0025_Bartok2010} or finger printing techniques.\cite{0107_Duvenaud2015, 0250_Li2017e, 0253_Lim2018, 0317_Rogers2010, 0332_Ehmki2019, 0334_Schmidt2019, 0404_Weininger1988, 0405_Weininger1989, 0041_Behler2011, 0045_Behler2016}
Alternatively, an NN model infers the representation from data.\cite{0336_schutt2017quantum,0338_schutt2018schnet,Unke2018,0107_Duvenaud2015} 
To date, none of the proposed approaches are without compromise, which is why the optimal choice of descriptor depends on the learning task at hand.

\subsubsection{Training}

The training process is the key step that ties together the dataset and model architecture. Through the choice of the model architecture, we implicitly define a function space of possible solutions, which is then conditioned on the training dataset by selecting suitable parameters. This optimization task is guided by a loss function that encodes our two somewhat opposing objectives: (1) achieving a good fit to the data, while (2) keeping the parametrization general enough such that the trained model becomes applicable to data that is not covered in the training set (see the two terms in Eq.~\ref{eq:empirical_risk_min}). Satisfying the latter objectives involves a process called \emph{model selection} in which a suitable model is chosen from a set of variants that have been trained with exclusive focus on the first objective. Depending on the model architecture, more or less sophisticated optimization algorithms can be applied to train the set of model candidates.

\paragraph{Kernel-based learning algorithms} are typically linear in their parameters $\vec{\alpha}$ (see Eq.~\ref{SVM_eq}). Coupled with a quadratic loss function, $\mathcal{L}(\hat{f}(\mathbf{x}), \mathbf{y}) = (\hat{f}(\mathbf{x}) - \mathbf{y})^2$, they yield a convex optimization problem. Convex problems can be solved quickly and reliably due to only having a single solution that is guaranteed to be globally optimal. This solution can be found algebraically by taking the derivative of the loss function and setting it to zero. For example, KRR and GPs then yield a linear system of the form
\begin{equation}
\nabla_{\vec{\alpha}} \mathcal{L}(\hat{f}(\mathbf{x}), \mathbf{y}) = (\mathbf{K}+\lambda \mathbb{I}) \vec{\alpha} - \mathbf{y} = 0
\label{eq:krr_closed_form}
\end{equation} 
which is typically solved in a numerically robust way by factorizing the kernel matrix $\mathbf{K}$. There exist a broad spectrum of matrix factorization algorithms  such as the \emph{Cholesky decomposition} that exploit the symmetry and positive definiteness properties of kernel matrices.\cite{murray2009gaussian,wilson2016deep,gardner2018product,gardner2018gpytorch,wang2019exact} Factorization approaches are however only feasible if enough memory is available to store the matrix factors, and this can be a limitation for large-scale problems. In that case, numerical optimization algorithms provide an alternative: they take a multi-step approach to solve the optimization problem iteratively by following the gradient:
\begin{equation}
\vec{\alpha}^{t}=\vec{\alpha}^{t-1}-\gamma \underbrace{
\nabla_{\vec{\alpha}} \mathcal{L}(\hat{f}(\mathbf{x}), \mathbf{y})
}_{
\text{e.g. } (\mathbf{K}+\lambda \mathbb{I}) \vec{\alpha}^{t-1}-\mathbf{y}
},
\label{eq:gradient_d_krr}
\end{equation} 
where $\gamma$ is the step size (or learning rate). Iterative solvers follow the gradient of the loss function until it vanishes at a minimum, which is much less computationally demanding per step, because it only requires the evaluation of the model $\hat{f}$. In particular, kernel models can be evaluated without storing $\mathbf{K}$ (see Eq.~\ref{eq:gradient_d_krr}).

\paragraph{Neural networks} are constructed by nesting non-linear functions in multiple layers, which yields non-convex optimization problems. Closed-form solutions similar to Eq.~\ref{eq:krr_closed_form} do not exist, which means that NNs can only be trained iteratively, i.e. analogous to Eq.~\ref{eq:gradient_d_krr}. Several variants of this standard gradient descent algorithm exist including stochastic or mini-batch gradient descent, where only an $n$-sized portion of the training data $(\mathbf{x}, \mathbf{y})_{i:i+n}$ is considered in every step. Due to multiple local minima and saddle points on the loss surface, the global minimum is exponentially hard to obtain such that these algorithms usually converge to a local minimum. 
However, thanks to the strong modelling power of NNs, local solutions are usually good enough.\cite{lecun2012efficient}

\paragraph{Hyper-parameters}

In addition to the parameters that are determined when fitting a ML model to the dataset (i.e. the node weights/biases or regression coefficients), many models contain so-called hyper-parameters that need to be fixed before training. Two types of hyper-parameters can be distinguished: ones that influence the model, such as the type of kernel or the NN architecture, and ones that affect the optimization algorithm, e.g. the choice of regularization scheme or the aforementioned learning rate. Both tune a given model to the prior beliefs about the dataset and thus play a significant role in model effectiveness. Hyper-parameters can be used to gauge the generalization behavior of a model.

\begin{figure*}
	\centering
	\includegraphics[width=\textwidth]{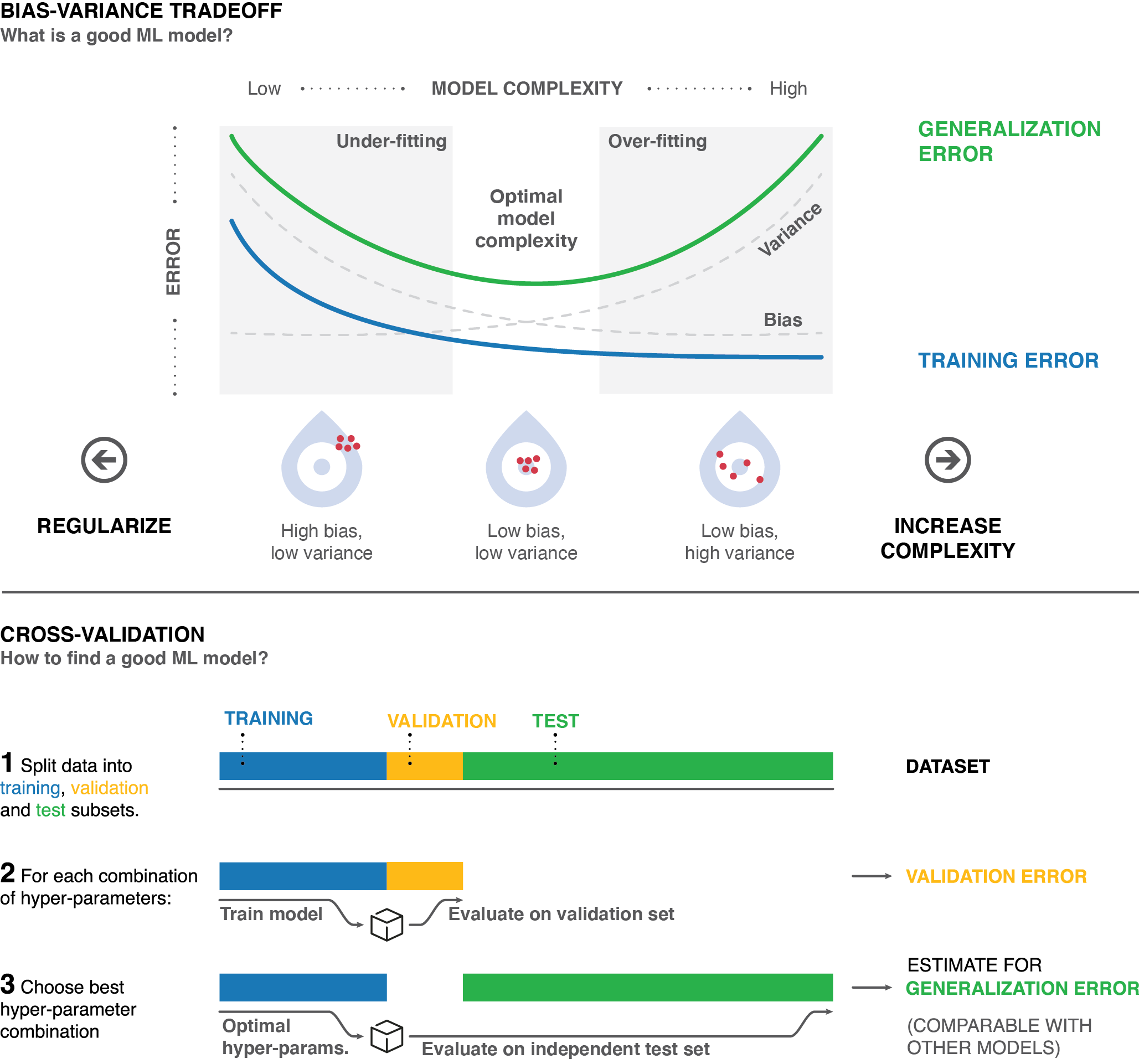}
	\caption{Supervised learning algorithms have to balance two sources of error during training: the bias and variance of the model. A highly biased model is based on flawed assumptions about the problem at hand (under-fitting). Conversely, a high variance causes a model to follow small variations in the data too closely, therefore making it susceptible to picking up random noise (over-fitting). The optimal bias-variance trade-off minimizes the generalization error of the model, e.g. how well it performs on unknown data. It can be estimated with cross-validation techniques.}
	\label{fig:cross_validation} 
\end{figure*}

Hyper-parameter spaces are often rather complex: certain parameters might need to be selected from unbounded value spaces, others could be restricted to integers or have interdependencies. This is why they are usually optimized using primitive exhaustive search schemes like grid or random search in combination with educated guesses for suitable search ranges. Common gradient-based optimization methods typically cannot be applied for this task. Instead, the performance of a given set of hyper-parameters is measured by evaluating the respective model on another training dataset called the validation dataset (see Fig.~\ref{fig:cross_validation}). This process is also referred to as model selection.

\paragraph{Model selection}
\label{sec:model_selection}

Cross-validation or out-of-sample testing is a technique to assess how a trained ML model will generalize to previously unseen data.\cite{scholkopf2002learning, hastie2001elements} For a reasonably complex model, it is typically not challenging to generate the right responses for the data known from the training set. This is why the training error is not indicative of how the model will fulfill its ultimate purpose of predicting responses for new inputs. Alas, since the probability distribution of the data is typically unknown, it is not possible to determine this so-called \emph{generalization error} exactly. Instead, this error is often estimated using an independent test subset that is held back and later passed through the trained model to compare its responses to the known test labels. If the model suffers from over-fitting on the training data, this test will yield large errors. It is important to remember not to tweak any parameters in response to these test results, as this will skew this assessment of the model performance and will lead to overfitting on the test set.\cite{0441_hansen2013assessment}

Besides cross-validation, there are alternative ways to estimate the generalization error, for example via maximization of the marginal likelihood in Bayesian inference.\cite{mackay1992evidence,mackay1992practical,kwok2000evidence} Some well-defined learning scenarios even allow the computation of rigorous upper bounds for the generalization error.\cite{vapnik1995, akaike1974new, schwarz1978estimating, murata1994network}

%% file: Sections/4-whyML4Chem.tex
\newpage
\section{Applications of Machine Learning to Chemical Systems}

We now discuss ways that CompChem methods described in Section 2 and ML methods in Section 3 can be implemented as CompChem+ML approaches for insights into chemical systems.
We often notice the lack of details about \emph{why} an ML model is used and \emph{how} it actually contributes to worthwhile and scientific insights.
Thus, we will summarize the underlying attributes of conventional CompChem+ML efforts and then explain why these attributes are important for specific applications.

To begin, consider molecules or materials in a dataset, and any entry will be related to another based on an abstract concept of ``similarity''.
While similarity is an application-dependent concept, it should go hand in hand with CPI. 
For instance, physical properties of chemical systems can be attributed to the structure and/or composition of the chemical fragments within those systems. 
Thus, if chemical structures and/or compositions of two entries in the database were similar, then their physical properties would also likely be similar.  

For CompChem+ML using a supervised algorithm, a CompChem prediction might be made on a hypothetical system, pinpointed by an ML model that was trained to identify chemical fragments that correlate with labeled physical properties. 
This would be a direct exploitation of chemical similarity.
Alternatively, for CompChem+ML using an unsupervised algorithm, the ML model would identify an underlying distribution or key features based on the similarity between pairs of entries in the dataset without labeled data.
This would be a more nuanced leveraging of chemical similarity. 
In both cases the accuracy, efficiency and reliability of the ML models depends strongly on the how similarity is defined and measured.

In this section we will first describe state-of-the-art descriptors and kernels for atomic systems that can be used to quantify the similarity between chemical systems. 
We will then explain the essential attributes of good atomic descriptors.
Lastly for this section, we will elucidate why and how specific combinations of these descriptors and ML algorithms are beginning to revolutionize the field of CompChem.

\subsection{Representing chemical systems}

In CompChem, molecules and materials are usually represented by the Cartesian coordinates and the chemical elements of all the atoms.  Thus, the size of the vector representation containing the coordinates and charges will be $\{ \mathcal{R}^{3N} \text{ and } \mathcal{Z}^N \}$, respectively, for a system of size $N$.
Even though these atomic coordinates provide a complete description of the system, they are hardly ever used as the input of a ML model because this vector would introduce substantial superfluous redundancy.
For instance, an ML model might treat two identical molecules that are rotated or translated as different molecules, and that in turn might cause the ML model to predict different physical properties for the two otherwise indistinguishable molecules. 
There are further difficulties when comparing molecules having different numbers of atoms.
To work around these problems, atomic coordinates are usually converted into an appropriate representation $\psi$ that is suitable for a particular task.
Such conversions are useful because they allow the incorporation of physical invariances. 
Mathematically speaking, the representation fulfills
\begin{equation}
    \psi(\{ \mathcal{R}^{3N}, \mathcal{Z}^N \}) 
    = \psi(S(\{ \mathcal{R}^{3N}, \mathcal{Z}^N \})),
    \label{eqn:sym}
\end{equation}
where $S$ indicates a symmetry operation, e.g. a rigid rotation about an axis $C_i$, an exchange of two identical atoms, or a translation of the whole system in the Cartesian space, etc.
It can also be advantageous to adopt a coarse-grained representation of the system.~\cite{wang2019machine,wang2020ensemble} 
For example, dihedral angles of a peptide might be accounted for without the positions of the side-chains, positions of ions in a solution might be accounted for without the explicit coordinates of solvents, or just the center of mass for a water molecule might be accounted for in place of the full three-centered atomistic representation.
The choice of these coarse-grained representations provides a way to incorporate prior knowledge of the data,
or such representations can be learned from an unsupervised learning step.~\cite{bonati2020data} 

\subsubsection{Descriptors}

Atomistic systems can be represented in a myriad of ways. Some descriptions are designed to emphasise particular aspects of a system, while others aim to disambiguate similar chemical or physical principles across a wide range of molecules or materials. The set of desirable properties in a representation thus depends on the task at hand. The following overview gives a coarse characterization of the most popular representations.

Table \ref{tab:descriptors} summarizes representations for chemical systems.~\cite{behler2011atom,  huo2017unified,0025_Bartok2010,0444_Bartok2013,faber2018alchemical,sadeghi2013metrics,barker2017lc,0161_Hansen2015}
All respect the aforementioned physical symmetries and invariances needed for chemical systems. 
Many have similar theoretical foundations that can be understood as the basis onto which the atomic density is projected,\cite{willatt2019atom}
and the connection between them has been summarised in a recent review.~\cite{musil2021physics}
The descriptors in Table~\ref{tab:descriptors} can be classified into two categories: 
global and atomic, as indicated under the column ``globality''.
Traditional descriptors used in cheminformatics are global descriptors based on the covalent connectivity of atoms. 
These include simple valence counting and common neighbor analysis,~\cite{tsuzuki2007structural} the presence or absence of predefined atomic fragments (e.g. the Morgan fingerprints\cite{0317_Rogers2010}), pairwise distances between atoms (e.g. Coulomb Matrix,\cite{rupp2012fast} Sine Matrix,\cite{faber2015crystal} Ewald Sum Matrix,\cite{faber2015crystal} Bag of Bonds (BoB)\cite{0161_Hansen2015}), etc.
However, atomic descriptors~\cite{behler2011atom,0444_Bartok2013,christensen2020fchl,faber2018alchemical,huo2017unified,drautz2019atomic,Gastegger2018} are generally more popular than the global ones in ML and CompChem.
In this case, a chemical system is described as a set of atomic environments, $\mathcal{X}_1, \ldots \mathcal{X}_i \ldots \mathcal{X}_N$, and
each consists of the atoms (chemical species and position) within a sphere of radius $r_\mathrm{cut}$ centered at a specific atom $i$.
One needs to combine the set of atomic descriptors of all environments to construct a descriptor for the entire atomic structure.
The most straightforward way to do this is to average the atomic descriptors,
\begin{equation}
    \Phi (A) = \dfrac{1}{N_A}
    \sum_{i \in A}^{N_A} \mathbf{\psi} (\mathcal{X}_i),
\label{eq:fp-global}
\end{equation}
where the sum runs over all $N_A$ atoms $i$ in structure $A$ and $\mathcal{X}_i$ is the environment around atom $i$. 
When there are multiple chemical species, the descriptors for the local environments of different species can either be included in the single sum, or the averaging can be performed for the environments of each species separately and the species-specific averaged local descriptors can be concatenated. 
This can be done by considering the Root Mean Square Displacement (RMSD),~\cite{sadeghi2013metrics}
the best match between the environments of the two structures (best-match),~\cite{de2016comparing}
or by combining local descriptors using a regularized entropy match (REMatch).~\cite{de2016comparing}

\begin{table}[htbp!]
\caption{ML descriptors found in the literature$^{\text{[a]}}$}
\label{tab:descriptors}
\scriptsize
\begin{tabular}{p{0.25\linewidth} p{0.2\linewidth} cc|c|c|c|cc}

\multirow{2}{*}{\textbf{Descriptors}} & 
\multirow{2}{*}{\textbf{Comp. efficiency}$^{\text{[b]}}$} & 
\multirow{2}{*}{\textbf{Periodic}} & 
\multirow{2}{*}{\textbf{Unique}} &
\multicolumn{3}{|p{2cm}|}{\textbf{Invariances}:$^{\text{[c]}}$} &
\multirow{2}{*}{\textbf{Global}} & 
\multirow{2}{*}{\textbf{Smooth}} \\ \cline{5-7}
 & & & & \textbf{T} & \textbf{R} & \textbf{P} & & \\

\toprule
Atom-centered Symmetry Functions (ASCF)~\cite{behler2011atom}  & 
\circled{B} 1,2,3-body terms, cut-off & 
\CheckmarkBold{} & 
\ding{56} & 
\CheckmarkBold{} & 
\CheckmarkBold{} & 
\CheckmarkBold{} & 
\ding{56} & 
\CheckmarkBold{} \\ 

Smooth Overlap of Atomic Positions (SOAP)~\cite{0444_Bartok2013} & 
\circled{B} density based, $\text{SO}_3$ integration & 
\CheckmarkBold{} & 
\ding{56} & 
\CheckmarkBold{} & 
\CheckmarkBold{} & 
\CheckmarkBold{} & 
\ding{56} & 
\CheckmarkBold{} \\ 

Coulomb Matrix (CM)~\cite{rupp2012fast} & 
\circled{A} 1,2-body terms & 
\ding{56} & 
\CheckmarkBold{} & 
\CheckmarkBold{} & 
\CheckmarkBold{} & 
\ding{56} & 
\CheckmarkBold{} & 
\CheckmarkBold{} \\ 

Sine Matrix~\cite{faber2015crystal} & 
\circled{A} 1,2-body terms & 
\CheckmarkBold{}  & 
\CheckmarkBold{} & 
\CheckmarkBold{} & 
\CheckmarkBold{} & 
\ding{56} & 
\CheckmarkBold{} & 
\CheckmarkBold{} \\ 

Ewald Sum Matrix~\cite{faber2015crystal} & 
\circled{A} 1,2-body terms & 
\CheckmarkBold{}  & 
\CheckmarkBold{} & 
\CheckmarkBold{} & 
\CheckmarkBold{} & 
\ding{56} & 
\CheckmarkBold{} & 
\CheckmarkBold{} \\ 

Bag of Bonds (BoB)~\cite{0161_Hansen2015} & 
\circled{A} 1,2-body terms &
\ding{56} & 
\ding{56} & 
\CheckmarkBold{} & 
\CheckmarkBold{} & 
\Circle{} & 
\CheckmarkBold{} & 
\ding{56} \\ 

Faber-Christensen-Huang-Lilienfeld (FCHL)~\cite{christensen2020fchl} & 
\circled{C} 1,2,3-body terms &
\CheckmarkBold{} & 
\ding{56} & 
\CheckmarkBold{} & 
\CheckmarkBold{} & 
\CheckmarkBold{} & 
\ding{56} & 
\CheckmarkBold{} \\ 

Spectrum of London and Axilrod-Teller-Muto potential (SLATM)~\cite{faber2018alchemical} & 
\circled{D} 1,2,3,4-body terms &
\CheckmarkBold{} & 
\ding{56} & 
\CheckmarkBold{} & 
\CheckmarkBold{} & 
\CheckmarkBold{} & 
\ding{56} & 
\CheckmarkBold{} \\ 

Many-body Tensor Representation (MBTR)~\cite{huo2017unified} & 
\circled{C} 1,2,3-body terms &
\ding{56} & 
\ding{56} & 
\CheckmarkBold{} & 
\CheckmarkBold{} & 
\CheckmarkBold{} & 
\CheckmarkBold{} & 
\CheckmarkBold{} \\ 

Atomic cluster expansion~\cite{drautz2019atomic} & 
\circled{A} 1,2-body terms &
\CheckmarkBold{} & 
\ding{56} & 
\CheckmarkBold{} & 
\CheckmarkBold{} & 
\CheckmarkBold{} & 
\CheckmarkBold{} & 
\CheckmarkBold{} \\ \\ 

Invariant many-body interaction descriptor (MBI)~\cite{pronobis2018many} & 
\circled{B} 1,2,3-body terms &
\ding{56} & 
\ding{56} & 
\CheckmarkBold{} & 
\CheckmarkBold{} & 
\CheckmarkBold{} & 
\ding{56} & 
\CheckmarkBold{} \\ \\ 

\toprule
\multicolumn{9}{l}{\textbf{Neural Network Architectures}} \\
\bottomrule

Deep Potential - Smooth Edition (DeepPot-SE)~\cite{zhang2018deep,zhang2018end} & 
\circled{B} 1,2,3-body terms, cut-off &
\CheckmarkBold{} & 
\ding{56} & 
\CheckmarkBold{} & 
\CheckmarkBold{} & 
\CheckmarkBold{} & 
\ding{56} & 
\CheckmarkBold{} \\ 

MPNN,SchNet~\cite{0336_schutt2017quantum,0338_schutt2018schnet} & 
\circled{A}/\circled{B} 1,2-body terms, hierarchical &
\CheckmarkBold{} & 
\ding{56} & 
\CheckmarkBold{} & 
\CheckmarkBold{} & 
\CheckmarkBold{} & 
\ding{56} & 
\CheckmarkBold{} \\ 

Cormorant~\cite{anderson2019cormorant} & 
\circled{B} 1,2-body terms, hierarchical &
\ding{56} & 
\ding{56} & 
\CheckmarkBold{} & 
\CheckmarkBold{} & 
\CheckmarkBold{} & 
\ding{56} & 
\CheckmarkBold{} \\ 

Tensor Field Networks~\cite{thomas2018tensor} & 
\circled{B} 1,2-body terms &
\CheckmarkBold{} & 
\ding{56} & 
\CheckmarkBold{} & 
\CheckmarkBold{} & 
\CheckmarkBold{} & 
\ding{56} & 
\CheckmarkBold{} \\ \\ 

\toprule
\multicolumn{9}{l}{\textbf{Similarity metrics}} \\
\bottomrule

Root mean square deviation of atomic positions (RMSD)~\cite{sadeghi2013metrics} & 
\circled{A} 1,2-body terms, input matching &
\ding{56} & 
\ding{56} & 
\Circle{} & 
\Circle{} & 
\ding{56} & 
\CheckmarkBold{} & 
\ding{56} \\ 

Overlap Matrix~\cite{sadeghi2013metrics} & 
\circled{A} 1,2-body terms, input matching &
\ding{56} & 
\ding{56} & 
\CheckmarkBold{} & 
\CheckmarkBold{} & 
\CheckmarkBold{} & 
\CheckmarkBold{} & 
\ding{56} \\ 

REMatch~\cite{de2016comparing} & 
\circled{C} 1,2-body terms, input matching &
\ding{56} & 
\ding{56} & 
\CheckmarkBold{} & 
\CheckmarkBold{} & 
\CheckmarkBold{} & 
\CheckmarkBold{} & 
\ding{56} \\ 

sGDML~\cite{Chmiela2018} & 
\circled{A}  1,2-body terms &
\CheckmarkBold{} & 
\CheckmarkBold{} & 
\CheckmarkBold{} & 
\CheckmarkBold{} & 
\Circle{}$^{\text{[c]}}$ & 
\CheckmarkBold{} & 
\CheckmarkBold{} \\ \\ 
\bottomrule
\\
\multicolumn{9}{l}{[a] `\textbf{\CheckmarkBold{}}' = satisfies condition; `\Circle{}' = partially satisfies condition; `\textbf{\ding{56}}' = does not satisfy condition} \\
\multicolumn{9}{p{\linewidth}}{[b] Computational efficiency ranks with grades $\circled{A} - \circled{D}$ in descending order. The efficiency class reflects the extent that the descriptor requires expensive operations (e.g. a hierarchical processing or matching of inputs)} \\
\multicolumn{9}{l}{[c] `\textbf{T}' = translational; `\textbf{R}' = rotational; `\textbf{P}' = permutational; } \\
\multicolumn{9}{l}{[d] Only invariant permutations represented in the training data} \\
\end{tabular}
\end{table}
\newpage

\subsubsection{Representing local environments}

We will now describe the Smooth Overlap of Atomic Positions (SOAP) descriptors~\cite{0444_Bartok2013} since many other descriptors based on the atomic density are similar and differ mainly by how the density if projected onto basis functions.\cite{willatt2019atom, drautz2019atomic}
To construct SOAP descriptors, one first considers an atomic environment $\mathcal{X}$ that contains only one atomic species, and a Gaussian function of width $\sigma$ is then placed on each atom $i$ in $\mathcal{X}$ to make an atomic density function:

\begin{equation}
    \rho_{\mathcal{X}_i} (\mathbf{r}) = \sum_{i \in \mathcal{X} } \exp \left[-\frac{\left| \mathbf{r}- \mathbf{r}_{i}\right|^2}{2\sigma^2} \right] f_{\mathrm{cut}}(|\mathbf{r}|).
\end{equation}

Here, $\mathbf{r}$ denotes a point in Cartesian space,  $\mathbf{r}_{i}$ is the position of atom $i$ relative to the central atom of $\mathcal{X}$, and the cutoff function $f_{\mathrm{cut}}$ smoothly decays to zero beyond the radius $r_\mathrm{cut}$.
This density representation ensures invariance with respect to translations and permutations of atoms of the same species but not rotations.
To obtain a rotationally-invariant descriptor, 
one expands the density in a basis of spherical harmonics, $Y_{lm}(\mathbf{\hat{r}})$, and a set of orthogonal radial functions, $g_n(r)$, as
\begin{equation}
    \rho_{\mathcal{X}}(\mathbf{r}) =
    \sum_{nlm} c_{nlm} g_n(\left|\mathbf{r}\right|) Y_{lm} (\mathbf{\hat{r}}),
\end{equation}
to construct the {\em power spectrum} of the density using the expansion coefficients:
\begin{equation}
    \psi_{nn'l} (\mathcal{X}) =
    \sqrt{\dfrac{8}{2l+1}}
    \sum_m (c_{nlm})^* c_{n'lm}.
    \label{eqn:partial-k}
\end{equation}
One then obtains a vector of descriptors $\mathbf{\psi} = \{\psi_{nn'l}\}$ by considering all components  $l \leq l_\text{max}$ and $n,n' \leq n_\text{max}$ that act as band limits that control the spatial resolution of the atomic density. 
The generalization to more than one chemical species is straightforward:\cite{de2016comparing}  
one constructs separate densities for each species $\alpha$ and then computes the power spectra $\psi_{nn'l}^{\alpha \alpha'} (\mathcal{X})$ for each pair of elements $\alpha$ and $\alpha'$ 
where the two species indices correspond to the $c^*$ and $c$ coefficients, respectively.
The resulting vectors corresponding to each of the $\alpha$ and $\alpha'$ pairs are then concatenated to obtain the descriptor vector of the complete environment.

Atom-centered Symmetry Functions (ACSFs) or sometimes called Behler-Parrinello Symmetry Function\cite{behler2011atom} descriptors differ from SOAP descriptors in that it projects the atomic densities over selected 2-body or 3-body symmetry functions.
Faber–Christensen–Huang–Li\-lien\-feld (FCHL)~\cite{faber2018alchemical} descriptors follow similar principles while also considering the correlations between the atomic densities coming from different chemical species. 
The Many-Body Tensor Representation (MBTR)\cite{huo2017unified} approach involves taking the histograms of atom counts, inverse pair-wise distances, and angles.
Atomic cluster expansion (ACE) descriptors~\cite{drautz2019atomic} first express atomic densities using spherical harmonics and then generate invariant products by contracting the spherical harmonics with the Clebsch-Gordan coefficients.

\paragraph{Length-scale hyper parameters}

Most atomic descriptors use length-scale hyper parameters specifically chosen for a given problem and system.\cite{behler2011atom,  huo2017unified,0025_Bartok2010,0444_Bartok2013,faber2018alchemical,sadeghi2013metrics,barker2017lc,0161_Hansen2015}
There are several ways to automate hyper parameter selections.
Ref.~\citenum{cheng2020mapping} introduced general heuristics 
for choosing the SOAP hyper-parameters for a system with arbitrary chemical composition based on characteristic bond lengths.
Ref.~\citenum{0183_Imbalzano2018} adopts the strategy to first generate a comprehensive set of ACSFs and then select a subset using the sparsification methods such as farthest point sampling (FPS)~\cite{schlomer2011farthest} and CUR matrix decomposition.\cite{mahoney2009cur}

\paragraph{Incompleteness of atomic descriptors}
A structural descriptor is complete when there is
no pair of configurations that produces the same descriptor.\cite{pozdnyakov2020completeness} 
For atomic descriptors, this means that different atomic environments --- after considering the invariances of rotation, translation and permutation of identical atoms --- should adopt distinct descriptors.
Without completeness, any ML model using the descriptors as input will give identical predictions of physically different systems.
Ensuring completeness while preserving the invariances is non-trivial, however.
One of the simplest descriptors is based on permutationally-invariant pairwise atomic distances (2-body descriptors), and Ref.~\citenum{0444_Bartok2013} demonstrated that these are generally not complete since one can construct two distinct tetrahedra using the same set of distances.
Many have assumed that permutationally-invariant 3-body atomic descriptors uniquely specify atomic environments due to the tremendous success of ML models for chemical systems and particularly MLPs. 
However, Refs.~\citenum{vonLilienfeld2015} and \citenum{ pozdnyakov2020completeness} exemplify that structural degeneracies can be found even when using 3- or 4-body descriptors.
This underscores an important shortcoming of state-of-the-art 3-body descriptors such as ACSF,~\cite{behler2011atom} SOAP,~\cite{0444_Bartok2013} FCHL,~\cite{faber2018alchemical} and MBTR.\cite{huo2017unified}
Atomic cluster expansion~\cite{drautz2019atomic} should be a complete descriptor of local environments,
but its reliance on spherical harmonic expansion and the subsequent contraction makes their evaluations expensive.
Hence, there are still opportunities to develop improved atomic descriptors.

\subsubsection{The locality approximation}\label{sec:locality}

Representing a many-body chemical system in terms atomic environments brings physical significance since
certain extensive physical properties (e.g. the total energy, total electrostatic charge, and polarizability of a system) can be approximated by the sum of the 
atomic contributions coming from each atomic environment, e.g. $\Theta = \sum_i \theta(\mathcal{X}_i)$.
This approximation is valid because the atomic contribution associated with a central atom is largely determined by its neighbors, and long-range interactions can be approximated in a mean-field manner without explicitly considering distant atoms.
Such ``locality'' is tacitly assumed in many ML models for CompChem, and it is a crucial necessity for most common atomistic potentials and MLPs (Section \ref{sec:FF}).
Most MLPs (e.g. BPNN,\cite{0032_Behler2007a} GAP,\cite{0025_Bartok2010} and DeepMD\cite{zhang2018end}) approximate the total energy of a system as sums of local atomic energies.

\begin{figure*}[hbt]
\includegraphics[width=0.6\textwidth]{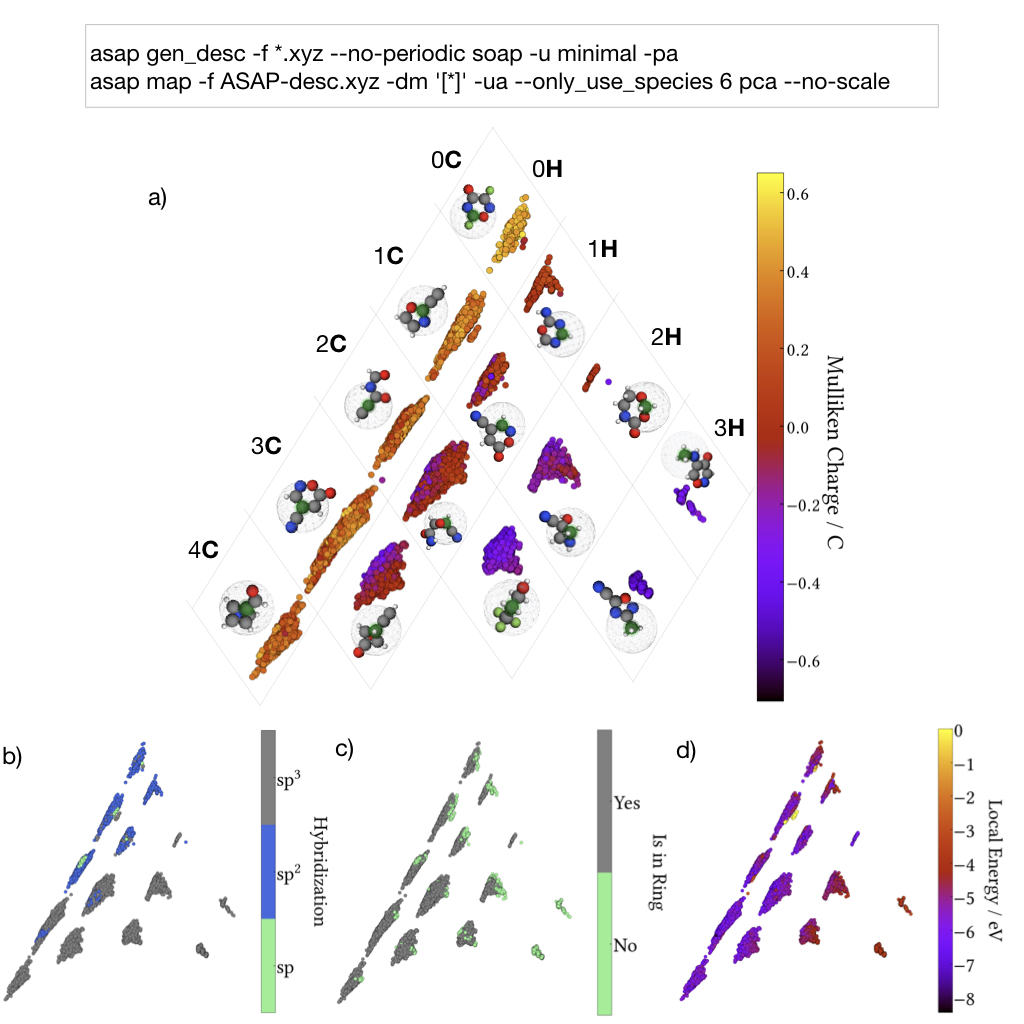}
\caption{KPCA maps of carbon atom environments in the QM9 database. Maps are color-coded according to Mulliken charges (a), hybridization (b), whether atoms are found in rings (c) and according to local energies predicted by a machine learning potential (d). 
Reprinted with permission from Ref.~\citenum{cheng2020mapping}. Copyright (2020) American Chemical Society.
}
\label{fig:qm9-atom}
\end{figure*}

Figure \ref{fig:qm9-atom} illustrates locality by showing a KPCA map of the atom environments of carbon in the QM9 set (see Section 4.3 for more detailed descriptions of the data set). 
By color-coding the KPCA plot with the local energies from a SOAP-based GAP model trained on QM9 energies,\cite{bart+17sa} one observes a systematic and smooth trend in energies across clusters. 
The total molecular energy can then be accurately predicted by the sum of local energies, which means the total energy can be approximated on the basis of all the local environments contained in the molecule.
For example, an NN potential trained on liquid water simulations can predict the densities, lattice energies, and vibrational properties of diverse ice phases because the local atomic environments found in liquid water span the similar environments as those observed in ice phases.\cite{monserrat2020extracting}
Another GAP potential of carbon trained on amorphous structures and other crystalline phases predicted novel carbon structures in random structure searches as well as approximate reaction barriers.\cite{Deringer2017, rowe2020accurate}

The locality approximation is typically rationalized based on the multiscale nature of interatomic interactions in chemical systems. It is generally expected that shorter interatomic distances correspond to stronger interactions, such that a cutoff may be imposed after a certain radial distance $d$ given a certain energy accuracy threshold $\epsilon$. The multiscale nature of interactions underlies the usual classification of chemical interactions, from strong covalent bonds and ionic interactions to weaker non-covalent hydrogen bonds and van der Waals interactions. However, our understanding of non-covalent interactions in large molecules and materials is still emerging~\cite{Stohr2019} and no general rules-of-thumb exist to define the cutoff distance $d$ corresponding to a defined $\epsilon$. Hence, for systems having long-range interactions (which includes most chemical systems), the locality assumption needs revision. Models such as (s-)GDML~\cite{0080_chmiela2017machine,Chmiela2018} that learn global interactions can be better suited in these cases. Global models tend to be more data-efficient because they are more specialized toward learning a full PES for a particular molecule or material, but this restricts the use of a single ML model to only the system it was trained upon.

\subsubsection{Advantages of built-in symmetries}

Built-in symmetry in ML models substantially compresses the dimensionality of atomic representations and ensures that physically equivalent systems are predicted to have identical properties. 
One of the most rigorous ways of imposing symmetry onto a model $f$ is via the invariant integration over the relevant group $\mathcal{S}$
\begin{equation}
f_{\mathrm{sym}}(\mathbf{x})=\int_{\pi \in \mathcal{S}} f\left(\mathbf{P}_{\pi} \mathbf{x}\right),
\end{equation}
where $\mathbf{P}_{\pi} \mathbf{x}$ is a permutation of the input. 
However, the cardinality of even basic symmetry groups is exceedingly high, which makes this operation prohibitively expensive. 
This combinatorial challenge can be solved by limiting the invariant integral to the physical point group and fluxional symmetries that actually occur in the training dataset, as done in sGDML.\cite{Chmiela2018} 
Alternative approaches such as parameter sharing~\cite{0140_Gilmer2017, 0336_schutt2017quantum, 0337_Schutt2017a, 0033_Kocer2019} or density representations~\cite{0025_Bartok2010} have also proven effective. 
For example, the DeepMD potential has two versions,
the Smooth Edition (DeepPot-SE) explicitly preserves all the natural symmetries of the molecular system, and the other version that does not.\cite{zhang2018end}
The DeepPot-SE offers much improved stability and accuracy.\cite{Chmiela2018, zhang2018end}

For ML predictions of scalar properties, the rotationally-invariant atomic descriptor framework described earlier is appropriate.
One may wish to predict vectorial or tensorial properties including dipole moments, polarizability, and elasticity.
A co-variant version of descriptors may be advantageous, and this can be expressed as
\begin{equation}
    S(\psi(\{ \mathcal{R}^{3N} \}) )
    = \psi(S(\{ \mathcal{R}^{3N} \})),
    \label{eqn:sym-rv}
\end{equation}
where $S$ indicates a symmetry operation such as a rigid rotation about an axis.
Ref.~\citenum{Glielmo2017} proposed a general method for transforming a standard kernel for fitting scalar properties into a co-variant one.
Ref.~\citenum{Grisafi2018} derived a rotational-symmetry-adapted SOAP kernel, which can be understood as using the angular-dependent SOAP vectors based on spherical harmonics expansions as the descriptors. 
Note that the SOAP kernels for learning scalar properties introduced in Ref.~\citenum{0444_Bartok2013} remove angular dependencies by summing up the SOAP vectors in separate spherical harmonics channels.

Symmetry can be further exploited into ``alchemical'' representations that incorporate similarity between chemical species that are relatable by changing one atom into another.
The FCHL~\cite{faber2018alchemical} representation considers the similarity between elements in the same row and columns of the periodic table and performs very well on chemical compounds across chemical space.
Ref.~\citenum{Willatt2018} compiled a data-driven periodic table of the elements by fitting to an elpasolite data set using an alchemical representation.


\subsubsection{End-to-end NN representations}

All descriptors introduced above rely on a suitable set of hyperparameters (e.g. length scales, radial and angular resolution).
Determining an optimal set of hyperparameters can be a tedious process, especially when heuristics are unavailable or fail due to the structural and compositional complexity of the system.
A poor choice of descriptors can limit the accuracy of the final ML model, e.g. when certain interatomic distances can not be resolved.

End-to-end NN representations follow a different strategy to learn a representation directly from reference data.
Using atom types and positions of a system as inputs, end-to-end NNs construct a set of atom-wise features $\mathbf{x}_i$.
These features are then used to predict the property of interest, e.g. the energy as a sum of atom-wise contributions.
Unlike static descriptors, the representation is also optimized as part of the overall training process.
This way end-to-end NNs can adapt to structural features in the data and the target properties in a fully automatic fashion to eliminate the need for extensive feature engineering from the practitioner.

The deep tensor NN framework (DTNN)\cite{0336_schutt2017quantum} introduced a procedure to iteratively refine a set of atom-wise features $\{\mathbf{x}_i\}$ based on interactions with neighboring atoms.
Higher order interactions can then be captured in an hierarchical fashion.
For example, a first information pass would only capture radial information, but further interactions would recover angular relations and so on.
In DTNN, a learnable representation depending only on atom types $\mathbf{x}^0_i=\mathbf{e}_{Z_i}$ serves as an initial set of features.
These are then refined by successive applications of an update function depending on the atomic environment that takes the general form:
\begin{equation}
\mathbf{x}^{l+1}_i = \mathrm{F}^l \left( \mathbf{x}^l_i + \sum_j \mathrm{G}^l\left[\mathbf{x}^l_i, \mathbf{x}^l_j, \mathbf{g}^l\left(|\mathbf{r}_i-\mathbf{r}_j| \right) \right] \mathrm{f}_\mathrm{cut}\left(|\mathbf{r}_i-\mathbf{r}_j| \right) \right) \label{eq:dtnn}
\end{equation}
Here, $l$ indicates the number of overall update steps.
The sum runs over all atoms $j$ in the local environment, and a cutoff function $\mathrm{f}_\mathrm{cut}$ ensures smoothness of the representation.
Each feature is updated with information from all neighboring atoms through the interaction function $\mathrm{G}$.
Apart from the neighbor features $\mathbf{x}_j$, $\mathrm{G}$ also depends on the interatomic distance $|\mathbf{r}_i-\mathbf{r}_j|$, which is usually expressed in the form of a radial basis vector $\mathbf{g}$.
After the update, an atom-wise transformation $\mathrm{F}$ can be applied to further modulate the features.
Since each update depends only on the interatomic distances and the summation over neighboring atoms is commutative, end-to-end NNs of this type automatically achieve a representation that is invariant to rotation, translation and permutations of atoms.
Using these atom-type dependent embeddings compactly encodes elemental information.
This is advantageous for systems comprised of many different chemical elements.
Such multi-component systems can be problematic to treat with pre-defined descriptors (e.g. ACSFs or SOAP), as these typically introduce additional entries for each possible combination of atom types, resulting in a large number of descriptor dimensions.

Since the introduction of DTNN, many different types of end-to-end NNs have been developed, and these vary by the choice for the functions $\mathrm{F}$ and $\mathrm{G}$.
For example, SchNet\cite{0338_schutt2018schnet} uses continuous convolutions inspired by convolutional neural networks (CNNs) to describe the interatomic interactions. 
In this case, the update in Eq.~\ref{eq:dtnn} takes the form
\begin{equation}
\mathbf{x}^{l+1}_i = \mathbf{x}^l_i +\mathrm{MLP}^l_\mathrm{tr}\left( \sum_j \mathbf{x}^l_j \mathrm{MLP}^l_\mathrm{rad}(\mathbf{g}^l\left(|\mathbf{r}_i-\mathbf{r}_j| \right) \mathrm{f}_\mathrm{cut}\left(|\mathbf{r}_i-\mathbf{r}_j| \right) \right) \label{eq:schnet}
\end{equation}
where the feature transformation ($\mathrm{MLP}_\mathrm{tr}$) and the radial dependence ($\mathrm{MLP}_\mathrm{rad}$) are both modeled as trainable ML potentials.

Other ML models introduce additional physical information. 
The hierarchical interacting particle NN (HIP-NN)\cite{Lubbers2018} enforces a physically motivated partitioning of the overall energy between the different refinement steps, while the PhysNet architecture\cite{unke2019physnet} introduces explicit terms for long range electrostatic and dispersion interactions.
In Ref.~\citenum{0140_Gilmer2017}, Gilmer \emph{et al.} categorize graph networks of this general type as message-passing NNs (MPNNs) and introduce the concept of edge updates.
These make it possible to use interatomic information beside the radial distance metric in the refinement procedure, and they have since been adapted for other architectures.\cite{jorgensen2018neural}
Another interesting extension are end-to-end NNs incorporating higher-order features beside the scalar $\mathbf{x}_i$ used in the original DTNN framework.
These are equivariant features that encode rotational symmetry and can be based on angles, dipole moment vectors, or features that can be expressed as spherical harmonics with $l>0$. 
This enables the exchange of only radial information between atoms in each interaction pass and instead include higher structural information, such as dipole-dipole interactions or angular information.
In addition, equivariant end-to-end NNs can also be used to predict vectorial or tensorial properties in a manner similar to the rotational-symmetry-adapted SOAP kernel.
Examples include TensorField networks,\cite{thomas2018tensor} Cormorant,\cite{anderson2019cormorant} DimeNet,\cite{klicpera_dimenet_2020} PiNet,\cite{shao2020pinn} and FieldSchNet\cite{gastegger2020machine}.

\subsection{From descriptors to predictions}

After a descriptor vector for each chemical structure is defined, one can then construct the design matrix and the kernel matrix for a set of structures.
These matrices can then be used as the input of ML models.
As described in Section 3, supervised ML methods such as NNs and GPs
can be used to approximate non-linear and high-dimensional functions, particularly when massive amounts of training data become available.
Thus, one should expect that using CompChem would be very useful for generating a large amount of almost noise-free training data of specific systems or atomic configurations, as long as a physically accurate method is being applied in the right way with appropriate computational resources.
In contrast, experimental observations can be difficult to measure and reproduce precisely. 
Note that the aim of most CompChem+ML efforts have a similar scope as decades-old quantitative structure activity/property relationship (QSAR/QSPR) models that are often based on experiments or CompChem modeling.~\cite{0378_Tropsha2010,0439_Muratov2020, Katritzky2010} 
Thus, researchers in CompChem+ML should  be aware of potentially relatable work done by the QSAR/QSPR communities, and to what extent questions being posed have been sufficiently answered. 
On the other hand, ML usually provides higher accuracy than non-ML statistical models, and so QSAR/QSPR efforts have been turning toward  ML models as well.\cite{0263_Ma2015}  

Depending on the types of chemical problems being solved, different CompChem methods suitable for treating different chemical interactions will be used for training ML models having different approximations for underlying high-dimensional functions.
For example, research efforts are towards learning electron densities,\cite{grisafi2018transferable} density functionals,\cite{0060_Brockherde2017} and molecular polarizabilities.\cite{wilkins2019accurate}

Besides these direct learning strategies,
ML has been used to enhance the performance and suitability of CompChem models.
As mentioned in Section 2, the $\Delta$-ML~\cite{Ramakrishnan2015} approach is now a common technique for adapting an ML model that improves the quality of a theoretically insufficient but computationally affordable method. 
This approach has been used to learn many body corrections for water molecules to allow a relatively inexpensive KS-DFT approach like BLYP to more accurately reproduce CCSD(T) data.\cite{0026_Bartok2013}
Along similar lines, Shaw and coworkers used CompChem features along with a neural network to reweight terms from an MP2 interaction energy to provide ML-enhanced methods with increased performance.\cite{0274_McGibbon2017} 
Miller and coworkers have developed ML-models where molecular orbitals themselves are learned to generate a density matrix functional that provides CCSD(T)-quality potential energy surfaces with a single reference calculation.\cite{Cheng2019} 
von Lilienfeld and coworkers have investigated how the choice of regressors and molecular representations for ML models impacts accuracy, and their finding suggest ways that ML models may be trained to be more accurate and less computationally expensive than hybrid DFT methods.\cite{0115_Faber2017} 
Burke and co-workers have studied how ML methods can result in improved understanding and more physical exact KS-DFT\cite{0362_Snyder2012,0436_Hollingsworth2018,0243_Li2016, 0384_Vu2015} and OFDFT functionals.\cite{0363_Snyder2013} 
Brockherde \emph{et al.} have presented an approach where ML  models can directly learn the Hohenberg-Kohn map from the one-body potential efficiently to find the functional and its derivative.\cite{0060_Brockherde2017,bogojeski2020quantum} 
Akashi and coworkers have also reported the out-of-training transferability of neural networks that capture total energies, which shows a path forward to generalizable methods.\cite{0028_Nagai2018} 

Besides the above-mentioned applications of supervised learning, one can exploit the ``universal approximator'' nature of ML architectures to find a function that gives the best solution in a variational setting.
For instance, using restricted Boltzmann machines~\cite{carleo2017solving} or deep neural networks as a basis representation of wavefunctions~\cite{manzhos2009improved, hermann2020deep,pfau2020ab} in Quantum Monte Carlo calculations.

\subsection{CompChem data}

We have laid the general framework for CompChem+ML studies, but this direction would not be complete without more details about training data (i.e., garbage in, garbage out).
We now review the landscape of data sets in CompChem and how they will likely evolve over time.
The past decade has seen continually increasing usefulness and availabilty of ``big data'' from CompChem that include community-wide data repositories comprised of millions of atomistic structures along with diverse physical and chemical properties.\cite{quantummachine, de2019new, materialsproject, nomad}
Such repositories are becoming the norm, and it is more customary for different users to deposit raw or processed simulation data there for the benefit of the research community.
This brings the possibility of robust validation tests for ML models, but it also necessitates approaches that are well-equipped to handle large and complex data sets.
Typical data sets may come from diverse origins such as MD trajectories from \emph{ab initio} simulations, data sets of small molecules and molecular conformers, or other training sets used for developing ML and non-ML forcefields for specific applications.
As the data sets grow, so do the scope of publications that involve ML as shown in Fig. \ref{fig:heatmap}. 
 
\subsubsection{Benchmark data sets}

\begin{table}[h!]
    \caption{ML databases for CompChem}
    \centering
    \renewcommand{\arraystretch}{1.5}
    {\scriptsize
    \begin{tabular}{p{0.10\linewidth} p{0.5\linewidth} p{0.3\linewidth}}
    \hline
        Database & Description & Location  \\
        \hline
        AFLOWLIB & Databases containing calculated properties of over 625k materials\cite{0065_Calderon2015} & \url{http://www.aflowlib.org} \\ 
        ANI-1 & Large computational DFT database, which consists of more than 20M off equilibrium conformations for ~57.5k small organic molecules.\cite{smith2017ani,0359_Smith2017} & \url{https://github.com/isayev/ANI1_dataset} \\
        ANI-1x / ANI-1ccx & ANI-1x contains multiple QM properties from 5M DFT calculations, while ANI-1ccx contains 500k data points obtained with an accurate CCSD(T)/CBS extrapolation.\cite{0029_Smith2020}  & \url{https://github.com/aiqm/ANI1x_datasets} \\
        BindingDB & Measured binding affinities focusing on interactions of proteins considered to be candidates as drug-targets. 1,200,000 binding data for 5,500 proteins and over 520,000 drug-like molecules.\cite{Liu2007} & \url{http://www.bindingdb.org} \\
        Clean Energy Project & Contains $\sim$10,000,000 molecular motifs of potential interest which cover small molecule organic photovoltaics and oligomer sequences for polymeric materials.\cite{0156_Hachmann2011} & \url{http://cepdb.molecularspace.org} \\
        CoRE MOF & Database containing over 4,700 porous structures of metal-organic frameworks with publically available atomic coordinates. Includes important physical and chemical properties.\cite{0083_Chung2014} & \url{http://dx.doi.org/10.11578/1118280} \\
        FreeSolv & Experimental and calculated hydration free energies for neutral molecules in water.\cite{0282_Mobley2014} & \url{http://www.escholarship.org/uc/item/6sd403pz} \\
        GDB & GDB-11, GDB-13 and GDB-17. Together these databases contain billions of small organic molecules following simple chemical stability and synthetic feasibility rules.\cite{0121_Fink2007} & \url{http://gdb.unibe.ch/downloads/} \\
        Hypothetical zeolites & Contains approximately 1M zeolite structures.\cite{0109_Earl2006} & \url{http://www.hypotheticalzeolites.net/} \\
        Materials Project & Contains computed structural, electronic, and energetic data for over 500k compounds.\cite{0190_Jain2013} & \url{https://www.materialsproject.org} \\
        MD17 & Datasets in this package range in size from 150k to nearly 1M conformational geometries. All trajectories are calculated at a temperature of 500 K and a resolution of 0.5 fs.\cite{0080_chmiela2017machine} & \url{http://www.sgdml.org} \\
        MoleculeNet & Contains data on the properties of over 700k compounds.\cite{0411_Wu2018} & \url{http://moleculenet.ai} \\
        Open Catalyst Project & 1.2 M molecular relaxations with results from over 250 M DFT calculations relevant for renewable energy storage.\cite{zitnick2020introduction} & \url{https://opencatalystproject.org/index.html} \\
        OQMD & Consists of DFT predicted crystallographic parameters and formation energies for over 200k experimentally observed crystal structures.\cite{0324_Saal2013} & \url{http://oqmd.org} \\
        PubChemQC & Provides ca. 3 million molecular structures optimized by density functional theory (DFT) and excited states for over 2 million molecules using time-dependent DFT (TDDFT).\cite{0024_Nakata2017} & \url{http://pubchemqc.riken.jp/} \\
        QM7-X & Comprehensive dataset of 42 physicochemical properties for $\sim$4.2 M equilibrium and non-equilibrium structures of small organic molecules with up to seven non-hydrogen (C, N, O, S, Cl) atoms.\cite{hoja2020qm7} & \url{https://zenodo.org/record/4288677#.X9jHNC2ZNTY} \\
        QM9 & Geometric, energetic, electronic, and thermodynamic properties for 134k stable small organic molecules out of GDB-17.\cite{0308_Ramakrishnan2014} & \url{https://figshare.com/collections/Quantum_chemistry_structures_and_properties_of_134_kilo_molecules/978904} \\
        Synthesis Project & Collection of aggregated synthesis parameters computed using the text contained within over 640,000 journal articles.\cite{0212_Kim2017} & \url{ www.synthesisproject.org} \\
        quantum-machine.org & A repository of diverse datasets, including valence electron densities, chemical reactions, solvated protein fragments and molecular Hamiltonians.& \url{ http://quantum-machine.org/datasets/} \\
        \hline
    \end{tabular}}
    \label{tab:DATABASES}
\end{table}

ML model must be validated before they can be trusted for predictions. Validations of descriptors or model trainings are performed on benchmark data sets, and the most popular ones are summarized in Table~\ref{tab:DATABASES}.
These allow ML models to be compared on the same ground and provide large amounts of data for robust training.
Their availability to the public also ensures that the data sets can evolve with time and be extended as a part of community efforts.\cite{0373_Tetko2017} 

\begin{figure}
\includegraphics[width=0.85\textwidth]{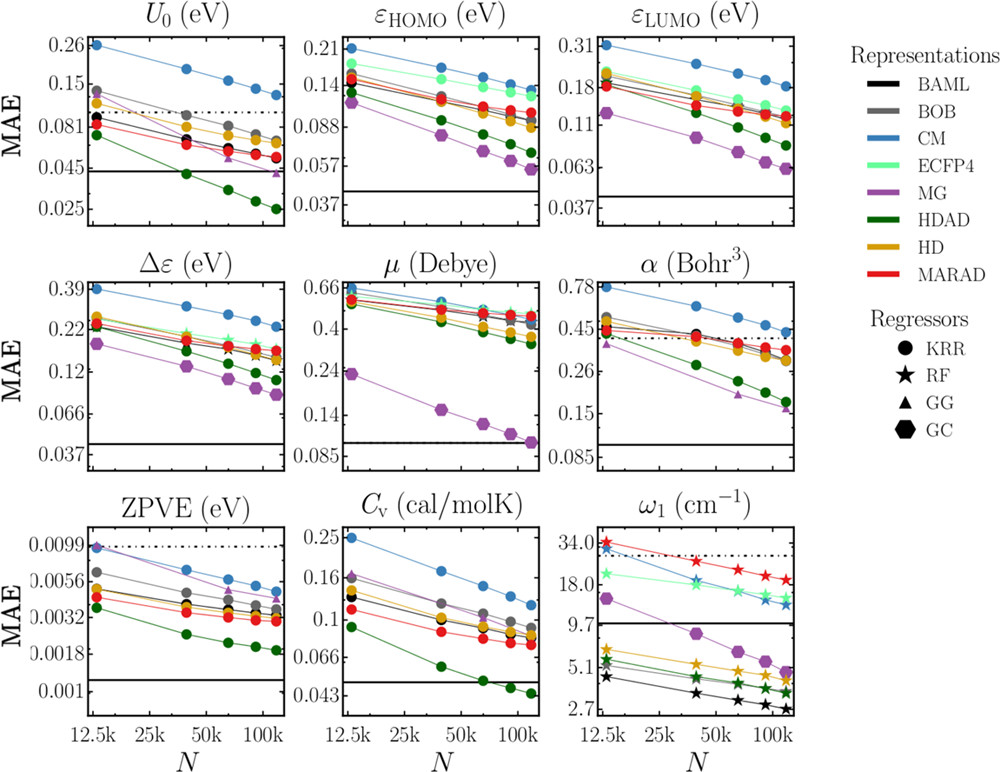}
\caption{Models shown differ by representation and architecture. 
The learning curves of various properties as a function of training sizes of the QM9 set.
Reprinted with permission from Ref.~\citenum{0115_Faber2017}. Copyright (2020) American Chemical Society.
}
\label{fig:qm9-challenge}
\end{figure}

Amongst the entries in Table~\ref{tab:DATABASES}, the most often used one is the QM9 set, which consists of approximately 134,000 of the smallest organic molecules that contain up to 9 heavy atoms (C, O, N, or F; excluding H) along with their CompChem-computed molecular properties such as total energies, dipole moments, HOMO-LUMO gaps, etc.
Several ML studies have already been published using this dataset (see Fig.~\ref{fig:qm9-challenge}, Ref.~\citenum{0115_Faber2017}).
A popular challenge is to develop a next-generation ML model that learns the electronic energies of random assortments of organic molecules with higher accuracy and less required training data than other existing models.
Doing so tests next generation molecular representations and training algorithms.
The next significant advance will potentially be due to a combination of supervised and unsupervised learning models.

\subsubsection{Visualization of data sets}

\begin{figure*}[hbt]
\includegraphics[width=1.\textwidth]{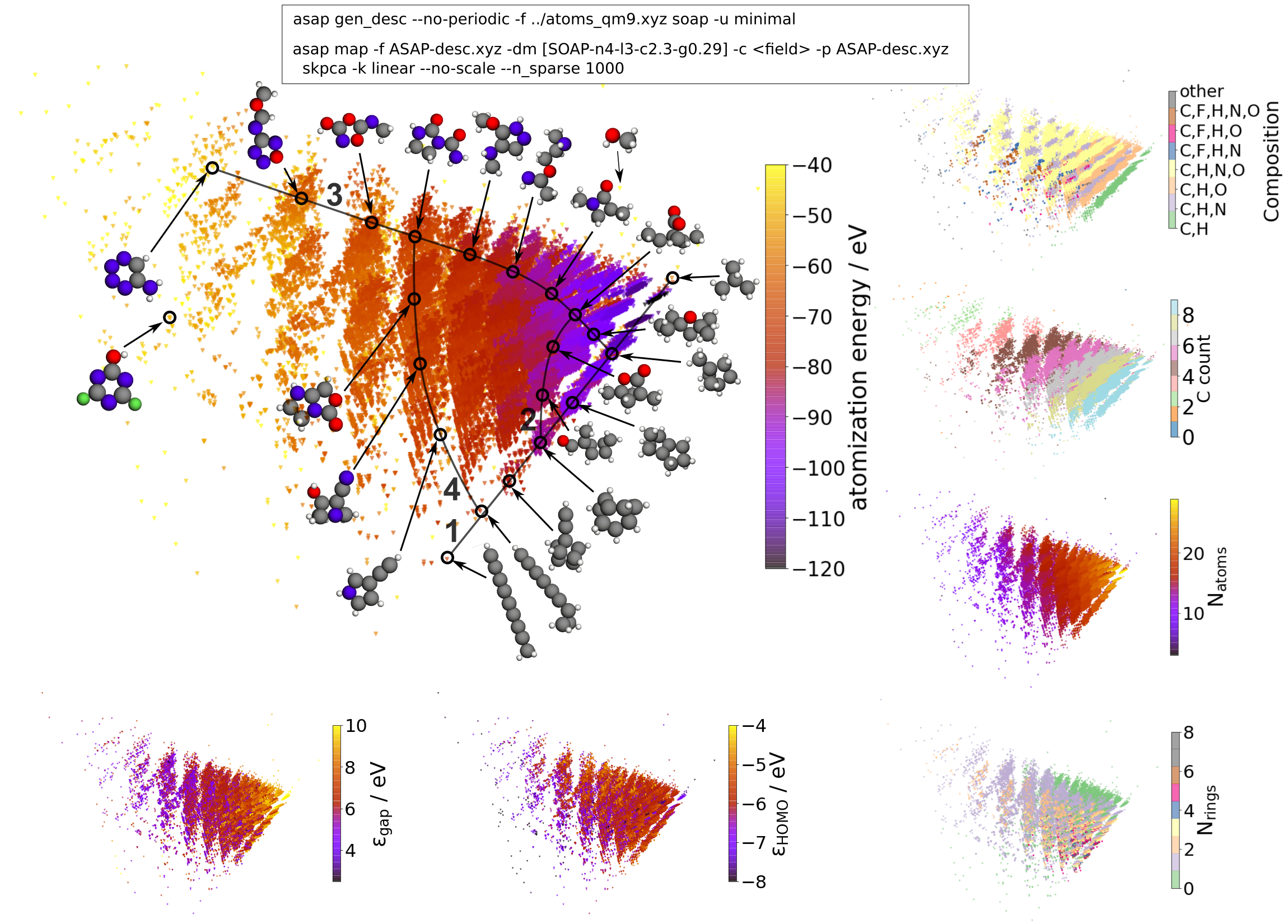}
\caption{KPCA maps of the QM9 database using a global SOAP kernel. The frames are color-coded according to structural descriptors (b,c,d,g) and quantum mechanical properties (a,e,f).
Reprinted with permission from Ref.~\citenum{cheng2020mapping}. Copyright (2020) American Chemical Society.
}
\label{fig:qm9-overview}
\end{figure*}

As the structural data sets grow it becomes infeasible to manually identify hidden patterns or curate the data.
Data-driven and automated frameworks for visualizing these data sets become increasingly popular.\cite{Montavon2013,0184_Isayev2015,ceri19jcp}
Dimensionality reduction effectively translates the high dimensional data (i.e. the xyz-coordinates for molecules or materials in different atomic configurations) into a low-dimensional space easily visualized on paper or a computer screen. 
In this way, entries such as those in the QM9 set can be shown (see Fig.~\ref{fig:qm9-overview}).
This map thus is a useful tool to help navigate the QM9 set by showing how molecules of different compositions (represented as colored dots) are similar or dissimilar based on molecular properties (i.e. atomization energies) along the principal axes.
Similarly, Ref.~\citenum{Basdogan2020Machine} used SOAP-sketchmaps in conjunction with quasi-chemical theory to show an unsupervised learning procedure for identifying local solvent environments that significantly impact solvation energies of small ions.

These data-driven maps are generated by processing the design matrix (or kernel matrix) associated with a data set using dimensionality reduction techniques introduced in Section ~\ref{ssec:learning_types}. 
A simple option is to use the ASAP code,~\cite{cheng2020mapping} a Python-based command line tool, that automates analysis and mapping.
Fig.~\ref{fig:qm9-atom} and ~\ref{fig:qm9-overview} were generated using ASAP using only two commands that are displayed in the figure.
Data sets can also be explored in an intuitive manner using
interactive visualizers~\cite{frau+20joss} that run in a web browser and display 3D-structures corresponding to each atomistic structure in the data set. 

\subsubsection{Text and data mining for chemistry}

Conventional publications are an essential part of CompChem knowledge base, and ML is becoming useful at accelerating information extraction from the scientific literature via text mining.\cite{0301_Qiang2006,0409_Wu2008,0435_Li2019} 
This topic was previously comprehensively reviewed in the context of cheminformatics.\cite{0471_Tshitoyan2019,0257_Lo2018} 
Natural language processing has already driven text-mining efforts materials science discovery\cite{0471_Tshitoyan2019} and experimental synthesis conditions of oxides.\cite{0212_Kim2017,0213_Kim2017a} 
CompChem+ML can also amplify existing efforts in chemometrics,\cite{0226_Kowalski1975} the science of data-driven extraction of chemical information.\cite{Swain2016} 
This area has also branched into related disciplines of data mining for specific classes of materials\cite{0364_Sparks2016} and catalysis informatics.\cite{0278_Medford2018} 
These approaches have great promise, especially for deriving information and knowledge from data, but it remains challenging to implement these in ways that achieve insight (and true impact).

Some have shown paths forward for doing so. 
For example, ML models can obtain knowledge from failed experimental data more reliably than humans who are more susceptible to survivor bias,\cite{0305_Raccuglia2016} and it can also be used to distill physical laws and fundamental equations using experimental\cite{schmidt2009distilling} and computational data.\cite{xie2019functional}    
ML models can also be used to reliably predict SMILES representations that allow encoded information to be derived from low-resolution images found in the literature.\cite{0366_Staker2019} 
ML models can interpret experimental X-ray absorption near edge structure (XANES) data and predict real space information about coordination environments.\cite{0375_Timoshenko2017} 
Likewise, scanning tunneling microscopy (STM) data can be used to classify structural and rotational states on surfaces,\cite{0429_Ziatdinov2017} and name indicators can be used to predict in tandem mass spectrometry (MS/MS) properties.\cite{0425_Zhou2017} 
In closing, we see exciting opportunities for future applications that complement data and text mining to chemometrics through chemical space. 

\subsection{Transforming atomistic modelling}

We previously mentioned that ML can handle large data sets and extract insights while circumventing the high cost of quantum-mechanical calculations by statistical learning.
CompChem+ML also has great potential in developing MLPs.
Car and Parrinello proposed running MD using electronic-structure methods in 1985.~\cite{car-parr85prl}
These are now mainstream but also quite computationally demanding and normally restricted to small system sizes ($\sim$ 100 atoms) and short simulation time ($\sim 10^{-12}~\mathrm{s}$).
Alternatively, accurate atomistic potentials introduced in Section \ref{sec:FF} have been developed to allow Monte Carlo and MD calculations, but sufficiently accurate potentials are sometimes not available.
MLPs have thus emerged as way to achieve as high accuracy as KS-DFT or correlated wavefunction methods but with a fraction of the cost.
MLPs have been constructed for far-reaching systems from small organic molecules to bulk condensed materials and interfaces.\cite{0045_Behler2016,Butler2018, Deringer2019}
Several of the co-authors of the current review have also written separate review focused more narrowly on this topic,\cite{unke2020} and so we only provide a brief overview here.  

Training an MLP to reproduce a system's PES usually requires generating diverse and high quality CompChem data points that cover the relevant temperature and pressure conditions, reaction pathways, polymorphs, defects, compositions, etc.\cite{0105_Dral2017,0159_Handley2010,0198_Jiang2016,0222_Kolb2016,0351_Shao2016, 0252_Li2019, 0125_Fu2018,Ballard2017} 
After data points comprised of atomic configurations, system energies, and forces are obtained, different methods for constructing MLPs employ either different descriptors (see a list of examples in Table~\ref{tab:descriptors}) or different ML architectures to perform interpolations of the full PES.
Again, smoothness is an essential feature for any PES,
so special considerations are needed to avoid numerical noise that would result in discontinuities. \cite{0075_Chen2013,0061_Brown1996}
Kernel method-based MLPs such as GAP\cite{0025_Bartok2010, 0017_Bernstein2019} and sGDML\cite{0080_chmiela2017machine,Chmiela2018,chmiela2019sgdml} ensure smoothness by relying on smoothly varying basis functions, but the scaling of kernel-based methods with respect to the number of training points is challenged without reduction mechanisms.~\cite{scholkopf1999input,Partay2010}
As a much more efficient but somewhat less accurate alternative to GAP, Spectral Neighbor Analysis Potential (SNAP)~\cite{Thompson2015} uses the coefficients of the SOAP descriptors and assumes a linear or quadratic relation between energies and the SOAP bispectrum components.\cite{Zuo2020}
The most popular MLPs are currently NN-based due to their flexibility and capacity to train based on large amounts data.
Amongst these,
ANI~\cite{smith2017ani, 0029_Smith2020} and BPNN\cite{0045_Behler2016,0032_Behler2007a,0043_Behler2014} potentials use ACSF descriptors as inputs,
while Deep neural networks such as SchNet\cite{0337_Schutt2017a,0338_schutt2018schnet,0339_Schutt2019} and DeepMD\cite{0158_Han2017} use the coordinates and nuclear charges of atoms.
We now focus on a few example applications.

\subsubsection{Predicting thermodynamic properties}

Many CompChem efforts focus on predicting thermodynamic properties at finite temperatures,
such as heat capacity, density, and chemical potential. 
Although many physical properties are already accessible from MD simulations, doing estimations of free energies that establish the relative stability of different states using electronic structure methods remains difficult.
The configurational part of the Gibbs free energy of a bulk system that has $N$ distinguishable particles with atomic coordinates $\textbf{r}=\{ \textbf{r}_{1\ldots N}\}$, and the associated potential energy $U(\textbf{r})$ can be expressed as:
\begin{equation}
    G(N,P,T) = 
    -k_B T \ln \int d \textbf{r}
  \exp\left[-\dfrac{U(\textbf{r})+PV}{k_B T}\right],
    \label{eq:gibbs}
\end{equation}
integrated over all possible  coordinates $\textbf{r}$,
where $k_B$ is the Boltzmann constant.
In order to rigorously determine $G$,
one must exhaustively sample the configuration space that has relatively high weight $\exp\left[-\dfrac{U(\textbf{r})+PV}{k_B T}\right]$.
This normally requires thermodynamic integration or enhanced sampling methods (e.g. umbrella sampling,~\cite{torrie1977nonphysical} metadynamics,~\cite{laio-parr02pnas} transition path sampling\cite{bolhuis2002transition}), that require simulation times and scales far beyond what is accessible with MD simulations based on KS-DFT or correlated wavefunction methods.

However, MLPs have unleashed both limits on the time scale and system size.
An early example,~\cite{mora+16pnas} used an MLP with umbrella sampling~\cite{torrie1977nonphysical} and the free energy perturbation method~\cite{tuckerman2010statistical} to reveal the influence of van der Waals corrections on the thermodynamic properties of liquid water.
Later, the combination of an MLP trained from hybrid DFT data and free energy methods 
reproduced several thermodynamic properties of water from quantum mechanics, including the density of ice and water, the difference in melting temperature for normal and heavy water, and the stability of different forms of ice.\cite{cheng2019ab,Reinhardt2021}
Ref.~\citenum{Niu2020} employed the DeepMD approach to study the relatively long time-scale nucleation of gallium.
MLPs for high-pressure hydrogen provided evidence on how hydrogen gradually turns into a metal in giant planets.~\cite{cheng2020evidence} 
In all these examples, high accuracy and long timescales were required to model the specific phenomena and reveal physical insights, and it is precisely the combination of CompChem+ML that enables both.

\subsubsection{Nuclear quantum effects}

As mentioned in Sec.~\ref{sec:nqes}, NQEs of chemical systems having light elements bring challenges for atomistic modelling because the added mobility of lighter atoms in dynamics simulations requires higher computational cost to treat.
To make the matter even more complicated, many atomistic potentials (see Sec.~\ref{sec:FF}), particularly the ones for water or organic molecules, cannot be used to model NQEs, because they often describe colavent bonds as rigid, and thus cannot describe the fluctuations of the bond lengths and angles.
As a remedy, several studies have been performed by training an MLP using higher rungs of KS-DFT (e.g. hybrid-DFT or meta-GGA) and this use this potential in PIMD simulations.\cite{chen+16jpcl,morawietz2018interplay,cheng2019ab,Ko2019}
The study of water that was mentioned in the previous section and used MLPs trained from hybrid DFT revealed that NQEs were critical for promoting the hexagonal packing of molecules inside ice that ultimately lead to the six-fold symmetry of snowflakes.\cite{cheng2019ab}
Highly data efficient ML potentials can even trained on reference data at the computationally very expensive quantum-chemical CCSD(T) level of accuracy. For example, the sGDML\cite{Chmiela2018,sauceda2019,sauceda2020dynamical} approach has been shown to faithfully reproduce such force fields for small molecules, which were then used to perform simulations with effectively fully quantized electrons and nuclei.

\subsection{ML for structure search, sampling, and generation}

Locating stationary points on the potential-energy surface (PES) is a frequent task in CompChem, since minima represent thermodynamically stable states and first order saddle points for transition states that define barrier heights that explaining reaction kinetics. 
Explorations for stationary points normally require many energy and force evaluations.
ML approaches are being implemented to dramatically accelerate minimum energy as well as saddle-point optimizations.\cite{DelRio2019,GarridoTorres2019,Peterson2016,Bisbo2020,meldgaard2018machine,Deringer2020,0017_Bernstein2019}
Bernstein \emph{et al.} proposed an automated protocol that iteratively explores structural space using a GAP potential.\cite{0017_Bernstein2019}
Bisbo and Hammer employed an actively-learned surrogate model of the potential energy surface to performs local relaxations while only performing single-point quantum-mechanical calculations for selected structures with high values of acquisition.\cite{Bisbo2020}
Refs.~\citenum{Peterson2016,GarridoTorres2019,Meyer2019,Koistinen2019} accelerate nudged elastic band (NEB) calculations by incorporating a surrogate ML models.

ML can also dramatically accelerate the challenge of efficiently sampling equilibrium or transition states by accelerating enhanced sampling methods such as umbrella sampling\cite{torrie1977nonphysical} and metadynamics.\cite{laio-parr02pnas} 
These procedures make use of collective variables (CVs) that define a reaction coordinate, and computing the associated free energy surface (FES) amounts to generating the marginal probability distribution in these CVs. 
Unfortunately, the choice of the CVs is not always clear for specific systems, and ML have shown some promise in guiding their determination.\cite{Chiavazzo2017,Rogal2019, Bonati2020cv}
Another direction is to exploit that ML models can be considered as universal approximators of free energy surfaces.\cite{Mones2016}
For example, there are reports of adaptive enhanced sampling methods using a Gaussian Mixture model,\cite{Debnath2020} using an NN architecture to represent the FES\cite{schneider2017stochastic} or the bias function in variational sampling simulations.~\cite{Bonati2019}

ML methods also offer fundamentally new ways to explore chemical compound and configuration space.
Generative models can learn the structural and elemental distribution underlying chemical systems, and once trained, these models can then be used to directly sample from this distribution.
It is furthermore possible to bias the generated structures towards exhibiting desired properties, e.g. drug activity or thermal conductivity.
As a consequence, generative models offer exciting new avenues in drug and materials design.\cite{vanhaelen2020advent,sanchez2018inverse}
%
Generative methods in CompChem include recurrent neural networks (RNNs), which can be used for the sequential generation of molecules encoded as SMILES strings (a string based representation of molecular graphs).\cite{gupta2017generative, segler2018generating, merk2018novo}
Segler \emph{et al.} demonstrated how such a recurrent model can first learn general molecular motifs and then be fine-tuned to sample molecules exhibiting activity against a variety of medical targets.\cite{segler2018generating}
Autoencoders (AE) are another frequently used ML method for molecular generation.
AEs learn to transform of molecular graphs or SMILES into a low-dimensional feature space and backwards.
The resulting feature vector represents a smooth encoding of the molecular distribution and can be used to effectively sample chemical space.\cite{liu2018constrained,jin2018junction,dai2018syntax,kusner2017grammar,jorgensen2018machine,jin2019learning}
By applying a variational AE to the QM9 and ZINC databases, Gomez-Bombarelli \emph{et al.} could generate several optimized functional compounds.\cite{0150_Gomez-Bombarelli2018}
An interesting extension to AEs are conditional AEs, which not only capture the distribution of molecular structures but also its dependence on various properties.\cite{0253_Lim2018,polykovskiy2018entangled}
This makes it possible to directly generate structures exhibiting certain property ranges or combinations without the need for biasing or additional optimization steps.
AEs can also form the basis of another approach for exploring chemical space called generative adversarial networks (GANs).\cite{blaschke2017auto,kadurin2017drugan}
In a GAN, a generator model (often an AE) attempts to create samples that closely match the underlying data, while a discriminator tries to distinguish true from generated samples.
These architectures can be enhanced by using reinforcement learning (RL) objectives.
RL learns an optimal sequence of actions (e.g. placement of atoms) leading to a desired outcome (e.g. molecule with certain property).
This makes it possible to drive generative process towards certain objectives, allowing for the targeted generation of molecules with particular properties.\cite{yu2016seqgan,guimares2017objective,de2018molgan,maziarka2020mol}
RL in general is a promising alternative strategy for generative models,\cite{You2018,Popova2018} and they offer the possibility for tightly integrating them into drug design cycles.\cite{Zhavoronkov2019}
Alternative approaches combine autoregressive models with graph convolution networks.\cite{li2018multi,li2018learning}
%

While these methods use SMILES or graphs to encode molecular structures, generative models have recently been extended to operate on 3D coordinates of molecules and materials.\cite{mansimov2019molecular,0134_Gebauer2018}
Gebauer \emph{et. al} proposed a auto-regressive generative model based on the SchNet architecture, called g-SchNet.\cite{gebauer2019gschnet}
Once trained on the QM9 dataset, g-SchNet was able to generate equilibrium structures without the need for optimization procedures.
It was further found, that the model could be biased towards certain properties.
In another promising approach, No{\'e} \emph{et al.} used an invertible neural network based on normalizing flows to learn the distribution of atomic positions (e.g. sampled from a molecular dynamics trajectory).
This network can then be used to directly sample molecular configurations by sampling from this distribution without performing costly simulations.\cite{noe2019boltzmann}

\subsection{Multiscale modeling}

Multiscale modeling is a term for including simulation or information from different scales (see Fig. \ref{fig:multiscale}).
ML has been introduced into QM/MM-like schemes that enable improved multiscale simulations,\cite{0020_Caccin2015,Zhang2019solvation,gastegger2020machine}
and on the side of coarse-graining.\cite{Gkeka2020}
Coarse-graining have been developed,~\cite{Saunders2013} but the inherent functional form for these potentials relies on CPI as well as trial-and-error procedures.
Several works used ML for constructing coarse-grained potentials by matching mean forces.\cite{John2017, Zhang2018a,wang2019machine,wang2020ensemble}
In closing, we see promise for experimental priors into ML models, for instance,
using experimental measurements to improving a ML potential energy surface by complementing with experimental data.
We are not aware of such efforts for developing highly accurate MLPs beyond the atomic scale, although much work has been done along this line to refine force fields of RNAs and proteins, often incorporating methods from ML, including the maximum entropy approach.\cite{cesari2016combining}

%% file: Sections/5-applications.tex
\newpage
\section{Selected applications and paths toward insights}

The central challenge posed at the beginning of this review was how to identify and make chemical compounds or materials having optimal properties for a given purpose.  
To do so would help address critical and broad issues from pollution, global warming to human diseases.
Traditional developments are often slow, expensive, and restricted by non-transferable empirical optimizations, and so efforts have turned to CompChem+ML to alleviate this.\cite{0144_Goh2017a,0146_Goh2018,0156_Hachmann2011,0024_Nakata2017} 

CompChem+ML are enabling searches through larger areas of chemical space much faster than before.\cite{0094_Curtarolo2013,0149_Gomez-Bombarelli2016, 0223_Kolb2017,von2020exploring,Tkatchenko2020NatCommun}
This section is not to extensively review the large amount of work using CompChem+ML in these different areas, but rather to highlight examples of applications that have resulted in notable insights so that others might use these notable works as templates for future efforts.  

\subsection{Molecular and material design}
Molecules and materials design is usually considered to be an optimization problem.\cite{0001_Agrawal2016,0150_Gomez-Bombarelli2018,jin2018junction,0229_Kuhn1996,0253_Lim2018} 
Thus, a comprehensive understanding of chemical space is needed to identify compounds with desired properties that are subject to certain required constraints (e.g. a specific thermal stability or a suitable optical gap for absorbing sunlight).
Those properties will also depend on many key variables (e.g. constitutive elements, crystal forms, geometrical and electronic characteristics, among others), which make the property prediction complex.\cite{0184_Isayev2015} 
CompChem calculations as explained in Section 2 should provide a continuous description of properties across a continuous representation (i.e. a descriptor or fingerprint) of molecules that is used to map molecular configurations to target properties, and vice-versa. 
ML methods then can be implemented to search large databases to extract structure-property relationships for designing compounds with specific characteristics.\cite{0184_Isayev2015, 0185_Isayev2017, 0229_Kuhn1996, 0424_Zhang2020} 
Optimizations would then be performed on the structure-based function learned from training configurations, and the composition of the chemical compound would then be recovered back from the continuous representation.

As a protoypical example of molecular design via high-throughput screening, Gomez-Bombarelli \textit{et al.}\cite{0149_Gomez-Bombarelli2016} showed a computation-driven search for novel thermally activated delayed fluorescence organic light-emitting diode (OLED) emitters. 
That work first filtered a search space of 1.6 million molecules down to approximately 400,000 candidates using ML to anticipate criteria for desirable OLEDs. 
For the purpose of evaluating candidates, they estimated an upper bound on the delayed fluorescence rate constant (k$_{\rm{TADF}}$).
Time-dependent DFT calculations were then used to provide refined predictions of specific properties of thousands of promising novel OLED molecules across the visible spectrum so that synthetic chemists, device scientists, and industry partners would be able to choose the most promising molecules for experimental validation and implementation. 
Notably, this example of CompChem+ML resulted in new devices that exhibited an external quantum efficiency of over 22$\%$. 
Fig.~\ref{fig:gomez_bombarelli} shows the high accuracy of ML in predicting useful properties for high-throughput screening of molecules and materials based on k$_{\rm{TADF}}$ calculations. 
This work exemplifies how ML can accelerate the design of novel compounds in such a way that could not be possible using traditional CompChem methods alone.

\begin{figure}
    \centering
    \caption{Neural network predictions compared to TD-DFT derived data of \textit{log k}$_{\rm{TADF}}$ (R$^{2}$=0.94). ML models computed molecular properties needed for screening with an accuracy comparable to CompChem calculations, but at a fraction of the computational cost. See Ref.~\bibpunct{}{}{,}{n}{,}{,}\cite{0149_Gomez-Bombarelli2016}. Reprinted by permission from Springer Nature Customer Service Centre GmbH: Springer Nature, Nature Materials. Design of efficient molecular organic light-emitting diodes by a high-throughput virtual screening and experimental approach, Gomez-Bombarelli, R., \textit{et al.}, COPYRIGHT (2016). Permission pending.}
    \label{fig:gomez_bombarelli}
\end{figure}

\begin{figure}
    \centering
    \includegraphics[width=0.85\textwidth]{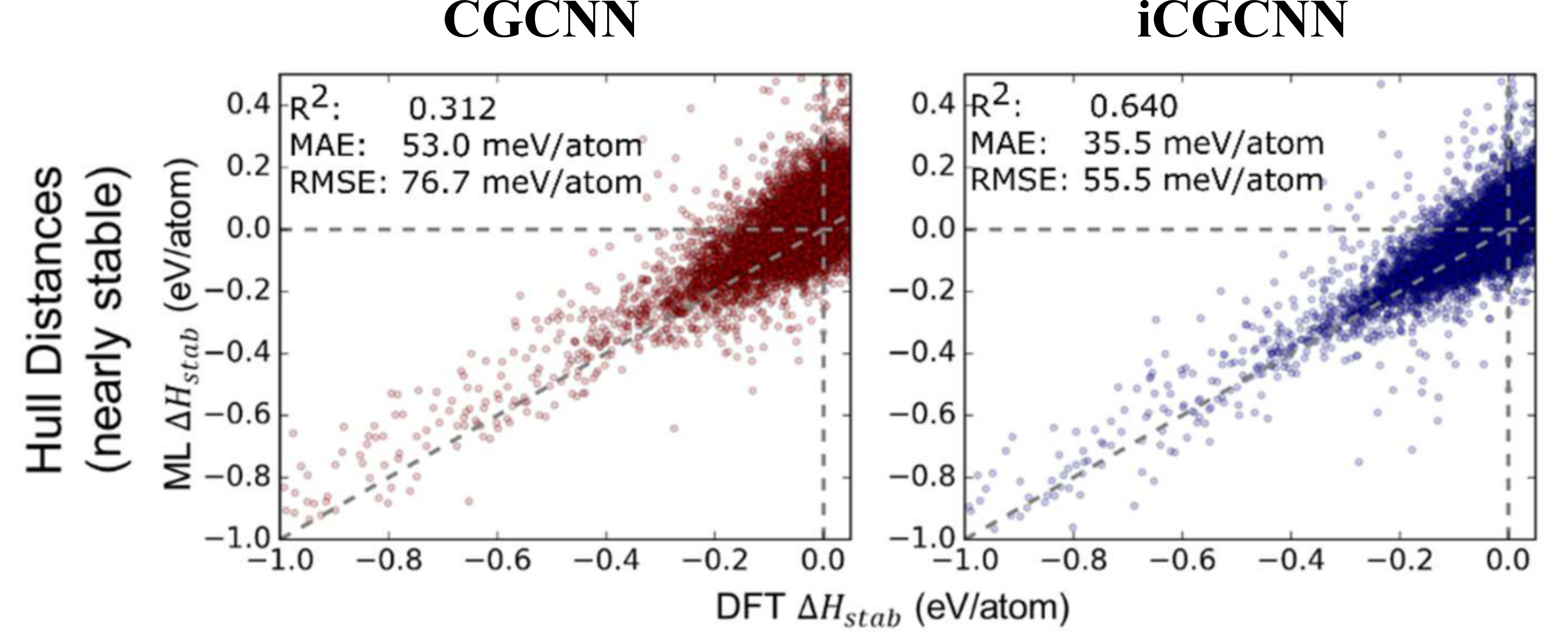}
    \caption{DFT vs ML predicted hull distances of nearly stable compounds (hull distances smaller than 50 meV/atom) for CGCNN and iCGCNN. The flexibility of ML approaches enable constructions of robust models tailored for specific target properties. See Ref.~\bibpunct{}{}{,}{n}{,}{,}\cite{Park2020a}. \protect\url{https://doi.org/10.1103/PhysRevMaterials.4.063801}.}
    \label{fig:park2020}
\end{figure}

One can also exploit the flexibility of ML models to increase the quality of the chemical insights. 
Integrations of features relevant to learning tasks allow one to improve the accuracy of ML predictions for a given target property. 
Park and Wolverton\bibpunct{}{}{,}{s}{,}{,}\cite{Park2020a} improved the performance of the crystal graph convolution neural network (CGCNN)\cite{Xie2018} by adding to the original framework information about Voronoi tessellated crystal structure, which are explicit 3-body correlations of neighboring constituent atoms, and an optimized representation of interatomic bonds. 
The new approach that was labeled as iCGCNN achieved a predictive accuracy 20$\%$ higher than that of the original CGCNN when determining thermodynamic stabilities of compounds (i.e. predictions of hull distances).
When used for high-throughput searches, iCGCNN exhibited a success rate higher than an undirected high-throughput search and higher than that of CGCNN. 
Fig.~\ref{fig:park2020} shows the improvement in predictions of nearly stable compounds after using more appropriate descriptors. 
This study showcases how descriptors can be tailored to further enhance the success of ML-aided high-throughput screening.


\subsection{Retrosynthetic technologies}

A grand challenge in chemistry is to understand synthetic pathways to desired  molecules.\bibpunct{}{}{,}{s}{,}{,}\cite{0102_Dewai1985, 0286_Nam2016} 
Retrosynthesis involves the design of chemical steps to produce molecules and materials that would be crucial to drug discovery, medicinal chemistry, and materials science. 
As a different kind of optimization problem, the general tactic is to analyze atomic scale compounds recursively, map them onto synthetically achievable building blocks, and then assemble those blocks into the desired compound.\cite{0346_Segler2018a, 0347_Segler2019, 0401_Warr2014} 

Three main issues make retrosynthesis a formidable intellectual challenge:\cite{0343_Schwaller2019} 
First, simple combinatorics make the space of possible reactions greater than the space of possible molecules. 
Second, reactants seldom contain only one reactive functional group, and thus require predictions of multiple functional groups. 
Third, one failed step in the route can invalidate the entire synthesis because organic synthesis is a multistep process.

Given these challenges, ML is becoming more established in determining reaction rules from CompChem data.\cite{0286_Nam2016} 
Computer-aided synthesis planning was actually first attempted in the 1960s.\cite{0090_Corey1969} 
Many have since attempted to formalize chemical perception and synthetic thinking using computer programs.\cite{0087_Coley2018, 0089_Cook2012, 0255_Liu2017} 
These programs are typically based on one of three possible algorithms:\cite{0255_Liu2017} 
\begin{packed_enum}
    \item Algorithms that use reaction rules (manually encoded or automatically derived from databases).
    \item Algorithms that use principles of physical chemistry based on \textit{ab initio} calculations to predict energy barriers.
    \item Algorithms based on ML techniques.
\end{packed_enum}

\begin{figure}
    \centering
    \includegraphics[height=0.5\textheight]{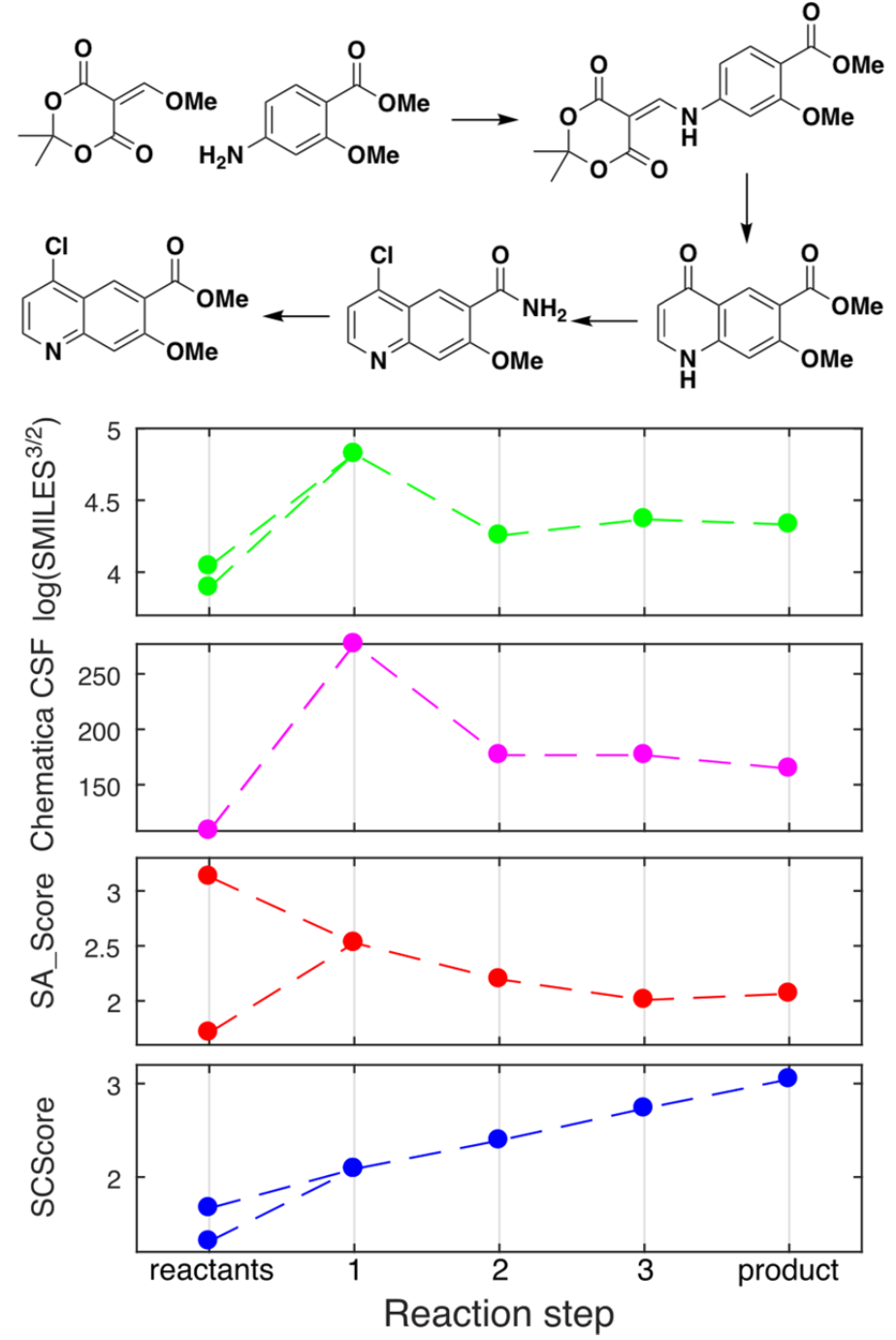}
    \caption{Use of different metrics to analyze the synthesis of a precursor to lenvatinib. Only the SCScore, a data-driven metric, correctly perceives a monotonic increase in complexity. ML models can give insights into which compounds are either reactants or products. See Ref.~\bibpunct{}{}{,}{n}{,}{,}\cite{0087_Coley2018}. Reprinted with permission from American Chemical Society, Accounts of Chemical Research. Machine Learning in Computer-Aided Synthesis Planning, Coley, C. W., \textit{et al.}, Copyright (2018) American Chemical Society.}
    \label{fig:coley2018}
\end{figure}

ML approaches are used to try to overcome the generalization issues of rule-based algorithms (that normally suffering from incompleteness, infeasible suggestions, and human bias) while also avoiding the high cost of CompChem calculations.
It is now possible to obtain purely data-driven approaches for synthesis planning, which are promoting a rapid advancement in the field. 
For example, Coley and co-workers\cite{Coley2018a} designed a data-driven metric, SCScore, for describing a real synthesis modeled after the idea that products are, on average, more synthetically complex than each of their reactants. 
The definition of a metric for selecting the most promising disconnections that produce easily synthesizable compounds is crucial for avoiding combinatorial explosion.
Fig.~\ref{fig:coley2018} shows that a data-driven metric, the SCScore, is more suitable than other heuristic metrics to perceive the complexity of each step in a given synthesis. 
This work offered a valuable contribution to the retrosynthesis working pipeline by providing a method that implicitly learns what structures and motifs are more prevalent as reactants.

Apart from isolated approaches or algorithms to deal with specific tasks within retrosynthesis, there is already software available to advance this field. 
One example is the Chematica program,\bibpunct{}{}{,}{s}{,}{,}\cite{0220_Klucznik2018} which has implemented a new module that combines network theory, modern high-power computing, artificial intelligence, and expert chemical knowledge to design synthetic pathways. 
A scoring function is used to promote synthetic brevity and penalize any reactivity conflicts or non-selectivities, thus allowing it to find solutions that might be hard for a human to identify.
Fig.~\ref{fig:klucznik2018}A shows the decision tree for one of the almost 50,000 reaction rules used in Chematica. 
Reaction rules can be considered as the allowed moves from which the synthetic pathways are built, and such moves lead to an enormous synthetic space (the number of possibilities within \textit{n} steps scales as 100$^{n}$) as the one shown by the graph in Fig.~\ref{fig:klucznik2018}B.
Chematica explores this large synthetic space by truncating and reverting from unpromising connections and drives its searches to the most efficient sequences of steps. 
Moreover, in the pathways presented to the user, each substance can be further analyzed with molecular mechanics tools.
This software was used to obtain insights into the synthetic pathways to eight targets (seven bioactive substances and one natural product). 
All of the computer-planned routes were not only successfully carried out in the laboratory, but they also resulted in improved yields and cost savings over previous known paths. 
This work opened an avenue for chemists to finally obtain reliable pathways from \textit{in silico} retrosynthesis.
For further reading we recommend the two-part reviews of Coley and co-workers.\cite{Coley2020,Coley2020a}

\begin{figure}
    \centering
    \caption{(A) Decision tree of one of the reaction rules within Chematica (double stereodifferentiating condensation of esters with aldehydes). The different conditions in the tree specify the range of admissible and possible substituents or atom types. (B) Reaction rules are used to explore the graph of synthetic possibilities (similar to the one shown here). Each node corresponds to a set of substrates. The combination of expert chemical knowledge, CompChem calculations and ML enables finding synthesizable paths. See Ref.~\bibpunct{}{}{,}{n}{,}{,}\cite{0220_Klucznik2018}. Reprinted from Chem, 4 (3), Klucznik, T., \textit{et al.}, Efficient Syntheses of Diverse, Medicinally Relevant Targets Planned by Computer and Executed in the Laboratory, 522-532, Copyright (2018), with permission from Elsevier. Permission pending.}
    \label{fig:klucznik2018}
\end{figure}

\bibpunct{}{}{,}{s}{,}{,}

\subsection{Catalysis}

Catalysis research requires multiscale approaches to determine chemical compounds that can influence barrier heights of reaction mechanisms to impact product yields and selectivities without otherwise being generated or consumed by the reaction.\cite{lindstrom2003brief}
Traditional catalysis is normally discussed in textbooks in terms of homogeneous (i.e. within a solution phase), heterogeneous (occurring at a solid/liquid interface), and biological (occurring within enzymes and riboenzymes), but it is best not to use these terms too strictly because actual reaction mechanisms can be quite complex and overall processes may sometimes exhibit characteristics (by design) of two or more of these classical processes.\cite{cui2018bridging,suen2017electrocatalysis,benson2009electrocatalytic}
Modern research in catalysis has been interested in studying chemical reactivity and reaction selectivity arising from stimuli from solar thermal energy,\cite{steinfeld2005solar,lee2019effect} electrochemical potentials,\cite{zeradjanin2016critical} photons,\cite{wenderich2016methods,hou2013review,julliard1983photoelectron,kavarnos1986photosensitization} plasmas,\cite{bogaerts20202020,van2008combining} or other external resonances.\cite{campos2019resonant}
Catalysis makes up roughly 35\% of the world's gross domestic product,\cite{ma2011heterogeneous} and it is important to guide toward the end goal of achieving greater sustainability with catalytic processes.\cite{anastas2013handbook,trost2002inventing,sheldon2007green}  

These reasons help make catalysis a fertile training ground for applying and developing theoretical models (e.g. Refs.~\citenum{Hammer2000, medford2015sabatier,calle2015introducing}) that can be used along with CompChem or CompChem+ML. 
The research field is also burgeoning with many reports and review articles\cite{0218_Kitchin2018,Toyao2020,0039_Grajciar2018,0278_Medford2018,Goldsmith2018,McCullough2020} that discuss perspectives and progress using ML methods for catalysis science; here, we will mention notable examples that present a broad range of ways that CompChem+ML can be used for insights.   
For example, CompChem+ML methods are enabling more data generation by allowing costly QC calculations to be run more efficiently, and more information means more comprehensive predictions of chemical and materials phase diagrams for catalysis\cite{Wexler2019, Ulissi2016} as well as stability and reactivity descriptors identified on the fly.\cite{Roling2017,Roling2019,Yan2018,Dean2019,Nunez2019} 
Figure \ref{fig:ulissiNature} shows examples of the palettes of insight available using state-of-the-art CompChem+ML modeling for identifying activity and selectivity maps, as well as visualizations of data using t-Distributed Stochastic Neighbor Embedding (t-SNE).\cite{JMLR:v15:vandermaaten14a}

\begin{figure}
    \centering
    \caption{CompChem+ML screening of hypothetical Cu and Cu-based catalyst sites. a) A two-dimensional activity volcano plot for CO$_2$ reduction. TOF, turnover frequency. b) A two-dimensional selectivity volcano plot for CO$_2$ reduction. CO and H adsorption energies in panels a and b were calculated using DFT. Yellow data points are average adsorption energies of monometallics; green data points are average adsorption energies of copper alloys; and magenta data points are average, low-coverage adsorption energies of Cu-Al surfaces. c) t-Distributed Stochastic Neighbor Embedding (t-SNE)\cite{JMLR:v15:vandermaaten14a} representation of approximately 4,000 adsorption sites on which we performed DFT calculations with Cu-containing alloys. The Cu-Al clusters are labelled numerically. d) Representative coordination sites for each of the clusters labelled in the t-SNE diagram. Each site archetype is labelled by the stoichiometric balance of the surface, that is, Al-heavy, Cu-heavy or balanced, and the binding site of the surface.
 Ref.~\bibpunct{}{}{,}{n}{,}{,}\cite{Zhong2020}. Permission pending.}
    \label{fig:ulissiNature}
\end{figure}

\begin{figure}
    \centering
    \includegraphics[width=\textwidth]{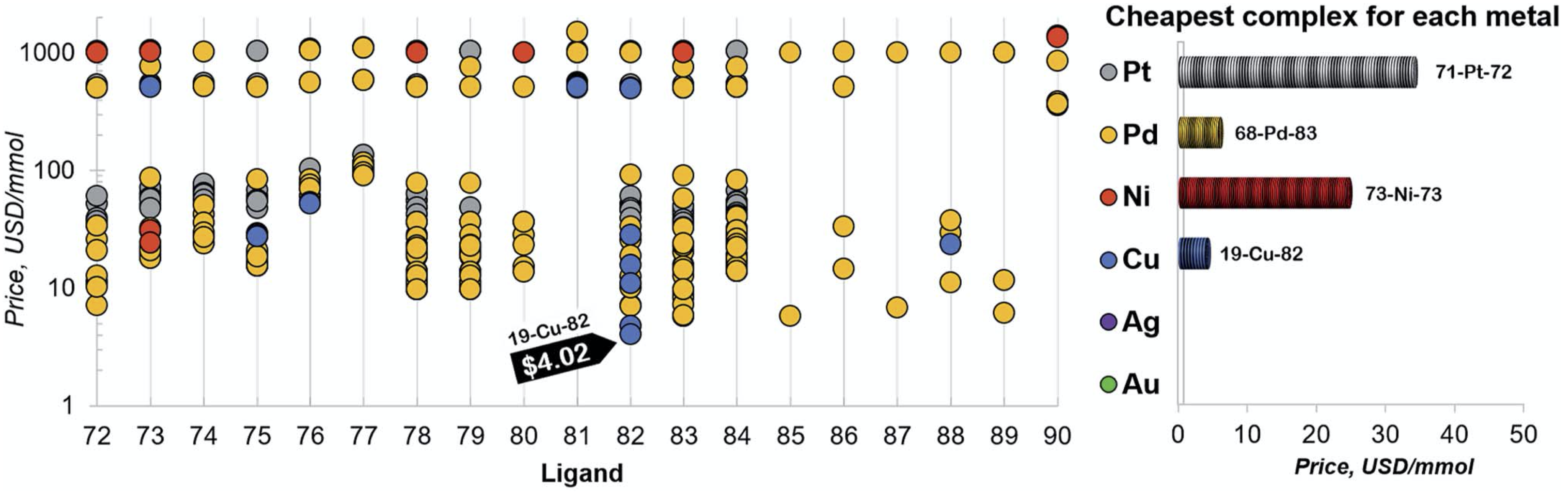}
    \caption{Estimated price (for one mmol in US dollars) of the catalysts in the selected range of $-$32.1/$-$23.0 kcal mol$^{-1}$ (for ligand no. 72 -- 90). The price is calculated as a summation of the commercial price of transition metal precursors (one mmol) and one mmol of each ligand. The cheapest complex for each metal is shown on the right. The estimated price of all the 557 catalysts is detailed in the reference: \citenum{Meyer2018} - Published by The Royal Society of Chemistry. Ref.~\bibpunct{}{}{,}{n}{,}{,}\cite{Meyer2018}}
    \label{fig:AvLCC}
\end{figure}

Regarding modeling of deeply complex chemical environments, Artrith and Kolpak developed MLPs for investigating the relationships between solvent, surface composition and morphology, surface electronic structure, and catalytic activity in systems composed of thousands of atoms interfaces.\cite{0007_Artrith2014}
We expect such simulations for electro- and photo-catalysis elucidation will continue to improve in size, scale, and accuracy. 
For other physical insights, new approaches by Kulik and Getman and co-workers have also focused on developing ML models appropriate for elucidating complex $d$-orbital participation in homogeneous catalysis.\cite{Nandy2019} 
Rappe and co-workers have used regularized random forests to analyze how local chemical pressure effects adsorbate states on surface sites for the hydrogen evolution reaction.\cite{Wexler2018} 
Almost trivially simple ML approaches can be used in catalysis studies to deduce insights into interaction trends between single metal atoms and oxide supports,\cite{OConnor2018} to identify the significance of features (e.g. adsorbate type or coverage) where CompChem theories break down,\cite{Griego2020} or they can be used to identify trends that result in optimal catalysis across multiple objectives such as activity and cost (Fig. \ref{fig:AvLCC}).\cite{Meyer2018} 

ML is also opening opportunities for CompChem+ML studies on highly detailed and complex networks of reactions.\cite{Boes2019, Ulissi2017b,Bort2020,Kayala2012,0403_Wei2016,Schwaller2020} 
Such models in principle can then significantly extend the range of utility of microkinetics modeling for predictions of products from catalysis.\cite{Blondal2019, Rangarajan2017} 
ML also enables studies of complicated reaction networks that can allow predictions of regioselective products based on CompChem data,\cite{Banerjee2018} asymmetric catalysis important for natural product synthesis,\cite{Reid2019,Zahrt2019} and biochemical reactions.\cite{Ravasco2020}  
Efforts to better understand `above-the-arrow' optimizations of reaction conditions relate back to the challenge of retrosynthetic challenges.\cite{Davies2019,Li2019c} 
Ideally, these efforts will continue while making use of rapid advances in CompChem+ML that enable predictive atomistic simulations to be run faster and more accurately.
We see reason for excitement for different approaches, but we again stress the importance of ensuring that models will provide unique and physical results (see Section \ref{section:ML} where we discuss the risk of ``clever Hans'' predictors\cite{lapuschkin2019unmasking}).

\subsection{Drug design}
The central objective for drug discovery is to find structurally novel molecules with precise selectivity for a medicinal function. 
This involves identifying new chemical entities and obtaining structures with different physicochemical and polypharmacological properties (i.e., combinations of beneficial pharmacological effects and/or adverse side-effects).\cite{Schneider2005,0313_Reker2015} 
Drug discovery involves the identification of targets (a property optimization task, as in material design) and the determination of compounds with good on-target effects and minimal off-target effects.\cite{0386_Wainberg2018} 
Traditionally, a drug discovery program may take around six years before a drug candidate can be used in clinical trials and additional six or seven years are required for three clinical phases. 
Thus, it is important to identify adverse effects as soon as possible to minimize time and monetary costs.\cite{0210_Khaket2014}  
Accelerating drug discovery relies on predicting how and where a certain drug binds to more than one protein, a phenomenon that sometimes results in polypharmacology. 
Researchers are developing ready-to-use tools aimed to facilitate research for drug discovery,\cite{blaschke2020reinvent} but CompChem+ML is expected to continue providing even more benefits to the drug development pipeline.\cite{Jimenez-Luna2020}

In a recent study, Zhavoronkov \textit{et al.}\cite{Zhavoronkov2019} developed a deep generative model for \textit{de novo} small-molecule design: the generative tensorial reinforcement learning (GENTRL) model that was used to discover potent inhibitors of discoidin domain receptor 1 (DDR1), a kinase target implicated in fibrosis and other diseases. 
The drug discovery process was carried out in only 46 days, beginning with the recollection of appropriate data for training and finishing with the synthesis and experimental test of some compounds (Fig.~\ref{fig:zhavoronkov2019}A). 
GENTRL was used to screen a total of 30,000 structures (some examples compared to the parent DDR1 kinase inhibitor are shown in Fig.~\ref{fig:zhavoronkov2019}B) down to only 40 structures that were randomly selected ensuring a coverage of the resulting chemical space and distribution of root-mean squared deviation values. 
Six of these molecules were then selected for experimental validation (see  Fig.~\ref{fig:zhavoronkov2019}C), with one of them demonstrating favorable pharmacokinetics in mice. 
The predicted conformation of the successful compound according to pharmacophore modelling was very similar to the one predicted to be preferred and stable by CompChem methods. 
This work illustrates the utility of CompChem+ML approaches to give insights into drug design by rapidly giving compound candidates that are synthetically feasible and active against a desired target.

\begin{figure}
    \centering
    \caption{(A) Workflow and timeline for the design of candidates employing GENTRL. (B) Representative examples of the initial 30,000 structures compared to the parent DDR1 kinase inhibitor. (C) Compounds found to have the highest inhibition activity against human DDR1 kinase. CompChem+ML methods can considerably accelerate the discovery of drugs that are effective against a desired target. See Ref.~\bibpunct{}{}{,}{n}{,}{,}\cite{Zhavoronkov2019}. Reprinted by permission from Springer Nature Customer Service Centre GmbH: Springer Nature, Nature Biotechnology. Deep learning enables rapid identification of potent DDR1 kinase inhibitors, Zhavoronkov, A., \textit{et al.}, COPYRIGHT (2019). Permission pending.}
    \label{fig:zhavoronkov2019}
\end{figure}

\begin{figure}
    \centering
    \includegraphics[width=0.9\textwidth]{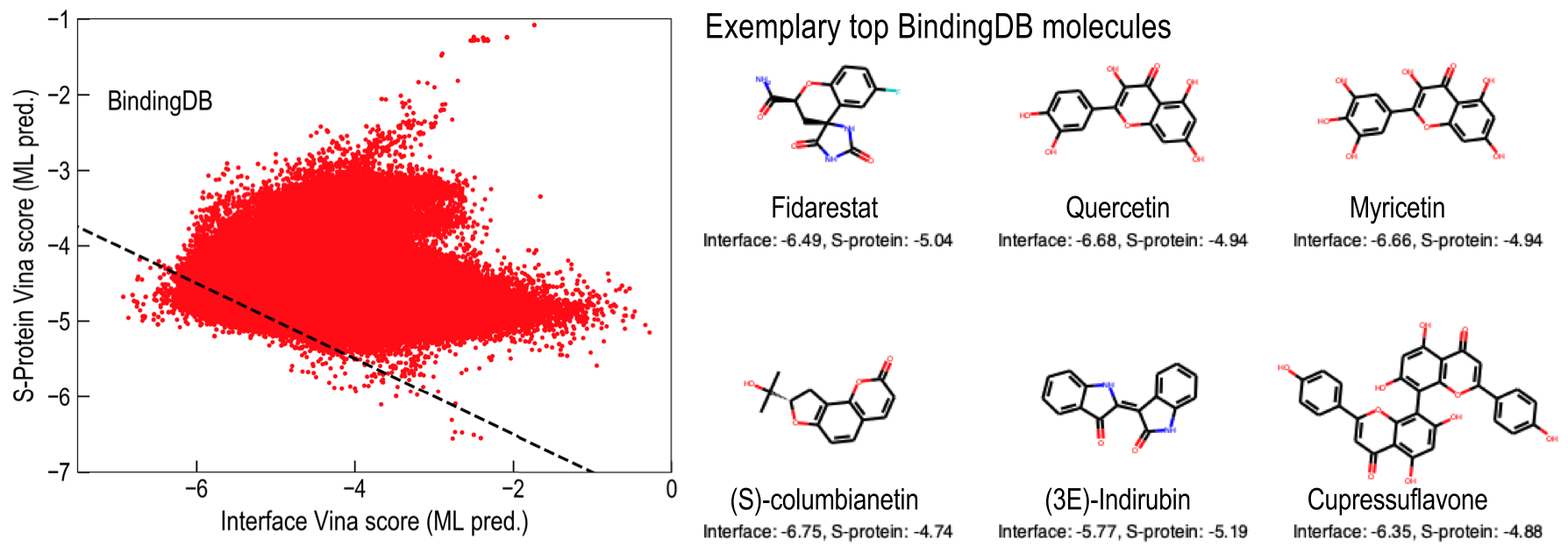}
    \caption{Vina scores predictions for the isolated protein (S-protein) and the protein-receptor complex (interface) for all the molecules in the BindingDB datasets and some exemplary top cases that satisfy the screening criteria. ML models trained on accurate CompChem databases are of upmost importance to efficiently gain insights into possible treatments, even for newly discovered diseases. Figure taken from Ref.~\bibpunct{}{}{,}{n}{,}{,}\cite{Batra2020}. \protect\url{https://doi.org/10.1021/acs.jpclett.0c02278}.}
    \label{fig:batra2020}
\end{figure}

Besides generating new chemical structures with favorable pharmacokinetics, ML methods are also used in pharmaceutical research and development for peptide design, compound activity prediction and for assisting scoring protein-ligand interaction (docking).\bibpunct{}{}{,}{s}{,}{,}\cite{0077_Chen2018, Piroozmand2020, Schneider2005, Fjell2012} 
An example of the latter was proposed by Batra \textit{et al.}\cite{Batra2020} for efficiently identifying ligands that can potentially limit the host-virus interactions of SARS-CoV-2. 
Those authors designed a high-throughput strategy based on CompChem+ML that involved high-fidelity docking studies to find candidates displaying high-binding affinities. 
The ML model was used to search through thousands of approved ligands by the Food and Drug Administration (FDA) and a million biomolecules in the BindingDB database.\cite{Liu2007} 
From these, insights were obtained for more than 19,000 molecules satisfying the Vina score (i.e. an important physicochemical measure of the therapeutic process of a molecule that is used to rank molecular conformations and predict free energy of binding). 
Fig.~\ref{fig:batra2020} shows the Vina score predictions that led to the selection of the best candidates, some of which are also illustrated in the figure.
The Vina scores for the top ligands were further confirmed using expensive docking approaches, resulting in the identification of 75 FDA-approved and 100 other ligands potentially useful to treat SARS-CoV-2. 
This study highlights a reasonable CompChem+ML strategy for making useful suggestions to aid expert biologists and medical professionals to focus in fewer candidates when performing either robust CompChem efforts or synthesis and trial experiments.

\bibpunct{}{}{,}{s}{,}{,}

%% file: Sections/6-conclusions.tex
\newpage
\section{Conclusions and Outlook}

Recent CompChem methods, algorithms, and codes have empowered new studies for a wealth of physical and chemical insights into molecules and materials. 
Today, the combination of CompChem+ML can be equipped to address new and more challenging questions in different domains of physics, materials science, chemistry, biology, and medicine. 
Productive research efforts in this direction necessitate interdisciplinary teams and increasing availability of high-quality data across appropriate regions of chemical compound space. 
Discovering new chemicals and materials requires thorough investigations.
One needs to predict reaction pathways and interactions between molecules, optimize environmental conditions for catalytic reactions, enhance selectivities that eliminate undesired side reactions and/or side effects, and navigate other system-specific degrees of freedom. 
Addressing this complexity calls for a statistical view on chemical design and discovery, and CompChem+ML provides a natural synergy for obtaining predictive insights to lead to wisdom and impact.  

This review provided an essential background of CompChem and ML and how they can be used together to make transformative impacts in the chemical sciences by amplifying insights available from CompChem methods. 
The successes of CompChem+ML are particularly visible in physical chemistry and include drastic acceleration of molecular and materials modeling, discovery and prediction of chemicals with desired properties, prediction of reaction pathways, and design of new catalysts and drug candidates. 
Nevertheless, we have only begun to scratch the surface of how successful applications of ML in chemistry can bring impact. 
There are many conceptual, theoretical, and practical challenges waiting to be solved to enable further synergies within the \emph{troika} of CompChem, ML, and CPI. 
Here we enumerate some of the challenges that we consider to be the most pressing and interesting at this moment:
\begin{enumerate}
    \item \emph{Reliance on ML in CompChem algorithms must be increased:} ML algorithms can be integrated into CompChem algorithms at almost any simulation level (Fig. \ref{fig:multiscale}). ML algorithms are already available to accelerate calculations of CompChem energies, navigations along reaction pathways, and sampling of larger regions of the PES, but the reluctance of their use impedes progress. In general, these algorithms must be made more effective, efficient, accessible, user-friendly, and reproducible to benefit fundamental and applied research (see for example, Ref. \citenum{Haghighatlari2020}). 
    
    \item \emph{More general ML approaches are needed:} ML methods must continue to evolve beyond now-common applications of learning a narrow region of a PES or identifying straightforward structure/property relationships. New ML methods should have the capacity to predict energetic and electronic properties and their more convoluted relationships across chemical space. Such approaches should grow toward uniformly describing compositional (chemical arrangement of atoms in a molecule) and configurational (physical arrangement of atoms in space) degrees of freedom on equal footing. Further progress in this field requires developing new universal ML models suitable for insights across diverse systems and physicochemical properties.
    
    \item \emph{ML representations must to include the right physics:} ML methods that are claimed to be accurate but incorrectly describe the true physics of a system will eventually fail to achieve meaningful insights while lowering the reputation of other work in the field. Current ML representations (descriptors) can successfully describe local chemical bonding, but few if any are treating long-range electrostatics, polarization, and van der Waals dispersion interactions that are critical for rationalizing physical systems, both large and small. Combining intermolecular interaction theory (a key focus of advanced CompChem methods) with ML is an important direction for future progress towards studying complex molecular systems. 
    
    \item \emph{CompChem + ML applications need to strive toward achieving realistic complexity:} Investigations using highly accurate CompChem methods normally require overly simplified model systems while more realistic model systems necessitate less accurate but computationally efficient CompChem methods. This compromise should no longer be necessary. 
    We are due for a paradigm shift in how thermodynamics, kinetics, and dynamics of systems in complex chemical environments (e.g. for multiscale biological processes like drug design and/or catalytic processes at solid liquid interfaces under photochemical excitations, etc.) can be treated more faithfully with less corner-cutting. 
    An emerging idea is to dispatch ML approaches into computationally efficient model Hamiltonians for electronic interactions based on correlated wavefunction, KS-DFT, tight-binding, molecular orbital techniques, and/or the many-body dispersion method. ML can predict Hamiltonian parameters and the quantum-mechanical observables would be calculated via diagonalization of the corresponding Hamiltonian. The challenge is to find an appropriate balance between prediction accuracy and computational efficiency to dramatically enhance larger scale simulations.    

    \item \emph{(Much) more experimental data is needed:} Validations of ML predictions require extensive comparisons with experimental observables such as reaction rates, spectroscopic observations, solvation energies, melting temperatures. Such experiments may have previously been considered too routine, too mundane, or not insightful enough alone, but all high quality brings great value for future CompChem+ML efforts that tightly integrate quantum mechanics, statistical simulations, and fast ML predictions, all within a comprehensive molecular simulation framework.~\cite{noe2020machine}
    
    \item \emph{(Much) more comprehensive data sets need to be assembled and curated:} Current CompChem+ML efforts have profited heavily by the availability of benchmark data sets for relatively small molecules that allow a comparison of existing models.\cite{rupp2012fast,0308_Ramakrishnan2014} While efforts fixated on boosting prediction accuracies and shrinking down requisite training set sizes for ML models have had their merits, it is time to move on as further improvements are meaningless if the ML models are not making useful and insightful predictions themselves. More useful predictions will require knowledge from larger datasets, and these will inevitably contain heterogeneous combinations of different levels of theory and/or experiments that must be analyzed, `cleaned', and uncertainties adequately quantified in order for models to productively learn.
    Such hybrid data sets may be the key to arrive at novel hypotheses in chemistry that could then be experimentally tested.
    
    \item \emph{Bolder and deeper explorations of chemical space are needed:} So far most efforts to generate chemical data have focused on exploring parts of chemical space for new compounds for a targeted purpose. This should change. Combining ML model uncertainty estimates across broader swaths of chemical space could open pathways for fruitful statistical explorations, say, in an active learning framework. This could lead to discovering new synergies between data that otherwise would not have been possible to enable advances in scientific understanding and improve ML models. 
    Generative models can bridge the gap between sampling and targeted structure generation imposing optimal compound properties, e.g. for inverse chemical design.\cite{0134_Gebauer2018,gebauer2019gschnet,schnorb} 
    
\end{enumerate}

This and other reviews\cite{von2018quantum,von2020exploring,vonLilienfeld2020,Tkatchenko2020NatCommun,noe2020machine,unke2020,strieth2020machine,QML-Book} have stated how ML has become instrumental for recent progress in CompChem. 
We would like to also mention inspirations that ML has drawn from being applied to physical and chemical problems.  

ML methods generally assume that data is subject to measurement noise while Comp\-Chem data is generally approximate but also noise-free from a statistical perspective. 
ML modeling still requires regularization, but regularizers should here particularly reflect the underlying physics of molecular and materials systems. 
ML models used in applications of vision contain discrete convolution filters that are suboptimal for chemical modeling, but recognition of this shortcoming has led to novel continuous convolution filters that are well suited for chemistry and have also become a popular novel architecture for core ML methods.\cite{0338_schutt2018schnet} 

Furthermore, invariances, symmetries, and conservation laws are key ingredients to physical and chemical systems. 
Incorporating them into ML has led to novel and useful models for chemistry since they can learn from significantly less data, which then makes it possible to build force fields at unprecedentedly high levels of theory.\cite{0080_chmiela2017machine,Chmiela2018,sauceda2020dynamical} 
Using these powerful ML techniques for computer vision, natural language processing, and other applications is currently being explored. 
Structural information from molecular graphs provide the basis for novel tensor neural networks or message passing architectures\cite{0336_schutt2017quantum,0140_Gilmer2017} as well as graph explanation methods.\cite{schnake2020xai} 

Many further challenges exist that have led or will lead to mutual {\em bidirectional} cross-fertilization between ML and chemistry.
These interdisciplinary efforts also initiate progress in respective application domains.
The power of this path is that solving a burning problem in chemistry with a novel crafted ML model may also result in unforeseen insights in how to better design core ML methods.   
Interestingly, the exploratory usage of ML for knowledge discovery in chemistry typically requires novel ML models and unforeseen scientific innovations, and this can lead to interesting insight that is not necessary limited to chemistry alone, rather it is likely to go beyond. 

To conclude, the past decade has shown that it has not been enough to {\em just} apply existing ML algorithms, but breakthroughs are happening by a handshaking of innovations resulting in novel ML algorithms and architectures driven by the pursuit of novel insights in chemistry {\em while} retaining a deep understanding about the underlying physical and chemical principles. 
Research programs that foster interdisciplinary exchange such as IPAM (www.ipam.ucla.edu) have seeded this progress, and these should be continued. 
Mixed teams with members educated in different aspects of physics, chemistry and ML have been instrumental. 
This also brings the need to  solve the new educational challenge of developing new generations of researchers with an academic curriculum that interweaves chemistry,
physics and computer science to enable a meaningful (multilingual)
research contribution to this exciting emerging field.

%% file: Sections/7-acknowledgement.tex
\begin{acknowledgement}
JAK was supported by the Luxembourg National Research Fund (INTER/MOBILITY/19/13511646) and the U.S. National Science Foundation (CBET-1653392 and CBET-1705592).

VVG acknowledges financial support from the Luxembourg National Research Fund (FNR) under the program DTU PRIDE MASSENA (PRIDE/15/10935404)

BC acknowledges funding from the Swiss National Science Foundation (Project P2ELP2-184408).

KRM was supported in part by Institute of Information \& Communications Technology Planning \& Evaluation (IITP) grants funded by the Korea Government (No. 2017-0-00451, Development of BCI based Brain and Cognitive Computing Technology for Recognizing User’s Intentions using Deep Learning) and funded by the Korea Government (No. 2019-0-00079,  Artificial Intelligence Graduate School Program, Korea University), and was partly supported by the German Ministry for Education and Research (BMBF) under Grants 01IS14013A-E, 01GQ1115, 01GQ0850, 01IS18025A, 031L0207D and 01IS18037A; the German Research Foundation (DFG) under Grant Math+, EXC 2046/1, Project ID 390685689.

AT acknowledges financial support from the European Research Council (ERC Consolidator Grant BeStMo and ERC-POC Grant DISCOVERER).

We gratefully acknowledge helpful comments on the manuscript by Hartmut Maennel. 

Address correspondence to: jakeith@pitt.edu, klaus-robert.mueller@tu-berlin.de, alexandre.tkatchenko@uni.lu
\end{acknowledgement}

%% file: Sections/8-bios.tex
\newpage
\section{Author Bios}

\begin{figure*}[h!]
\includegraphics[width=2in]{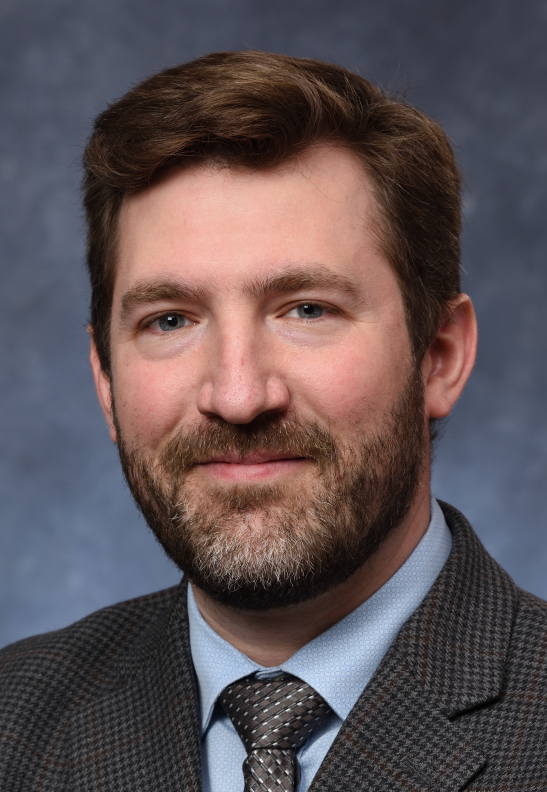}
  \caption{Picture of John A. Keith}
  \label{JAK_picture}
\end{figure*}

John A. Keith is an associate professor and R.K. Mellon Faculty Fellow in Energy at the University of Pittsburgh in the department of chemical and petroleum engineering. He obtained his bachelors' in chemistry at Wesleyan University and a Ph.D. degree in computational chemistry at Caltech in 2007. After an Alexander von Humboldt postdoctoral fellowship at the Universit{\"a}t Ulm, he was an Associate Research Scholar at Princeton University. He was a recipient of an NSF-CAREER award in 2017. His research interests lie in the applications and development of computational chemistry for engineering chemical reactions and materials for electrocatalysis, anticorrosion coatings, and the development of chemicals having less of an environmental footprint. He was a recipient of a Luxembourg Science Foundation INTER Mobility award in 2019-2020 to do a research sabbatical in Prof. Alexandre Tkatchenko's group at the University of Luxembourg. This review is a primary product of that visit.

\begin{figure*}[h!]
\includegraphics[width=2in]{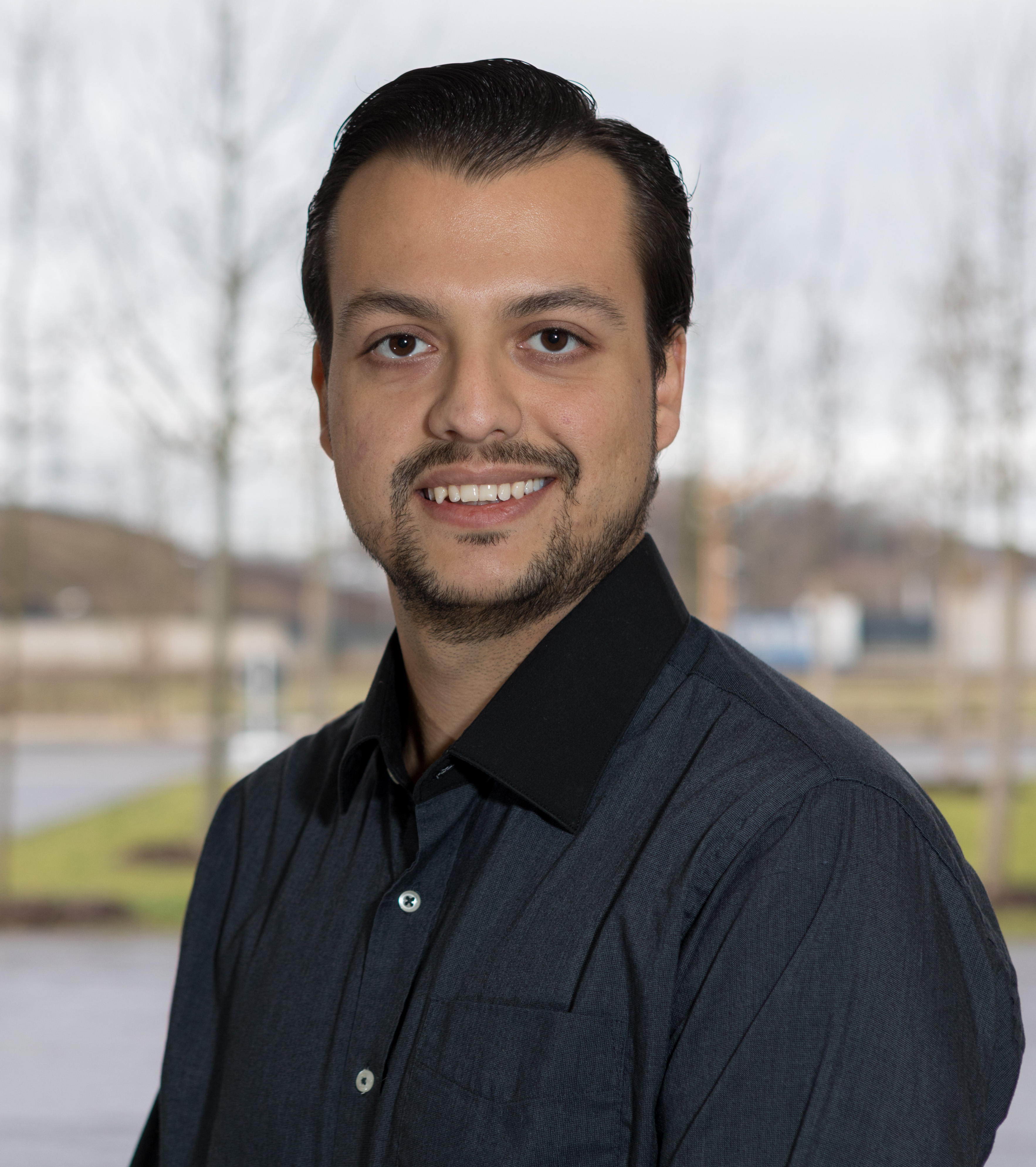}
  \caption{Picture of Valentin Vassilev-Galindo}
  \label{VVG_picture}
\end{figure*}

Valentin Vassilev-Galindo graduated with honors from University of Veracruz (Mexico) with a Bachelor's degree in Chemical Engineering in 2014. Then, he enrolled to the Master program in Physical Chemistry at Cinvestav-Mérida (Mexico) where he worked under the supervision of Professor Gabriel Merino until receiving the MSc. degree in 2017. He is currently pursuing a PhD degree at the University of Luxembourg in the research group of Professor Alexandre Tkatchenko. His research is mainly related to machine learning potentials.

\begin{figure*}[h!]
\includegraphics[width=2in]{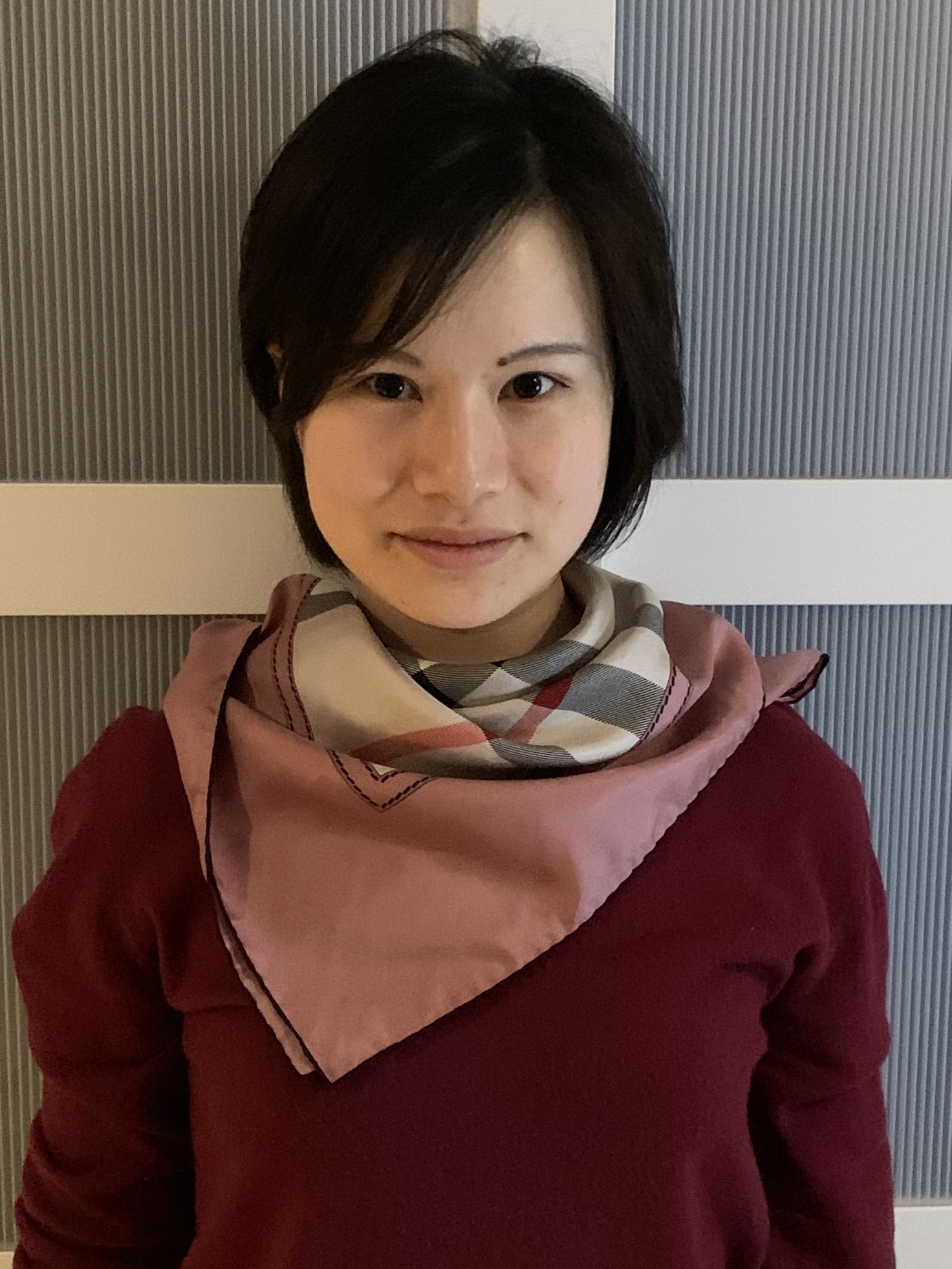}
  \caption{Picture of Bingqing Cheng}
\end{figure*}

Bingqing Cheng is a Departmental Early Career Fellow at the Computer Laboratory, University of Cambridge,
and 
a Junior Research Fellow at Trinity College.
She received her Ph.D. from the École polytechnique fédérale de Lausanne (EPFL) in 2019.
Her work focuses on theoretical predictions of material properties.

\begin{figure*}[h!]
\includegraphics[width=2in]{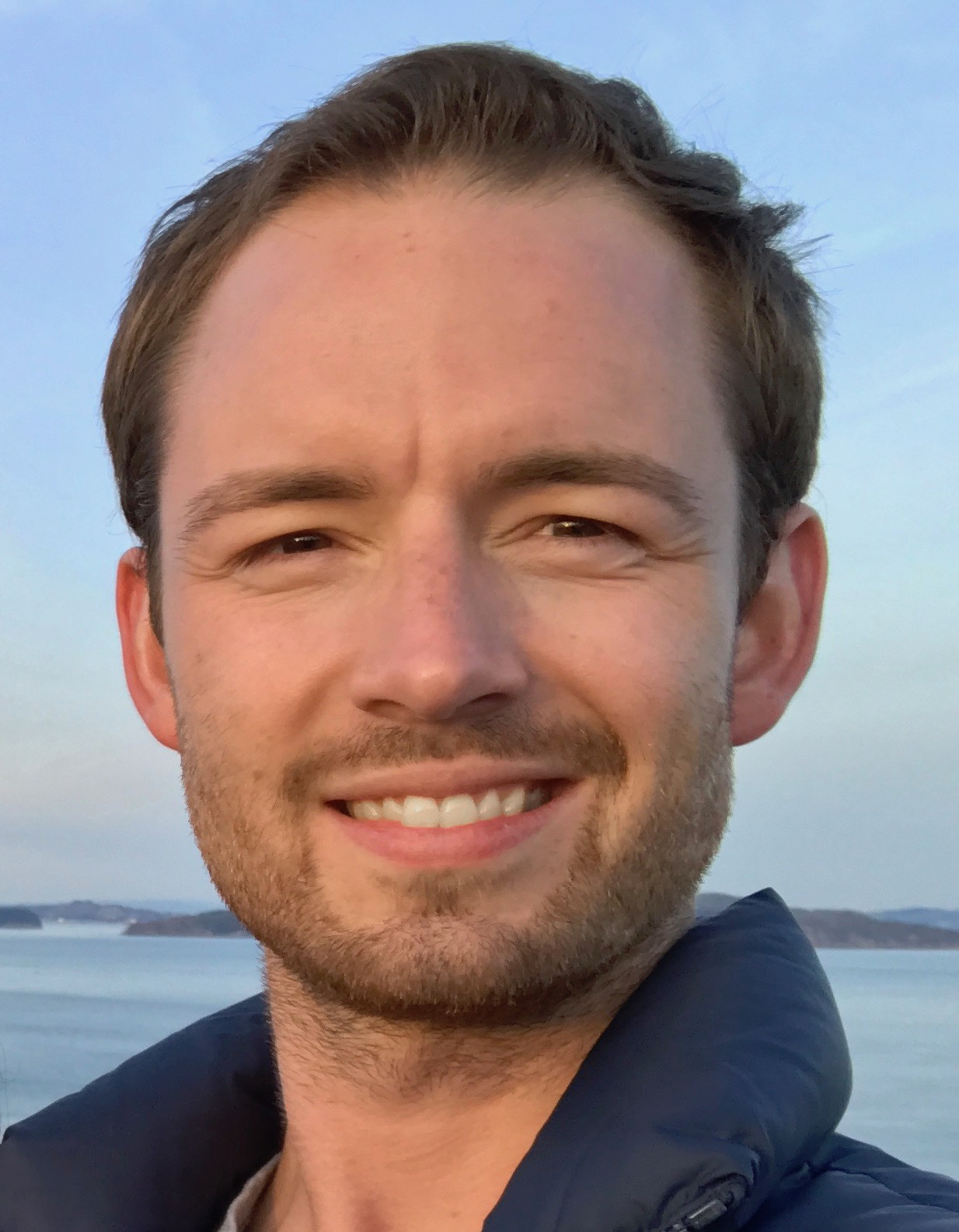}
  \caption{Picture of Stefan Chmiela}
  \label{Stefan_picture}
\end{figure*}

Stefan Chmiela is a senior researcher at the Berlin Institute for the Foundations of Learning and Data (BIFOLD). He received his Ph.D. from Technische Universit\"at Berlin in 2019. His research interests include Hilbert space learning methods for applications in quantum chemistry, with particular focus on data efficiency and robustness.

\begin{figure*}[h!]
\includegraphics[width=2in]{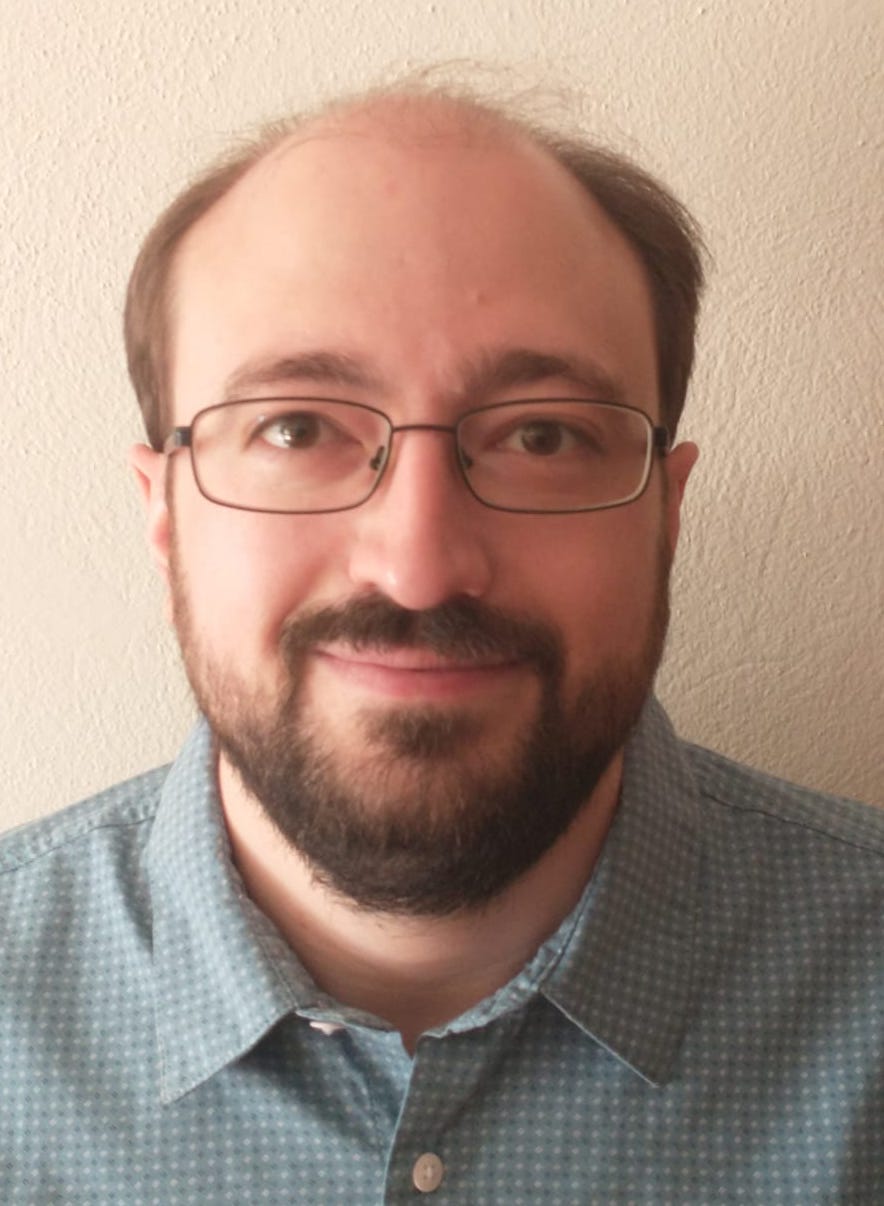}
  \caption{Picture of Michael Gastegger}
  \label{Michael_picture}
\end{figure*}

Michael Gastegger is a postdoctoral researcher in the BASLEARN project of the Machine Learning Group at Technische Universit\"at Berlin. He received his Ph.D. in Chemistry from the University of Vienna in Austria in 2017. His research interests include the development of machine learning methods for quantum chemistry and their application in simulations.

\begin{figure*}[h!]
\includegraphics[width=2in]{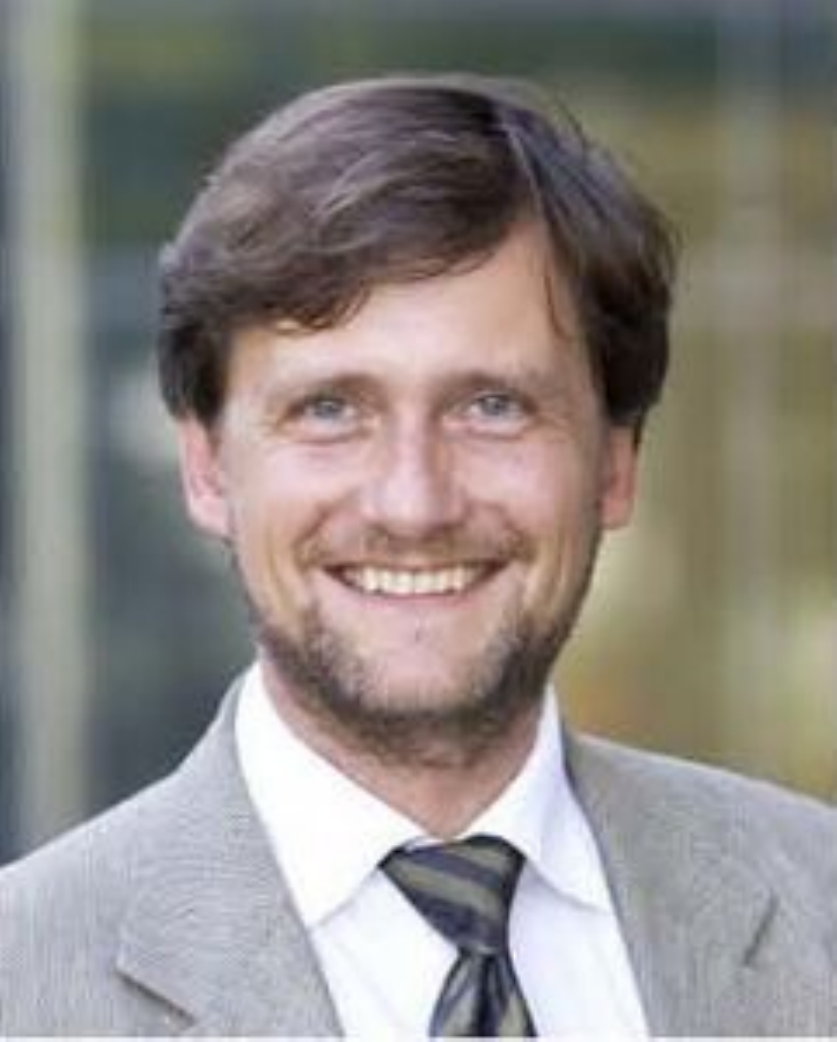}
  \caption{Picture of Klaus-Robert M\"{u}ller}
  \label{KRM_picture}
\end{figure*}

Klaus-Robert M{\"u}ller  has been a professor of computer science at Technische Universit{\"a}t Berlin since 2006; at the same time he is directing and co-directing the Berlin Machine Learning Center and the Berlin Big Data Center, respectively. He studied physics in Karlsruhe from 1984 to 1989 and obtained his Ph.D. degree in computer science at Technische Universit{\"a}t Karlsruhe in 1992. After completing a postdoctoral position at GMD FIRST in Berlin, he was a research fellow at the University of Tokyo from 1994 to 1995. In 1995, he founded the Intelligent Data Analysis group at GMD-FIRST (later Fraunhofer FIRST) and directed it until 2008. From 1999 to 2006, he was a professor at the University of Potsdam. He was awarded the Olympus Prize for Pattern Recognition (1999), the SEL Alcatel Communication Award (2006), the Science Prize of Berlin by the Governing Mayor of Berlin (2014), the Vodafone Innovations Award (2017). In 2012, he was elected member of the German National Academy of Sciences-Leopoldina, in 2017 of the Berlin Brandenburg Academy of Sciences and also in 2017 external scientific member of the Max Planck Society. In 2019 and 2020 he became a Highly Cited researcher in the cross-disciplinary area. His research interests are intelligent data analysis and Machine Learning in the sciences (Neuroscience (specifically Brain-Computer Interfaces), Physics, Chemistry) and in  industry.
\begin{figure*}[h!]
\includegraphics[width=2in]{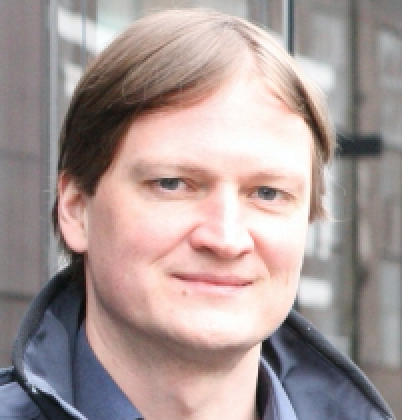}
  \caption{Picture of Alexandre Tkatchenko}
  \label{AT_picture}
\end{figure*}

Alexandre Tkatchenko is a Professor of Theoretical Chemical Physics at the University of Luxembourg and Visiting Professor at Technische Universit\"at Berlin. He obtained his bachelor degree in Computer Science and a Ph.D. in Physical Chemistry at the Universidad Autonoma Metropolitana in Mexico City. Between 2008 and 2010, he was an Alexander von Humboldt Fellow at the Fritz Haber Institute of the Max Planck Society in Berlin. Between 2011 and 2016, he led an independent research group at the same institute. Tkatchenko serves on editorial boards of two society journals: Physical Review Letters (APS) and Science Advances (AAAS). He received a number of awards, including elected Fellow of the American Physical Society, the 2020 Dirac Medal from WATOC, the Gerhard Ertl Young Investigator Award of the German Physical Society, and two flagship grants from the European Research Council: a Starting Grant in 2011 and a Consolidator Grant in 2017. His group pushes the boundaries of quantum mechanics, statistical mechanics, and machine learning to develop efficient methods to enable accurate modeling and obtain new insights into complex materials.